\documentclass{emulateapj}
\usepackage{graphicx}
\usepackage{epstopdf}

\newcommand{\Mjup}{\mbox{$M_\mathrm{Jup}$}}
\newcommand{\Msun}{\mbox{$M_{\odot}$}}
\newcommand{\Mearth}{\mbox{$M_{\earth}$}}

\begin{document}
\title{Planets Around Low-Mass Stars (PALMS). IV.  \\ The Outer Architecture of M Dwarf Planetary Systems*}
\author{Brendan P. Bowler,\altaffilmark{1,2,3,4} 
Michael C. Liu,\altaffilmark{2} 
Evgenya L. Shkolnik,\altaffilmark{5}
Motohide Tamura\altaffilmark{6}
\\ }
\email{bpbowler@caltech.edu}

\altaffiltext{1}{California Institute of Technology, Division of Geological and Planetary Sciences, 1200 E. California Blvd., Pasadena, CA 91101 USA.}
\altaffiltext{2}{Institute for Astronomy, University of Hawai`i; 2680 Woodlawn Drive, Honolulu, HI 96822, USA}
\altaffiltext{3}{Caltech Joint Center for Planetary Astronomy Fellow.}
\altaffiltext{4}{Visiting Astronomer at the Infrared Telescope Facility, which is operated by the University of Hawaii under 
Cooperative Agreement no. NNX-08AE38A with the National Aeronautics and Space Administration, Science Mission Directorate, Planetary Astronomy Program.}
\altaffiltext{5}{Lowell Observatory, 1400 W. Mars Hill Road, Flagstaff, AZ 86001}
\altaffiltext{6}{National Astronomical Observatory of Japan, 2-21-1 Osawa, Mitaka, Tokyo 181-8588, Japan}
\altaffiltext{*}{Some of the data presented herein were obtained at the W.M. Keck Observatory, which is operated as a scientific partnership 
among the California Institute of Technology, the University of California and the National Aeronautics and Space Administration. 
The Observatory was made possible by the generous financial support of the W.M. Keck Foundation.  This work was also based 
on data collected at Subaru Telescope, which is operated by the National Astronomical Observatory of Japan.}

\submitted{ApJS, Accepted (Nov 6 2014)}

\begin{abstract}

We present results from a high-contrast adaptive optics imaging search for giant planets and 
brown dwarfs ($\gtrsim$1~\Mjup) around 122 newly identified nearby ($\lesssim$40~pc) young M dwarfs.  
Half of our targets are younger than 135~Myr and 90\% are younger than the Hyades (620~Myr).  
After removing 44 close stellar binaries (implying a stellar companion fraction of $>$35.4~$\pm$~4.3\% within 100~AU), 
27 of which are new or spatially resolved for the first time, our remaining sample of 78 single   
M dwarfs makes this the largest imaging search for planets around young low-mass stars (0.1--0.6~\Msun) to date.
Our $H$- and $K$-band coronagraphic observations with Keck/NIRC2 and Subaru/HiCIAO 
achieve typical contrasts of 12--14~mag and 9--13~mag at 1$''$, respectively, which corresponds
to limiting planet masses of 0.5--10~\Mjup \ at 5--33~AU for 85\% of our sample.
We discovered four young brown dwarf companions: 
1RXS~J235133.3+312720~B (32~$\pm$~6~\Mjup; L0$^{+2}_{-1}$; 120~$\pm$~20~AU), 
GJ~3629~B (64$^{+30}_{-23}$~\Mjup; M7.5~$\pm$~0.5; 6.5~$\pm$~0.5~AU), 
1RXS~J034231.8+121622~B (35~$\pm$~8~\Mjup; L0~$\pm$~1; 19.8~$\pm$~0.9~AU), and 
  2MASS~J15594729+4403595~B (43~$\pm$~9~\Mjup; M8.0~$\pm$0.5; 190~$\pm$~20~AU).  
  Over 150 candidate planets were identified; we obtained follow-up imaging
  for 56\% of these but all are consistent with background stars.
Our null detection of planets enables strong statistical constraints on the 
occurrence rate of long-period giant planets around single M dwarfs. 
We infer an upper limit (at the 95\% confidence level) 
of 10.3\% and 16.0\% for 1--13~\Mjup \ planets between 10--100~AU 
for hot-start and cold-start (Fortney) evolutionary models, respectively.
Fewer than 6.0\% (9.9\%) of M dwarfs harbor massive gas giants in the 5--13~\Mjup \ range like 
those orbiting HR~8799 and $\beta$~Pictoris between 10--100~AU for a hot-start (cold-start) 
formation scenario.   The frequency of brown dwarf (13--75~\Mjup)  companions to single 
M dwarfs between 10--100~AU is 2.8$^{+2.4}_{-1.5}$\%.  Altogether we find that giant planets, 
especially massive ones, are rare in the outskirts of M dwarf planetary systems.
Although the first directly imaged planets were found around massive stars,
there is currently no statistical evidence for a trend of giant planet frequency with 
stellar host mass at large separations as predicted by the disk instability model of 
giant planet formation.

\end{abstract}

\section{Introduction}{\label{sec:intro}}

M dwarfs with masses between 0.1--0.6~\Msun \ constitute the peak of the initial mass function and vastly 
outnumber all earlier-type stars put together.
In the solar neighborhood they make up $\approx$75\% of stars (\citealt{Henry:2006p18988}; \citealt{Kirkpatrick:2012p24092}), 
which is a good estimate for their galactic-wide rate (\citealt{Bochanski:2010p23010}), and there is some evidence that M dwarfs 
represent even larger fractions of stellar populations in evolved galaxies (\citealt{vanDokkum:2010p23013}; \citealt{Conroy:2012p25477}).
Their abundance and relatively low close binary fractions 
($\approx$30\%; \citealt{Fischer:1992p18426}; \citealt{Delfosse:2004p24266}; \citealt{Janson:2012p23979}; \citealt{Dieterich:2012p24214})
mean that low-mass stars may also be the most common sites of planet formation (\citealt{Lada:2006p14060}).  

At small separations ($\lesssim$2~AU) where radial velocity and transit techniques are most sensitive,
the frequency of giant planets between $\sim$1--10~\Mjup \ has been found to be relatively low around single 
M dwarfs (2.5~$\pm$~0.9\%)  compared to high-mass A-type stars (11~$\pm$~2\%; \citealt{Johnson:2010p20950}).
This well-established trend between planet occurrence rate and stellar host mass  
(\citealt{Butler:2004p13884}; \citealt{Endl:2006p19580}; \citealt{Butler:2006p19581}; \citealt{Johnson:2007p169}; \citealt{Lovis:2007p17712}; 
\citealt{Cumming:2008p9188}; \citealt{Bowler:2010p19983}; \citealt{Bonfils:2013p25485}; \citealt{Gaidos:2013p25312})
lends support to the core accretion plus migration model of planet formation (\citealt{Pollack:1996p19730}; \citealt{Alibert:2005p17987}),
 which predicts fewer gas giants around M dwarfs as a result of lengthened timescales for planetesimal growth 
(\citealt{Laughlin:2004p19937}; \citealt{Ida:2005p18002}; \citealt{Kennedy:2008p18349}).  

On the other hand, recent radial velocity and transit surveys are showing that Earth- to Neptune-sized planets
not only exist in this stellar mass regime (e.g., \citealt{Udry:2007p25519}; \citealt{Mayor:2009p25518}; 
\citealt{Charbonneau:2009p19980}; \citealt{Muirhead:2012p23531}) 
but appear to be quite common (\citealt{Bonfils:2013p25485}; \citealt{Berta:2013p25285}).
In particular, \citet{Swift:2013p25517} and \citet{Dressing:2013p24825} find that the average rate of small planets 
from $Kepler$ is about one per star for 
periods shorter than 50 days, implying a vast galaxy-wide presence of rocky planets (\citealt{Morton:2014bf}).
This in turn has generated increasing interest in the habitability of planets around M dwarfs since
the nearest examples of habitable Earths may orbit low-mass stars (e.g., \citealt{Joshi:1997ig}; \citealt{Cantrell:2013it}; \citealt{Quintana:2014cc}).
 
Far less is known about planets at moderate separations of $\sim$2--10~AU.  Although microlensing 
probes the full range of planetary masses in this region (\citealt{Gould:1992p25589}), the masses and metallicities of the host
stars are usually poorly constrained with this technique and so are of limited value for statistical constraints.  
The lensing signal from the star itself becomes very weak beyond
projected separations of $\sim$10~AU (\citealt{Han:2006p23913}; \citealt{Han:2009p23894}),
leading to an ambiguity between isolated planetary-mass objects and bound planets on wide orbits (\citealt{Sumi:2011p22269}).
Nevertheless, initial statistical results point to a large reservoir of planets orbiting M dwarfs at moderate separations. 
\citet{Gould:2010p23021} find that the frequency of planets in the ice giant to gas giant range ($\gtrsim$0.05~\Mjup)
is a factor of 8 time larger than those from Doppler studies at small separations.
In a follow-up study, \citet{Cassan:2012p23478} measure a frequency of 17$^{+6}_{-9}$\% (52$^{+22}_{-29}$\%) 
for 0.3--10~\Mjup \ (10--30~$M_{\earth}$) planets between 
0.5--10~AU.  Across the entire range of sensitivity (10~$M_{\earth}$--10~\Mjup, 0.5--10~AU), 
these occurrence rates
imply that M dwarfs harbor on average 1.6$^{+0.7}_{-0.9}$ planets per star.
This result was recently bolstered by \citet{Clanton:2014hr}, who found that the total number
of 1--10$^4$ $\Mearth$ planets with periods of 1--10$^4$ days is 1.9~$\pm$~0.5 by
combining statistical results from radial velocity and microlensing surveys.

Another form of planet population statistical analysis in this intermediate-separation regime comes from combining long-baseline 
radial velocity monitoring with adaptive optics imaging.  
\citet{Montet:2014fa} apply this method to old M dwarfs in the field and find a frequency of 6.5~$\pm$~3\% for 1--13~\Mjup \ planets within 20~AU, which is consistent with microlensing results over the same region.

Beyond $\sim$10~AU, direct imaging is the best way to study the outer architecture of planetary systems.
Following the discoveries of planets orbiting the A-type stars HR~8799, Fomalhaut, 
and $\beta$~Pic (\citealt{Marois:2008p18841}; \citealt{Kalas:2008p18842}; 
\citealt{Marois:2010p21591}; \citealt{Lagrange:2010p21645}),
high-mass stars have received the most attention in direct imaging planet searches
(\citealt{Ehrenreich:2010p22443}; \citealt{Janson:2011p22503}; 
\citealt{Vigan:2012p24691}; \citealt{Rameau:2013it}; \citealt{Nielsen:2013jy}).
Yet despite their prevalence in the galaxy, imaging surveys have mostly neglected 
low-mass stars, so little is known about the 
the demographics of gas giants on wide orbits around M dwarfs.
This is largely due to a dearth of known nearby young M dwarfs, a population that
has been substantially enlarged over the past few years (\citealt{Shkolnik:2009p19565};
\citealt{Shkolnik:2012p24056}; \citealt{Schlieder:2012p25080}; \citealt{Malo:2013p24348}; \citealt{Rodriguez:2013fv}; \citealt{Malo:2014dk}).
Low mass stars are also optically faint and typically result in poorer AO performance
than their brighter, earlier-type counterparts.
Furthermore, few of the surveys that have incorporated M dwarfs expressly vetted close binaries from their statistical analyses, which is crucial
if the results are to be compared with radial velocity planet searches of single stars.
A handful of surveys sensitive to 1--10~\Mjup \ companions have targeted single, young, M0--M5 stars:  
\citet{Biller:2007p19401} observed 12 targets with VLT/MMT Simultaneous Differential Imaging (SDI), 
\citet{Lafreniere:2007p17991} imaged 16 stars with Gemini-North/NIRI,
\citet{Chauvin:2010p20082} imaged 16 single M dwarfs with VLT/NaCo, 
\citet{Delorme:2012p23484} targeted 12 stars with VLT/NaCo in $L'$ band, 
and \citet{Biller:2013fu} observed 35 single M dwarfs with Gemini-South/NICI.\footnote{Other 
imaging programs that have also observed single young M dwarfs 
with ground-based adaptive optics or the \emph{Hubble Space Telescope} 
have primarily been sensitive to brown dwarfs at wide separations, rarely reaching 1--5~\Mjup \ limits at small separations of $\sim$10~AU
(\citealt{McCarthy:2004p18279}; \citealt{Lowrance:2005p18287}; \citealt{Daemgen:2007p19541}; \citealt{Allen:2008p19542}).}

The aim of the Planets Around Low-Mass Stars (PALMS) survey is to find young giant planets and brown dwarfs for 
spectroscopic characterization and to measure the frequency of gas giants orbiting M dwarfs beyond 10~AU.
In \citet{Bowler:2012p23851} and \citet{Bowler:2012p23980} we discovered two new brown dwarf 
companions to young M dwarfs in our sample.\footnote{As part of a
complementary imaging survey targeting a much larger sample of young M dwarfs with shorter exposures, 
we have also discovered the young L-type companion 2MASS~J01225093--2439505~B
which has a mass at the deuterium-burning limit (\citealt{Bowler:2013p25491}).}
In this paper we present two additional substellar companion discoveries and the statistical analysis of our entire sample.
Below we describe our target selection, observations, processing pipeline, discoveries, survey statistical analysis,
and implications for giant planet formation around low-mass stars.

\section{Target Selection}

Our targets are selected primarily for their youth and proximity
in order to achieve the highest sensitivity to giant planets at small separations.
Previously known visual binaries with physical separations $\lesssim$100~AU have been excluded
since moderate-separation ($\sim$5--100~AU) binaries disperse protoplanetary disks
on rapid timescales (\citealt{Duchene:2010p25488}; \citealt{Kraus:2012p23303}), 
limiting the raw ingredients of planet formation 
and diminishing the region of dynamically-stable orbits in these systems.  
In addition, we have specifically designed our survey 
to compare with statistical results from radial velocity programs, which 
generally discard close binaries from their samples.
We have also prioritized targets not previously observed in direct imaging surveys to minimize target selection biases and
increase the chances of new discoveries.


\begin{deluxetable}{lccc}
\tabletypesize{\small}
\tablewidth{0pt}
\tablecolumns{4}
\tablecaption{Adopted Ages for Young Moving Group Members\label{tab:ymgages}}
\tablehead{
        \colhead{Moving Group}        & \colhead{Targets}       &   \colhead{Age}   &  \colhead{Age Ref}    
                 }   
\startdata
TWA            &    1       &    8 $\pm$ 2 Myr  & 1, 2, 3, 4, 5 \\
$\beta$ Pic  &  8        &   23 $\pm$ 3 Myr  &  6, 7, 8, 9 \\
Carina, Columba     &  5   &  30 $\pm$ 5 Myr &  10 \\
Tuc-Hor  &  3   &  35 $\pm$ 5 Myr &  10, 11, 12, 13 \\
Argus  &  6   &  40 $\pm$ 5 Myr  & 10, 14, 15 \\
AB Doradus  &  10  &  120 $\pm$ 10 Myr  &  16, 17, 18, 19 \\
Castor  &  3   &  400 $\pm$ 100 Myr  &  20, 21, 22, 23 \\
Ursa Major  & 6   &  500 $\pm$ 100 Myr & 24, 25, 26 \\
Hyades  &  4  &  620 $\pm$ 30 Myr  &  27, 28, 29, 30 
\enddata
\tablerefs{
(1) \citet{Webb:1999p23921}, 
(2) \citet{BarradoYNavascues:2006p22206}
(3) \citet{Mamajek:2005p19635}, 
(4) \citet{Torres:2006p19650},
(5) \citet{Weinberger:2013p25495}, 
(6) \citet{Yee:2010p21748},
(7) \citet{Binks:2014gd},
(8) \citet{Malo:2014bw},
(9) \citet{Mamajek:2014bf},
(10) \citet{Torres:2008p20087}, 
(11) \citet{Torres:2000p22894}
(12) \citet{Zuckerman:2001p25508}, 
(13) \citet{Kraus:2014ur}, 
(14) \citet{Torres:2003p25520}, 
(15) \citet{DeSilva:2013p25496},
(16) \citet{Zuckerman:2004p22744}, 
(17) \citet{Ortega:2007p22746}, 
(18) \citet{Luhman:2005p22437}, 
(19) \citet{Barenfeld:2013p24822},
(20) \citet{BarradoyNavascues:1998p23331}, 
(21) \citet{Torres:2002p23332}, 
(22) \citet{Ribas:2003p23325}, 
(23) \citet{Mamajek:2012p25026}, 
(24) \citet{Eggen:1983p25584}, 
(25) \citet{Soderblom:1993p25585}, 
(26) \citet{King:2003p22818}), 
(27) \citet{Perryman:1998p25578}, 
(28) \citet{Eggen:1998p25535}, 
(29) \citet{Lebreton:2001p25579},
(30) \citet{Degennaro:2009p18447}.
}
\end{deluxetable}


\begin{figure*}
  \hskip .2in
  \vskip -1.5in
  \resizebox{7in}{!}{\includegraphics{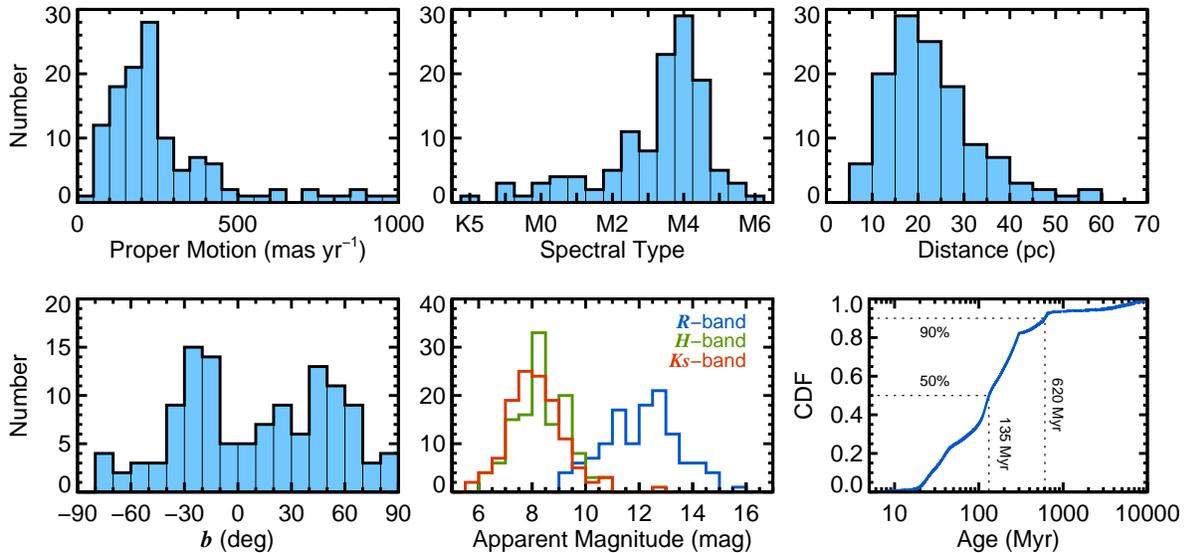}}
  \vskip -.6in
  \caption{Summary of our target sample.  The proper motions of our targets are high enough ($\gtrsim$50~mas/yr) to 
  distinguish comoving companions from background objects on short timescales ($\sim$1~yr).  Spectral types peak at M4 and range from K5 to M6.
  Most of the distances to our targets are between 10--40~pc, and we have prioritized high galactic latitudes to avoid fields with high background stellar densities.
  Nearly all targets are bright enough for NGS observations either at Subrau or Keck ($R$$<$15~mag).  A single target, NLTT~13844, was observed
  with LGS-AO.  The bottom right panel
  shows the cumulative distribution of ages for our sample.  50\% are younger than 135~Myr and 90\% of our targets
  are younger than 620~Myr.  \label{fig:targethist} }
\end{figure*}

Among our 122 targets, 69 originate from a recent search for nearby young M dwarfs
by \citet{Shkolnik:2009p19565} and \citet{Shkolnik:2012p24056}.
Motivated by the dearth of known low-mass members of young moving groups (YMGs),
\citet{Shkolnik:2009p19565}  identified 144 X-ray active M dwarfs with distances $\lesssim$30~pc
and ages of $\sim$10--300~Myr.  In a follow-up study,  \citet{Shkolnik:2012p24056} obtained
parallaxes for about half of these systems and found several dozen probable kinematic members of 
young moving groups.
Targets from \citet{Shkolnik:2009p19565} have been vetted for close spectroscopic binaries with few-day periods 
(\citealt{Shkolnik:2010p20931}), which also produce
activity as a result of rotationally-enhanced dynamo activity caused by tidal locking 
(e.g., \citealt{Torres:2002p25510}; \citealt{Kraus:2011p21976}).

Another 42 targets in our sample are drawn from an ongoing, complementary search for nearby young M dwarfs
using $GALEX$ data (\citealt{Shkolnik:2011p21923}; Shkolnik et al., in preparation).  
Among these, ten systems have been kinematically
tied to YMGs by \citet{Malo:2013p24348}, \citet{Lepine:2009p19553}, and \citet{Riedel:2014ce} and two
new candidate members are identified in this work (LHS~2613 and NLTT~48651).
Seven other systems  
(LHS1864~AB, NLTT~26359, LHS~2672, G~202-48, GJ~3997~AB; LP~447-38~AB; LHS~3321)
either show H$\alpha$ absorption or have red $NUV$--$W1$ colors ($>$13~mag) compared to YMG members (\citealt{Rodriguez:2013fv}).
These targets appear to be old inactive field stars that passed early $NUV$ selection cuts, so 
we adopt minimum ages from the activity-lifetime relations of \citet{West:2008p19562}.  
One system, 2MASS~J04220833--2849053~AB, 
has its age constrained from the detection of \ion{Li}{1}$\lambda$6708 absorption by \citet{Torres:2006p19650} (see Section~\ref{sec:indnotes}).
Similarly, the age of 2MASS~J15594729+4403595 is constrained from signatures of low gravity in
the spectrum of its substellar companion (Section \ref{sec:2m1559}).
The remaining 21 of these 42 targets show photometric and spectroscopic indications 
of youth similar to known young moving group members.
A detailed analysis of their ages, including a discussion of high resolution optical spectroscopy for these targets,
will be presented in a forthcoming paper (Shkolnik et al., in preparation).  
For this work we adopt conservative age ranges of 10--300~Myr similar to \citet{Shkolnik:2009p19565}
for the 21 targets without age estimates in the literature.

Finally, 11 targets are compiled from the literature from recent searches for M dwarf
members of YMGs.  Six originate from \citet{Schlieder:2012p23766}, \citet{Schlieder:2012p23477},  and \citet{Schlieder:2012p25080},
while another five are from \citet[TWA~30A]{Looper:2010p20583}, \citet[AP~Col]{Riedel:2011p22580}, \citet[L~449-1~AB]{Scholz:2005p22579}, 
\citet[TYC~7443-1102-1]{Lepine:2009p19553}, and \citet[GJ~354.1~B]{Lowrance:2005p18287}.

Where available, age estimates and YMG memberships have been taken from the literature.
Altogether, 46 targets (38\% of our total sample) are associated with YMGs.
Ages for YMG members (or likely members) are listed in Table~\ref{tab:ymgages}.\footnote{Recently the ages of several of the youngest 
moving groups have been called into question from Li depletion boundary measurements.  
For example, \citet{Binks:2014gd} find an older age of 21~$\pm$~4 for the $\beta$~Pic YMG from its Li-depletion boundary 
compared to its isochronal age of $\approx$12~Myr.  These results are bolstered by recent studies by \citet{Malo:2014bw}
and \citet{Mamajek:2014bf}.  Similarly, \citet{Kraus:2014ur} infer a Li-depletion age of $\approx$40~Myr for
the Tuc-Hor moving group, which is roughly 10~Myr older than its age from isochrone fitting.
Here we adopt the more recently-determined and internally consistent ages of 23~$\pm$~3~Myr for the $\beta$ Pic MG
and 35~$\pm$~5~Myr for Tuc-Hor.} 
Five systems are kinematically linked to young moving groups here for the first time: 
LHS~2613 (Argus), 1RXS~J022735.8+471021 (AB~Dor), NLTT 48651 (AB~Dor), GJ~354.1~B (Carina), and G~227-22 (UMa).

Figure~\ref{fig:targethist} and Table~2 summarize the properties of our sample.
Proper motions mostly originate from the UCAC4 database (\citealt{Zacharias:2013p24823}) 
and generally fall between 100--500~mas~yr$^{-1}$, which is high enough so that background stars
can be distinguished from \emph{bona fide} comoving companions on timescales of about one year.  
Spectral types are compiled from the literature and range from K5 to M6 ($\approx$0.2--0.6 \Msun), 
with most of the sample falling between M3 and M5.
69 targets (57\% of the sample) have parallactic distances.  For the rest, we have either adopted photometric distances (42 targets) or kinematic distances 
based on young moving group memberships (11 targets) from the literature (see Table~2 for details).
97 targets (80\% of the sample) are within 30~pc and 114 (93\% of the sample) are within 40~pc.

When possible we avoided stars with low galactic latitudes where background contamination rates are high.  
Targets near the galactic plane were generally only observed if an RA gap existed in the target list
for any particular night.
This preference is reflected in the relative dearth of targets for $|b| \lesssim$ 20$^{\circ}$ in Figure~\ref{fig:targethist}.
The distribution of $R$-band magnitudes ranges from $\approx$10--15~mag and is roughly divided 
into two bins according to observability with NGS-AO at Subaru ($\lesssim$13~mag) and Keck  ($\lesssim$15~mag).
The cumulative age distribution of our sample is shown in Figure~\ref{fig:targethist}: 50\% of stars
are younger than 135~Myr and 90\% are younger than 620~Myr.

\section{Observations}

\subsection{Keck~II/NIRC2 Adaptive Optics Imaging}

We carried out our observations at the Keck~II 10 m telescope with the facility near-infrared imaging camera NIRC2  
using natural guide star adaptive optics (NGS-AO; \citealt{Wizinowich:2000p21634}) between August 2010 and August 2013 (Table~3).
A single target, NLTT~13844 ($R$$\sim$14.8~mag), was observed with laser guide star AO
(LGS-AO; \citealt{Wizinowich:2006p25357}; \citealt{vanDam:2006hw}).
Most of our imaging was carried out with the narrow camera, which has 
a plate scale of 9.952~$\pm$~0.002 mas~pix$^{-1}$ (\citealt{Yelda:2010p21662}) 
and provides Nyquist sampling at the diffraction limit beyond $\sim$1.2~$\mu$m.
In this mode the field of view (FOV) across the array's 1024~$\times$~1024 pixels
is 10$\farcs$2~$\times$~10$\farcs$2.
When conditions were good (seeing below $\sim$1$''$) we used the Mauna Kea Observatory 
(MKO) $H$-band filter (\citealt{Simons:2002p20490}; \citealt{Tokunaga:2005p18542}) 
as a compromise between higher Strehl and increased sky background levels at longer wavelengths.
When conditions were below average, we used the $K_S$ filter to benefit from better AO correction.  

We first obtained short,
unsaturated images of each target to check for stellar multiplicity.  Binary systems were generally skipped, although
in a few cases close companions were only resolved in our second-epoch imaging.
For single stars we typically obtained 40~min of total on-source integration time 
(usually 40 frames each with 60-sec exposures and 1~coadd reading out with multiple correlated double sampling) 
in Angular Differential Imaging mode (ADI; \citealt{Liu:2004p17588}; \citealt{Marois:2006p18009})
after centering the target behind
the partly opaque ($\Delta$$H$$\sim$6~mag) 600~mas diameter coronagraph.
To avoid the lower-left quadrant of NIRC2, which suffers from elevated noise levels, we positioned the 
coronagraph at column 616  (the occulting spot is fixed in $y$ at row 430).
Raw images were first cleaned of bad pixels and cosmic rays then flat-fielded to remove pixel-to-pixel
sensitivity variations.

This results in an inner working angle (IWA) of 300~mas and an outer working angle (OWA) 
between $\approx$4$''$ (for complete spatial coverage)
and 8$\farcs$5 (for partial coverage).
The NIRC2 coronagraph is particularly useful for image registration and photometric calibration since the
star is visible behind the mask.  
Corrections for differential atmospheric refraction were applied during
most of the observations to keep the star centered behind the coronagraph.  
To further monitor the quality of AO correction, we 
also obtained a set of unsaturated frames immediately before and after our ADI sequences.

The scheduling of our ADI observations were optimized to maximize rotation at small separations ($\sim$0$\farcs$5)
but minimize blurring at modest separations ($\sim$3$''$).  This compromise ensures that physical companions
will have undergone enough rotation on the detector to avoid strong self-subtraction during post-processing.  
This strategy also reduces sensitivity losses
at several arcseconds caused by smearing when rotation rates are high, which can occur near transit for 
declinations near the observing site's latitude 
(+20$^{\circ}$ for Mauna Kea; see \citealt{Biller:2008p19421} for a detailed discussion of these effects).
In practice it is difficult to strictly adhere to these constraints, but most of the field-of-view rotations 
are near the desired values of $\sim$15--40$^{\circ}$.

Follow-up second-epoch observations were carried out in ADI mode, standard imaging mode with the telescope rotator on,
and with the wide (0$\farcs$04 pix$^{-1}$; 40$''$ FOV) and narrow NIRC2 camera modes depending on the 
candidate being recovered.
Integration times were generally much shorter than first-epoch exposures since it is often not
necessary to achieve the same limiting depth to recover faint candidates.

For the NIRC2 narrow camera, we adopt the plate scale measurement of 9.952~$\pm$~0.002 mas~pix$^{-1}$ and 
orientation of +0$\fdg$252~$\pm$~0$\fdg$009 found by \citet{Yelda:2010p21662}.  
Post-fit residuals for the NIRC2 narrow camera distortion solution made available by Keck Observatory 
are $\approx$0.6~mas (B. Cameron, 2007,  private communication).

\subsection{Subaru/HiCIAO Adaptive Optics Imaging}

Our NGS-AO observations at the 8.2-m Subaru Telescope were obtained with the AO188 adaptive optics system (\citealt{Hayano:2010p25015})
coupled with the High Contrast Instrument for the Subaru Next Generation Adaptive Optics 
(HiCIAO) imaging instrument (\citealt{Hodapp:2008p21642}; \citealt{Suzuki:2010p21643}; Table~3).
Our observing strategy with HiCIAO was similar to that with NIRC2.  ADI observations were
carried out with the star centered behind the 300~mas diameter opaque Lyot coronagraph.
The $H$ (MKO) and $K_S$ filters were used for our deep imaging 
with typical on-source integration times of 40~min (1~coadd $\times$ 60~s $\times$ 40 frames).
Sets of short unsaturated frames were taken before, in the middle, and after each ADI sequence
to monitor AO correction and photometrically calibrate our data.
The atmospheric dispersion corrector for the AO188 (\citealt{Egner:2010p25586}) was employed
to minimize drifting caused by changing airmass.
With a plate scale of 9.723$\pm$~0.011~mas~pix$^{-1}$ in $H$ band (Section~\ref{sec:append1}), 
the 2048~$\times$~2048 pixel HAWAII-2RG detector corresponds to a 
field of view of 20$\farcs$5.  
For our deep coronagraphic data the IWA is 0$\farcs$2 and the OWA
is $\approx$10$''$ ($\approx$14$''$) for full (partial) coverage.
Dome flats and bias frames were obtained at the start and end of each observing run.

The raw HiCIAO images suffer from horizontal and vertical electronic readout structure imprinted in each image, 
which corresponds to 32 readout channels with different voltages (\citealt{Suzuki:2010p21643}; 
\citealt{Brandt:2013p25443}).
To remove these random and changing bias stripes we use a procedure developed by the Subaru
Strategic Exploration of Exoplanets and Disks (SEEDS; \citealt{Tamura:2006p21640}) team
(R. Kandori 2011, private communication), which is based on measuring and subtracting these patterns in 
the science data itself.
To further remove residual patterns, we subtract the median-combined horizontal and vertical profiles
after masking out the science target in each image.  Together these eliminate nearly all systematic features
caused by the electronics.
After bias subtraction, cosmic rays and bad pixels are removed and the images are divided by a normalized flat field.

Seeing was poor (1--2$''$) during most of our HiCIAO observations.  This significantly degraded the
AO correction and the limiting contrasts for many of our targets.  While we attempted to re-observe the stars
with the worst data sets at Keck, a few of our observations only reach corresponding masses of $\sim$10--20~\Mjup.

For HiCIAO, we adopt the following plate scale measurements, which we found slightly vary with wavelength (see Appendix~\ref{sec:append1}): 
9.81~$\pm$~0.04 mas~pix$^{-1}$ for $Y$ band, 9.75~$\pm$~0.04 mas~pix$^{-1}$ for $J$ band, 
9.723~$\pm$~0.011 mas pix$^{-1}$ for $H$ band, and 9.67~$\pm$~0.03 mas~pix$^{-1}$ for $K_S$ band.
A constant plate scale orientation of 0$\fdg$0~$\pm$0$\fdg$1 is adopted for all of the filters.
Post-fit residuals from the distortion solution are $\approx$1~pix.

\subsection{IRTF/SpeX Near-Infrared Spectroscopy}

We obtained a near-infrared spectrum of the young, 5$\farcs$6-separation substellar companion 
2MASS~J15594729+4403595~B (Section~\ref{sec:2m1559}) with the Infrared Telescope Facility's SpeX spectrograph (\citealt{Rayner:2003p2588})
in short wavelength cross-dispersed (SXD) mode on 2012 August 11 UT in photometric conditions.  A slit width of 0$\farcs$5 yielded a 
resolving power ($R$~$\equiv$~$\lambda$/$\Delta \lambda$) of $\approx$1200 from 0.8--2.5~$\mu$m.
To avoid the host star we oriented the slit perpendicular to the primary star-companion position angle (PA).
We obtained a total of 36~min of integration time by nodding in an ABBA pattern along the slit.
Immediately after we observed the A0V standard 26~Ser at a similar airmass (1.22) for telluric correction.
Flats and arc lamps were acquired at the same sky position.  The spectra were extracted, median-combined,
and telluric-corrected using Spextool reduction package (\citealt{Vacca:2003p497}; \citealt{Cushing:2004p501}).
Table~4 summarizes our spectroscopic observations of three substellar companions in our sample.


\begin{deluxetable*}{lccccccc}
\tablenum{4}
\tabletypesize{\tiny}
\tablewidth{0pt}
\tablecolumns{8}
\tablecaption{Spectroscopic Observations \label{tab:specobs}}
\tablehead{
        \colhead{Object}   &  \colhead{Date}      &   \colhead{Telescope/}  &  \colhead{Filter}  &  \colhead{Slit Width}  & \colhead{Plate Scale}   &    \colhead{Exp. Time}   &    \colhead{Standard}   \\
                     \colhead{}  &  \colhead{(UT)}   &   \colhead{Instrument}       &  \colhead{}           &  \colhead{(")}               & \colhead{(mas pix$^{-1}$)}             &    \colhead{(min)}   &    \colhead{}             
        }   
\startdata
2MASS~J15594729+4403595 B    &    2012 Aug 11   &   IRTF/SpeX-SXD &   $\cdots$ &   0.5        &  150  &   36    &  26 Ser \\
GJ 3629 B                                            &    2013 Feb 1      &   Keck~I/OSIRIS    &  $Jbb$     & $\cdots$ &  20    &  20     &  HD 99960  \\
                                                               &    2013 Feb 1      &   Keck~I/OSIRIS    &  $Hbb$    & $\cdots$ &  20    & 13.5   &  HD 99960  \\
                                                               &    2013 Feb 1      &   Keck~I/OSIRIS    &  $Kbb$    & $\cdots$ &  20    &  10      &  HD 99960  \\
1RXS~J034231.8+121622~B         &    2013 Feb 2      &   Keck~I/OSIRIS     &  $Hbb$   & $\cdots$  & 20    &  24      & HD 31411  \\
                                                              &    2013 Feb 2      &   Keck~I/OSIRIS     &  $Kbb$    & $\cdots$  & 20    &  24      & HD 31411  
\enddata
\end{deluxetable*}

\subsection{Keck/OSIRIS Near-Infrared Spectroscopy}

On 2013 Feb 01 UT and 2013 Feb 02 UT we observed GJ~3629~B and 1RXS~J034231.8+121622~B with the 
OH-Suppressing Infrared Imaging Spectrograph (OSIRIS; \citealt{Larkin:2006p5570}) at the Keck~I telescope
using NGS-AO (Table~4).
Conditions were clear and both targets were observed at a low airmass of $\approx$1.1.
Since the companions are located $<$1$''$ from their host stars, we chose the 20~mas~pixel$^{-1}$ plate scale
to finely sample the PSF structure with a resolving power of $\approx$3800.  
These observations benefited from a new grating installed in OSIRIS in December 2012, increasing the average sensitivity
by a factor of 1.83 compared to its previous performance on Keck~II (\citealt{Mieda:2014dt}).

We acquired $Jbb$-, $Hbb$-, and $Kbb$-band spectra of GJ~3629~B with the 
long axis of the 0$\farcs$32$\times$1$\farcs$28 detector aligned with the primary-companion PA.  
The close separation of this system (0$\farcs$2) ensured that both components were on the detector.
The 1RXS~J034231.8+121622~A-B separation is 0$\farcs$83, so for this system we aligned the
long-axis of the detector perpendicular to the primary-companion PA so that the primary fell off of the
array.  We nodded the telescope along the detector by $\approx$1$''$ in an ABBA pattern 
for pair-wise sky subtraction.
A0V standards were observed in each filter at a similar airmass following both science targets.

Basic image reduction, wavelength calibration, assemblage of the 2D images into 3D spectral cubes, 
and pair-wise sky subtraction was carried out using version 3.2 of the OSIRIS Data Reduction Pipeline
with the latest rectification matrices from February 2013.
The companion and standard star spectra were then extracted from the cubes 
with aperture photometry.  For GJ~3629~B we used aperture radii of 2~spaxels with an annulus of 2.5--4.0~spaxels
to remove some contaminating flux from the nearby host star GJ~3629~A.  No local sky subtraction was applied for
1RXS~J034231.8+121622~B and the standards.  The spectra were corrected for telluric absorption
using the \texttt{xtellcor\_general} routine in the IRTF Spextool reduction package 
(\citealt{Vacca:2003p497}; \citealt{Cushing:2004p501}).  Finally, each band was flux calibrated using
photometry from \citet{Bowler:2012p23980} for GJ~3629~B and from Table~7 of this work for
1RXS~J034231.8+121622~B.


\begin{figure*}
  \vskip -.7in
  \hskip .7in
  \resizebox{6.5in}{!}{\includegraphics{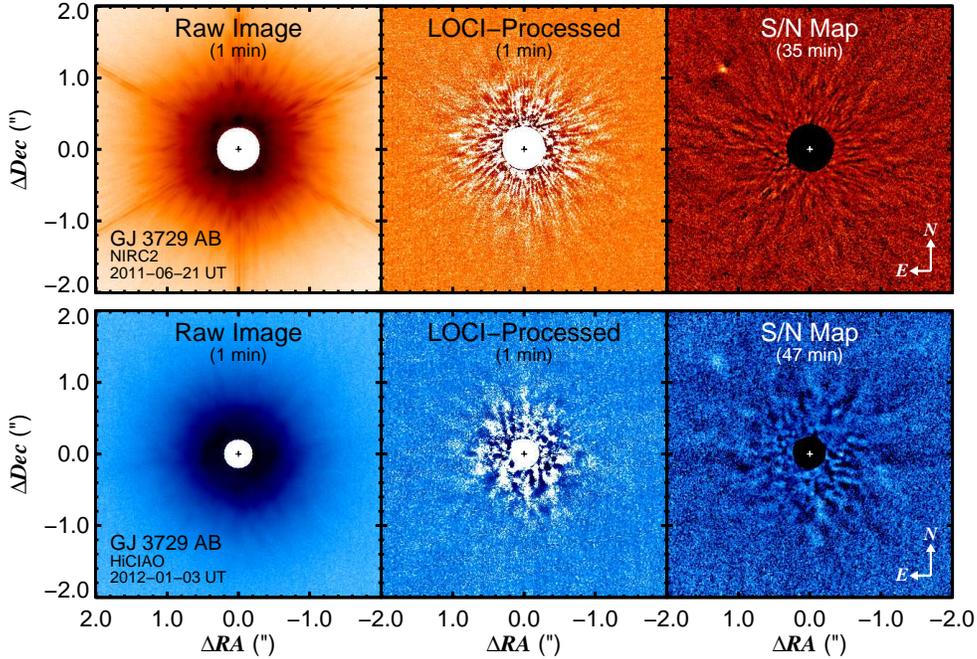}}
  \vskip -.4in
  \caption{Example of ADI processing to reveal faint companions.  The top panels show raw, LOCI-processed, and de-rotated and coadded
  images from NIRC2.  A candidate companion is clearly visible $\sim$2$''$ from the star GJ~3729~AB.  The bottom panel shows the same sequence for 
  observations with HiCIAO about six months later.    \label{fig:lociimgs} } 
\end{figure*}

\section{ADI Processing Pipeline}

We developed a processing pipeline for our ADI data to register the images, 
model and subtract the PSF and speckle pattern for each image, de-rotate and coadd the individual frames, 
identify point sources in the stacked images, compute contrast curves, and derive the sensitivity in 
mass and physical separation using information about the primary star coupled with evolutionary models.
Below we describe each step in detail for a typical ADI sequence consisting of forty 
60-second coronagraphic images and short unsaturated frames.

\subsection{Image Registration}{\label{sec:imgreg}}

For our NIRC2 images we correct for optical distortions using 
the narrow camera distortion solution
made available by Keck Observatory (B. Cameron 2007, private communication), which yields post-fit residuals of $\sim$0.6~mas
in the $x$ and $y$ directions.  For our HiCIAO data we derive distortion solutions from our own observations of the globular cluster M5
taken before and after the installation of a new camera lens in April 2011.  See Appendix~\ref{sec:append1} for details.  
The post-fit residuals across the entire array are $\sim$1~pix (10~mas).  

Images are then registered and assembled into cubes.  For NIRC2 we fit a two-dimensional (2D)
elliptical Gaussian to the star itself, which is visible behind the partly transparent coronagraph.  
For HiCIAO, which has an opaque coronagraph, we infer the position of the star by masking the 
coronagraph and fitting a 2D elliptical Gaussian to the PSF wings.
Sky values are measured, stored, and subtracted from each image after masking the science target.
Accurate accounting of the sky values are especially important for NIRC2. Since the coronagraph is
not completely opaque, photometric calibration using the apparent brightness of the star behind the
mask and the measured transmission of the coronagraph must also account for the background sky value. 
Once the stellar positions are measured, the images are assembled into a cube and aligned 
by shifting to a common position using sub-pixel resampling.
Parallactic angles and north orientations on the detector are stored for later processing.
For NIRC2, the parallactic angle is taken from FITS headers.  For HiCIAO, it is computed 
from the hour angle at the time of the observation, target declination, and latitude of Mauna Kea.

\subsection{PSF and Speckle Subtraction}

The adaptive optics PSF comprises a mixture of 
static structure from the diffraction pattern
and correlated, quasi-static speckle noise from  
imperfect wavefront correction and changing atmospheric conditions
(\citealt{Racine:1999p19415}; \citealt{Marois:2000p19404}; \citealt{Macintosh:2005p20425}; 
\citealt{Hinkley:2007p20367}; \citealt{Oppenheimer:2009p18850}).  
Together these conspire to make the detection of faint point sources difficult in the 
contrast-limited regime, and removing these features requires modeling and subtracting the 
PSF pattern while minimizing the subtraction of actual companions.
Observing strategies based on field of view (FOV) rotation (\citealt{Liu:2004p17588}; \citealt{Marois:2006p18009}) and/or
chromatic dependencies of the PSF pattern (e.g., \citealt{Sparks:2002p19396}; \citealt{Marois:2005p20463}; 
 \citealt{Thatte:2007p20429}; \citealt{Biller:2008p19421}; \citealt{Crepp:2011p21952})
together with more sophisticated processing techniques (\citealt{Lafreniere:2007p17998}; 
\citealt{Marois:2010p21652}; \citealt{Soummer:2011p22828}; \citealt{Pueyo:2012p23720})
are yielding contrasts $>$14~mag at 1$\arcsec$ 
(e.g., \citealt{Wahhaj:2013p25352}).

We experimented with a variety of PSF subtraction methods spanning a range of 
sky rotations and variable AO correction
(caused both by seeing conditions and target brightness).
Each method has advantages and disadvantages depending on the particular 
data set and instrument.  For example, for small sky rotations, aggressive use of the 
Locally-Optimized Combination of Images (LOCI) algorithm (\citealt{Lafreniere:2007p17998}) results in 
substantial self-subtraction of real point sources.  In these cases simply subtracting a median-combined
PSF model can result in a higher signal-to-noise ratio for companions.   
In other cases where AO quality changes substantially during an ADI sequence, LOCI generally performs better 
than other simpler methods.  We also tested a variation of LOCI described by \citet{Marois:2010p21652}
in which a small central portion of the region used to compute the correlation matrix is masked 
and then used to reconstruct the final image after calculation of the reference image weights.  
This method keeps noise levels across the processing region relatively constant and better preserves the photometry and astrometry of
known point sources, but we found that it was not the best technique to 
identify real objects in a first (blind) pass.

Altogether we adopt three PSF subtraction methods to homogeneously process
our survey data: subtraction of a scaled median-combined PSF model, a ``conservative'' application
of LOCI algorithm, and an ``aggressive'' form of LOCI.  
Each technique is applied to the inner
(contrast-limited) 3$''$ of our ADI sequences.   
In the background/read noise-limited regime beyond 3$''$, 
we subtract a median-combined sky frame created from the data set itself.

For the scaled subtraction method we first create a PSF model by median-combining images
in the ADI sequence.  For each science image the model is then scaled to the annulus 
spanning the IWA out to 3$''$ by computing the multiplicative factor $C$ that minimizes the $\chi^2$ value
over all $n$ pixels:

\begin{equation}
\chi^2 = \sum_{i=1}^{n}{\left( \frac{f_i - C \mathcal{F}_i} { \sigma_i} \right)^2},
\end{equation}

\begin{equation}
C = \frac{ \sum f_i \mathcal{F}_i/\sigma_i^2}{\sum \mathcal{F}_i^2/\sigma_i^2}.
\end{equation}

\noindent Here $f_i$ and $\mathcal{F}_i$ are the flux at pixel $i$ in the science and model images in units of DN,
and $\sigma_i$ is the uncertainty in the science flux.  In this flux-limited regime, $\sigma_i$$\propto$$\sqrt f_i$, 
in which case $C$ simplifies to $\sum \mathcal{F}_i / \sum (\mathcal{F}_i^2/ f_i) $.

Our implementation of LOCI follows the geometric regions described by \citet{Lafreniere:2007p17998}
with the following parameters: $N_A$=300, $g$=1, $dr$=2.  We perform two reductions with minimum
rotation parameter $N_{\delta}$ equal to 1.5 and 0.5 in units of PSF FWHM for the conservative and
aggressive cases, respectively.  If no reference frames satisfy the $N_{\delta}$ criterion
 for a  particular  annular subsection because of inadequate rotation at small separations then that
 section for that image is skipped.  This affects some of the conservative LOCI processing at small
 separations near the IWA, but rarely influences the aggressive reduction.

Point sources bright enough to bias the reduction are masked out of the images prior to PSF subtraction.
For the scaled median subtraction, masking reduces the influence of bright objects when computing
scale factors.  For the LOCI reduction, masking excludes these regions from the correlation matrix
to avoid influencing the reference image weights.
 
\subsection{Point Source Identification}

After PSF subtraction, the individual images are de-rotated to a common PA, median-combined, and 
oriented so that celestial north is at a PA of 0$^{\circ}$.  
A map of the noise is created by computing the standard deviation of flux values in 
concentric annuli with a width of 3~pixels after rejecting outlier pixels with a clipping threshold of 4-$\sigma$.
In addition to outlier rejection, bright point sources are manually masked from the coadded image to 
minimize their influence on the noise measurements, and,
from this, a signal-to-noise map is made to search for point sources in the images.

Automated point source identification is performed on the signal-to-noise maps using the \texttt{max\_search} routine
in the StarFinder AO imaging software package (\citealt{Diolaiti:2000p25588}).  \texttt{max\_search} identifies peak values above a 
threshold level relative to nearest neighbor pixel intensities.  Low threshold values ($\sim$3--5) tend to produce a large number of 
false positives near the star where the speckle density is high, so we adopted a 7-$\sigma$ limit.
However, visual inspection of each image ultimately proved to be the most robust way to identify fainter
point sources.  In general, we found that artificial point sources injected into the median-combined
images with peak values of 7-$\sigma$ are reliably recovered from visual inspection 
across the entire image.  

Figure~\ref{fig:lociimgs} shows a typical reduction sequence with NIRC2 and HiCIAO.  
(This particular case is for the 50~mas binary GJ~3729~AB, though it is not included in the final
survey statistics because of its binarity.)  The right-hand panels show the clear detection of a point source
in the NIRC2 S/N maps, which is recovered in the HiCIAO data six months later.  Multi-epoch astrometry
shows the candidate companion is a background object.


\begin{figure}
  \vskip -.6in
  \hskip -.6in
  \resizebox{5in}{!}{\includegraphics{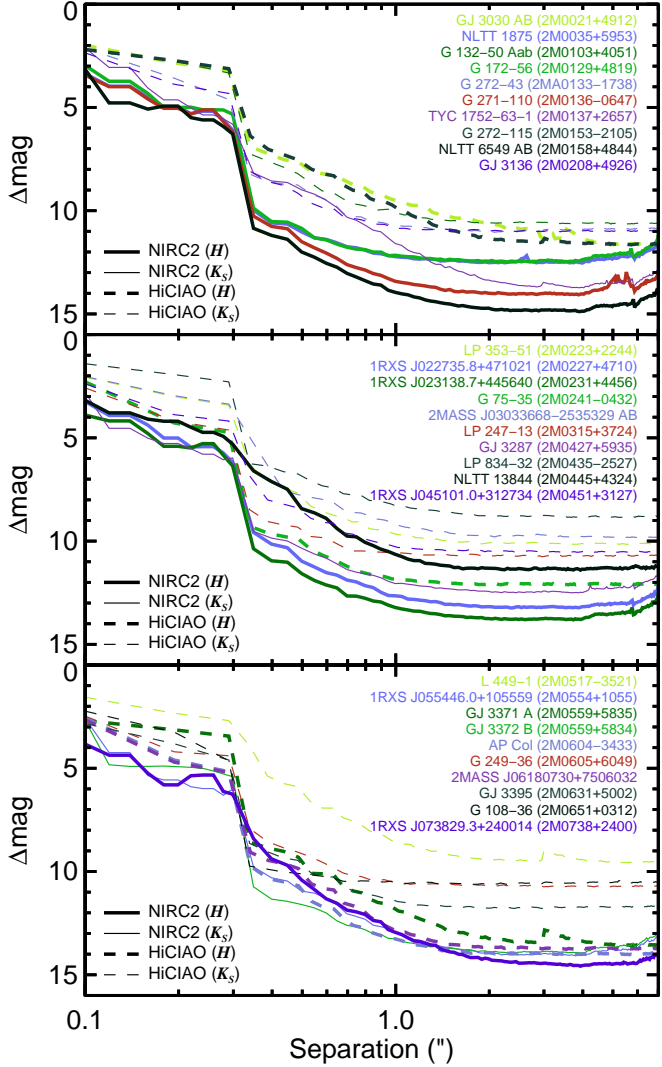}}
  \vskip -.6in
  \caption{7-$\sigma$ contrast curves from our survey.  Unsaturated frames are joined with the deep imaging at 0$\farcs$3.  PSF subtraction with LOCI is performed from 0.3--3$''$ and scaled median subtraction is applied beyond 3$''$.   \label{fig:contrast1} } 
\end{figure}


\begin{figure}
  \vskip -.6in
  \hskip -.6in
  \resizebox{5in}{!}{\includegraphics{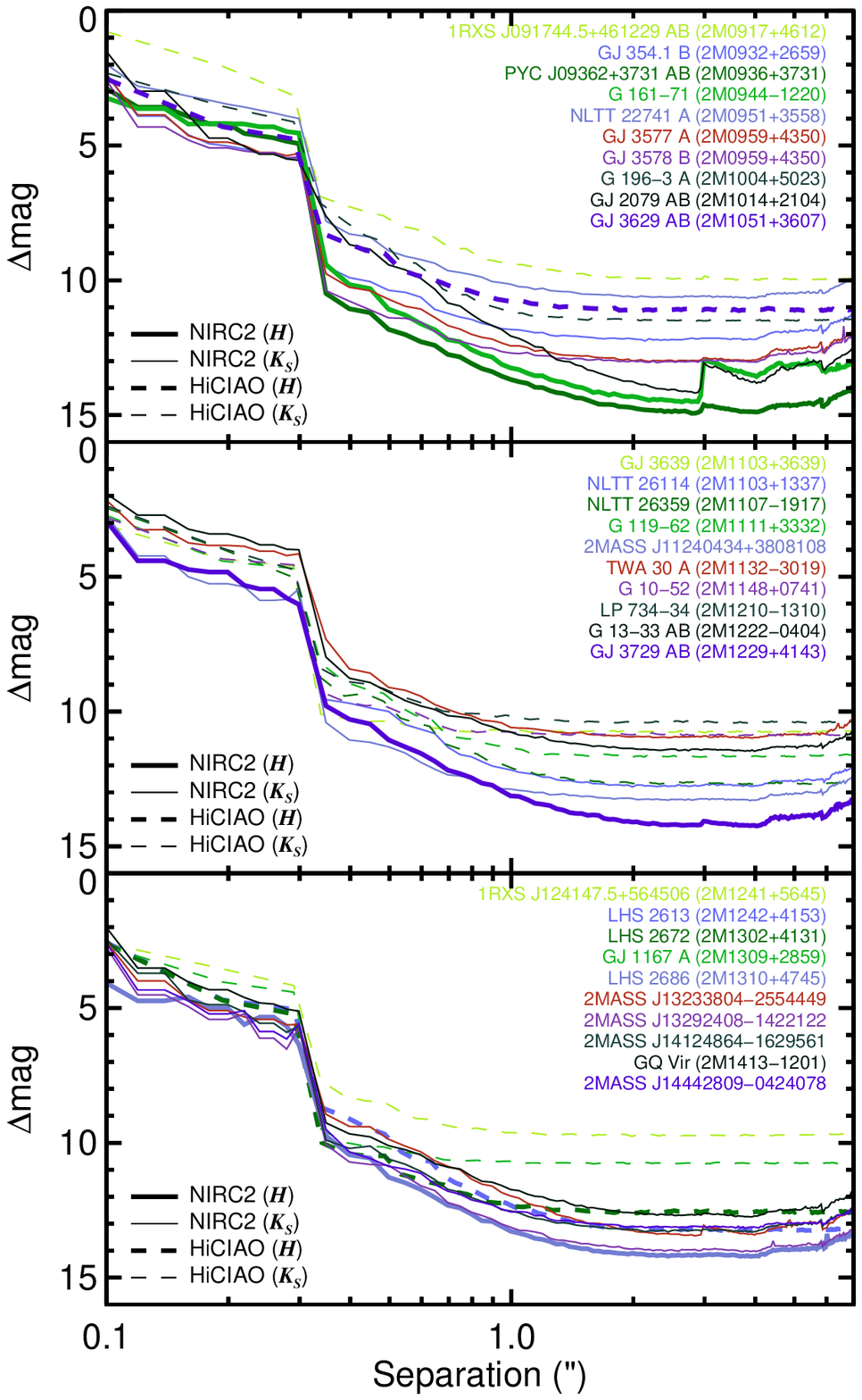}}
  \vskip -.6in
  \caption{7-$\sigma$ contrast curves from our survey (continued).   \label{fig:contrast2} } 
\end{figure}


\begin{figure}
  \vskip -.6in
  \hskip -0.6in
  \resizebox{5in}{!}{\includegraphics{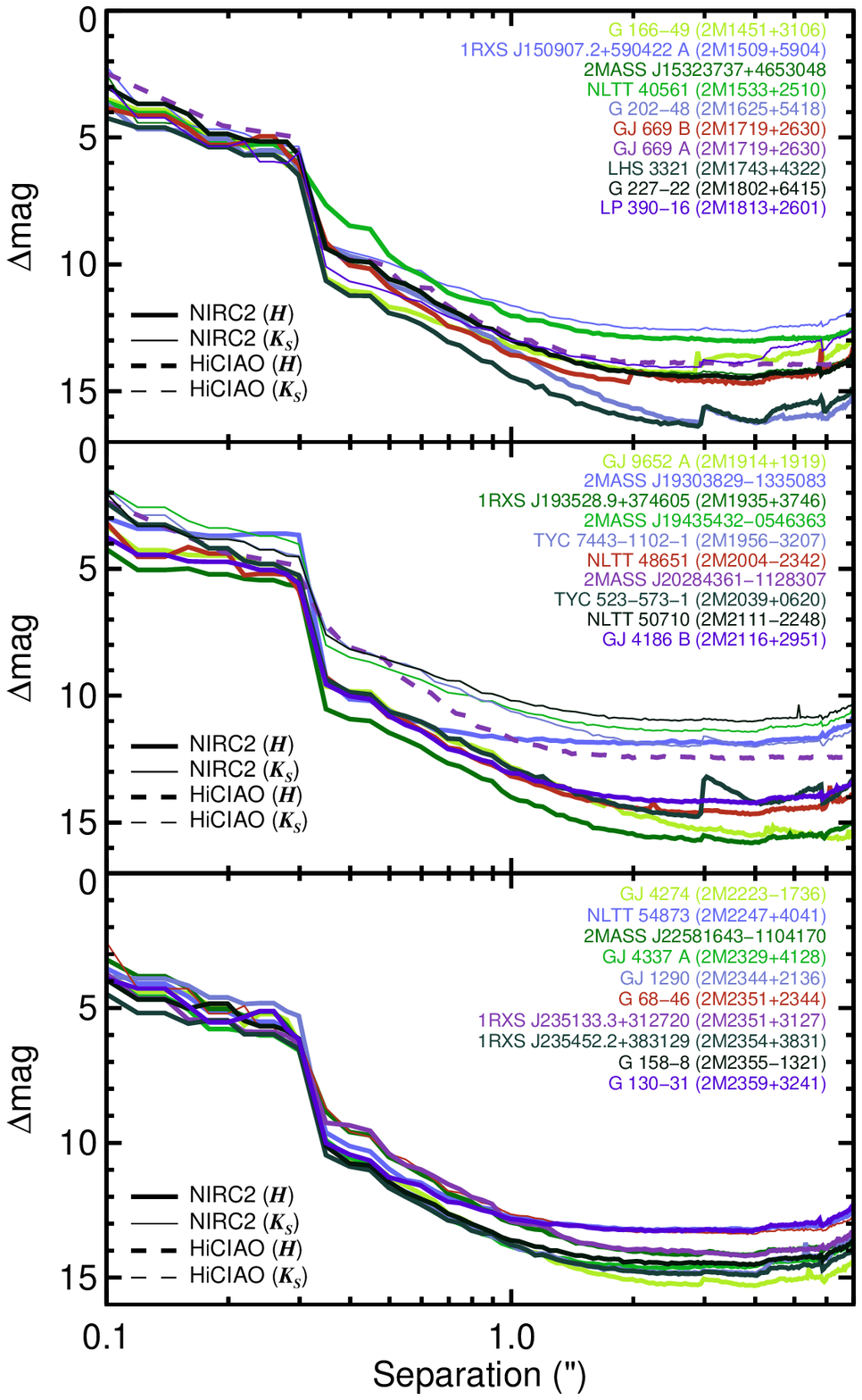}}
  \vskip -.6in
  \caption{7-$\sigma$ contrast curves from our survey (continued).   \label{fig:contrast3} } 
\end{figure}

\subsection{Contrast Curves}

Contrast curves are generated from the reduced images using the noise maps and the flux from
the primary star.  For NIRC2 we directly measure the star brightness behind the coronagraph.
This allows us to monitor AO correction throughout the ADI sequences and calibrate the detection limits using
the coronagraph throughput, which we have measured in the $H$ and $K_S$ filters using 
a binary star (Appendix~\ref{sec:nirc2_transmission}).
An important caveat is that the background sky level must be taken into account since the apparent
flux of the star behind the mask ($f_m$) is the filter-dependent attenuation of the sky-plus-stellar flux, 
rather than simply the attenuated stellar flux alone.  Since the sky level was subtracted from the raw images, it must also be taken into account to compute the 
corrected sky-subtracted stellar flux level ($f_c$):  $f_c = \frac{f_m + sky}{T_\mathrm{filt}}$ -- $sky$,
where $T_\mathrm{filt}$ is the filter-dependent transmission and $sky$ is the background sky level.
 This is particularly important for our 
target sample of M~dwarfs since the attenuated fluxes in 60~sec exposures can be comparable
to the (unattenuated) sky level in the raw frames.
Finally, we use the median-corrected, sky-subtracted peak flux of the primary star ($f_c$) from the
sequence together with the noise map to compute contrast curves at the desired $\sigma$ level.

In a comprehensive analysis of deep ADI and SDI observations from the Gemini NICI Planet-Finding Campaign (\citealt{Liu:2010p21647}),
\citet{Wahhaj:2013fq} shows that the 5-$\sigma$ threshold commonly used for contrast curves
in high-contrast imaging surveys corresponds to $\approx$0.2~mag brighter than the 95\% recovery
rate from Monte Carlo injections and extractions of artificial planets.  For this survey we therefore adopt 
a 7-$\sigma$ threshold as a robust threshold for our sensitivity limits.  

The FOV coverage changes for each observation based on the instrument, coronagraph position on
the detector, and total sky rotation of the ADI sequence.  To make full use of the data (i.e., corners of the array), 
we also compute the fractional
coverage as a function of radial separation from the star for each observation, which we incorporate into our sensitivity limits for the
statistical analysis.

For our HiCIAO data we use of the short unsaturated images taken in sets during the ADI sequences
to photometrically calibrate the coronagraphic frames.  For some Subaru data sets, neutral density filters with $\approx$1\% and $\approx$10\% 
throughputs were used to prevent saturating the detector.  We adopt the following transmission measurements for the neutral density
filters kindly provided by T. Kudo (2013, private communication): 9.740 $\pm$ 0.022\% for \emph{ND10} in $H$-band, 0.854 $\pm$ 0.002\% for 
\emph{ND1} in $H$-band,
10.460 $\pm$ 0.020\% for \emph{ND10} in $K_S$-band, and 1.142 $\pm$ 0.026\% for \emph{ND1} in $K_S$-band.

Unsaturated images are processed similarly to the coronagraphic frames: they are first registered, de-rotated 
to a common PA (which is
necessary for multiple sets taken during an ADI sequence), median combined, and north aligned.
7-$\sigma$ contrast curves are measured and joined with those from the 
deep imaging for separations inside the coronagraphic IWA.

\begin{figure*}
  \vskip -1in
  \hskip .1in
  \resizebox{6.8in}{!}{\includegraphics{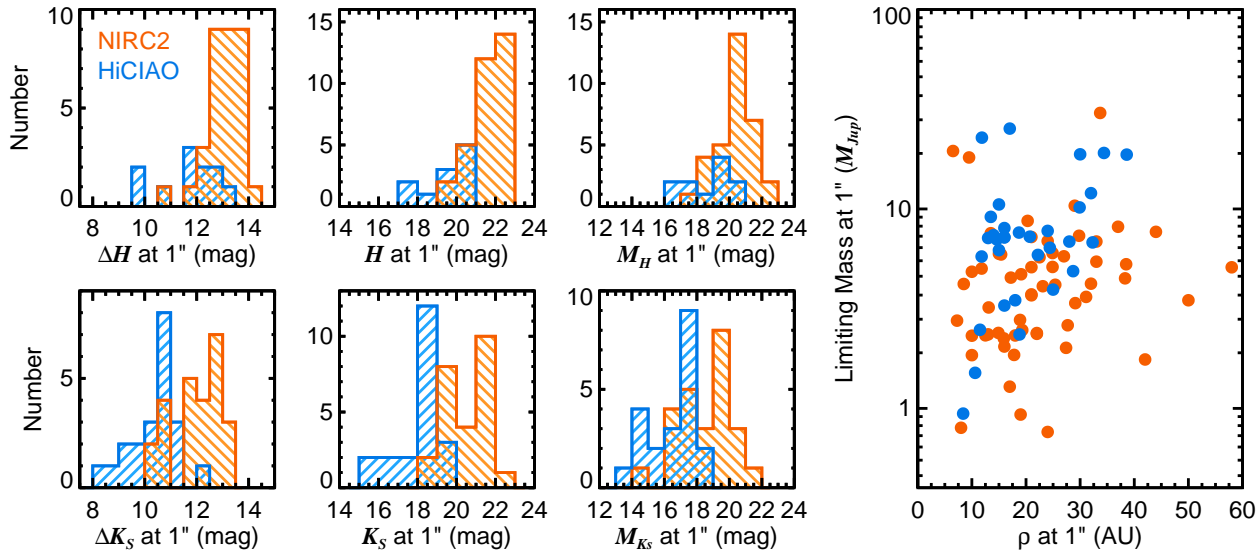}}
  \vskip -1in
  \caption{Summary of our measured contrasts, limiting apparent magnitudes, and limiting absolute magnitudes at 1$''$. 
  The right panel shows the limiting planet masses and projected physical separations at 1$''$ for NIRC2 (red) and HiCIAO (blue).
  We are sensitive to planets below 10~\Mjup \ at separations less than 33~AU at 1$''$ for 85\% of our sample.
   \label{fig:conthists} } 
\end{figure*}

Contrasts from the survey are shown in Figures~\ref{fig:contrast1}--\ref{fig:contrast3}
 and listed in Table~5.
Beyond $\approx$2$''$ the contrast curves flatten out to the background/read-noise level.
The AO performance at both Keck and Subaru is sensitive to seeing conditions and rapidly degrades when
seeing exceeds $\sim$1$\farcs$5.  Most of our observations with HiCIAO suffer from bad seeing, which is
reflected in the modest contrast limits.  Figure~\ref{fig:conthists} summarizes the survey sensitivity:
at 1$''$, our NIRC2 contrasts reach 11.5--14~mag, but the HiCIAO observations span a larger range from 8--13~mag.
The unsaturated frames reach between $\sim$3.5--5.5~mag at 0.3$''$.  
In limiting apparent magnitude, our observations at 1$''$ reach 15--23~mag, which translates to 13--23~mag in 
absolute magnitude.  Note that we have excluded contrasts in these histograms for the two substellar companion host
stars 1RXS~J034231.8+121622 and 2MASS~J15594729+4403595, for which we only obtained short
images when we discovered the companions (Section~\ref{sec:bddiscoveries}).  These contrast curves are, however, included in 
our statistical analysis.  

The right-most panel in Figure~\ref{fig:conthists} shows the limiting mass at 1$''$ as
a function of physical separation at 1$''$ for our entire sample.
The median limiting mass is 5.5 \Mjup \ and we are sensitive to masses below 10~\Mjup \ at 1$''$ for 85\% of our targets.  
The median physical separation at 1$''$ is 20.3 AU and 
85\%  of our targets correspond to physical separations less than 33 AU at 1$''$.


\subsection{Astrometry and Photometry}

\subsubsection{Stellar Binaries}

Astrometry for bright companions from our short unsaturated images is computed in the following manner.
After correcting each image for distortion, the separation and position angle are measured 
in one of two ways.  For companions with small separations ($\lesssim$1$''$) we fit each binary
component with a PSF composed of three elliptical Gaussians as described in \citet{Liu:2008p14548}.  This method 
performs well for both the NIRC2 and HiCIAO data in a variety of seeing conditions.
For separations wide enough to avoid contamination from the primary ($\gtrsim$1$''$), 
we use aperture photometry after subtracting the sky level from each image.
The mean of the separation and PA measured from individual images are adopted for our astrometry.  

Several sources contribute to the uncertainty in these values: random sub-pixel centroid 
measurement errors from image to image, 
systematic effects from residuals in the distortion solution, uncertainties in the
absolute sky orientation on the detector, and the finite precision of the measured plate scale.
These independent errors are propagated analytically to produce our final 
astrometric uncertainties.  The separation $\rho$ is therefore

\begin{equation}{\label{eqn:binsep}}
\rho = s \bar{\rho}_{meas}  \pm   s \bar{\rho}_{meas} \left( \left(\frac{\sigma_{s}}{s}\right)^2 + \left(\frac{\sigma_{\bar{\rho}, tot}}{\sigma_{\bar{\rho}, meas}}  \right)^2  \right)^{1/2},
\end{equation}

\begin{equation}
\sigma_{\bar{\rho}, tot}^2 = \sigma_{\bar{\rho}, meas}^2 + 2 \sigma_{d}^2
\end{equation}

\noindent where $s$ and $\sigma_{s}$ are the plate scale and associated uncertainty in mas~pix$^{-1}$, $\bar{\rho}_{meas}$ and 
$\sigma_{\bar{\rho}, meas}$ are the mean and standard deviation of the measured
separations for the individual images (in pixels), and $\sigma_{d}$ is the 
typical residual positional displacement after applying the distortion correction (in pixels, one for each 
binary component).  Likewise, the PA ($\theta$) is

\begin{equation}{\label{eqn:binpa}}
\theta = \bar{\theta}_{meas} + \theta_{North}  \pm    \left( \sigma_{\theta, meas}^2 + \sigma_{\theta, North}^2 \right)^{1/2},
\end{equation}

\noindent where $\bar{\theta}_{meas}$ is the mean PA of the individual images,  $\theta_{North}$ is the 
orientation of the detector relative to north, $\sigma_{\theta, meas}$ is the standard deviation of the
PA measurements, and $\sigma_{\theta, North}$ is the uncertainty in the sky orientation on the detector.

\subsubsection{Faint Point Sources}

Astrometry and photometry for faint point sources are measured from our final processed images.
The error budget for the separation measurement consists of positional uncertainties of the
star behind the mask, centroid errors for the candidate, and 
systematic errors in the distortion solution (Equation~\ref{eqn:binsep}).  For NIRC2, 
we adopt random measurement uncertainties ($\sigma_{\rho, meas}$) of 0.3~pix, 
which is typical of our binary star measurements.

The dominant source of astrometric uncertainty for our HiCIAO data are from the image registration process since
the peak of the PSF is hidden under the opaque 300~mas focal plane mask.
To estimate the typical magnitude of this uncertainty, we use our ADI sequence of the $\sim$2$''$ binary system 
GJ~3030~AB.  The $R$-band magnitude of this system from UCAC4 
($\sim$12.2~mag) is typical for our targets and, as with most of our HiCIAO observations in this program, these data were taken in unusually 
poor seeing conditions (1.5--2.0$''$).  Here GJ~3030~A was behind the mask and the following 
test was performed on the B component.
We first determine the centroid of the companion GJ~3030~B in each image,
then mask the central 150~mas, fit a 2-dimensional elliptical Gaussian to the wing of the PSF following our general method
for HiCIAO image registration (see Section~\ref{sec:imgreg}),
and compute the relative difference of the inferred to the actual measured stellar position.
The inferred position is generally $\sim$1--2.5~pix from the actual center (the mean value is 1.6~pix for this representative sequence), 
so we conservatively adopt an uncertainty of 2~pix, or $\sim$20~mas, for the HiCIAO registration term ($\sigma_{\bar{\rho}, meas}$).
For  the random error term $\sigma_{\theta, meas}$ we adopt characteristic uncertainties from our moderate-contrast binary star 
astrometry of 0$\fdg$2 and 0$\fdg$1 for NIRC2 and HiCIAO, respectively.

Position angle uncertainties originate from 
centroid errors for the candidate companion, systematic errors in the north alignment, and, for HiCIAO, uncertainties associated
with image registration.  In this latter case we add an additional term, $\sigma_{reg}$, in quadrature with the random and systematic errors
in Equation \ref{eqn:binpa}.  $\sigma_{reg}$ refers to the angular uncertainty associated with a positional error (from image registration) 
orthogonal to the primary-candidate companion direction; for HiCIAO we adopt a 2~pix positional uncertainty, so $\sigma_{reg}$ (in deg)
scales as arcsin(2~pix/$\rho_{meas}$).

Many of the wide-separation ($\gtrsim$5$''$) candidates identified in first epoch imaging were followed up with the
NIRC2 wide camera mode, which has a plate scale of 39.884~$\pm$~0.039~mas~pix$^{-1}$ (\citealt{Pravdo:2006p22531})
and field of view ($\approx$40$''$$\times$40$''$).  
For these observations, we use the distortion solution from \citet{Fu:2014hz}
and estimate a residual positional uncertainty $\sigma_{d}$ of 1~pix from the residual map.\footnote{http://www2.keck.hawaii.edu/inst/nirc2/dewarp.html}
Even behind the partly opaque coronagraph, the primary stars usually saturated in this wide camera mode.
Although this had a minimal impact on 
relative astrometry, it prevented accurate relative photometry for these observations.

Aperture photometry is measured for all point sources to derive contrasts relative to the star.  
We use aperture radii of 6~pix for candidates identified in our NIRC2 data.  For our HiCIAO observations, which
typically suffered from poor seeing and modest AO correction, we use larger aperture radii of 15~pix.
Photometric errors incorporate measurement
errors computed at the source location in the noise map, uncertainties in the coronagraph
transmission (for NIRC2), uncertainties in the neutral density filters (for HiCIAO, when applicable),
and uncertainties in the measured flux of the primary star itself.  This last term is the standard deviation
of flux measurements from the star (behind the mask for NIRC2, and from unsaturated images
for HiCIAO).

\section{Survey Results}

Out of 122 targets imaged in our survey, 44 are close stellar binaries, 27 of which are new or spatially resolved for the first time.
We discovered four new young brown dwarf companions confirmed to be comoving with their host stars.  
Over 150 planet candidates were
identified out to projected separations of $\approx$400~AU; we recovered the majority (56\%) of these in second-epoch imaging,
and all of these are background stars.  Below we describe these results in detail.

\subsection{Substellar Companions from the PALMS Survey}{\label{sec:bddiscoveries}}

Our four brown dwarf discoveries span masses of 30--70~\Mjup \ at projected separations of 6--190~AU.
Two have already been published as part of this series 
(1RXS J235133.3+312720~B in \citealt{Bowler:2012p23851}; GJ~3629~B in \citealt{Bowler:2012p23980})
and here we present two new companions, 
1RXS J034231.8+121622~B  and 2MASS~J15594729+4403595~B.
Note that the young stars G~196-3~A (2MASS~J10042148+5023135) and NLTT~22741~A 
(2MASS~J09510459+3558098; LP~261-75) 
 in our sample have previously known wide-separation brown dwarf companions with L spectral types 
 (\citealt{Rebolo:1998p19498}; \citealt{Reid:2006p22856}).  However, neither companion falls in the
  FOV of our observations so they are not included in our statistical analysis (Section~\ref{sec:stats}).
Below we summarize our discoveries and present new photometry and astrometry for each system.

\subsubsection{1RXS J235133.3+312720~B (2MASS~J23513366+3127229)}

1RXS~J235133.3+312720~B is a 32~$\pm$~6~\Mjup \ companion to the active M2.0 star 1RXS~J235133.3+312720~A
and was the first discovery from our PALMS survey (\citealt{Bowler:2012p23851}).
\citet{Shkolnik:2012p24056}, \citet{Schlieder:2012p23477}, and \citet{Malo:2013p24348} independently
identify the primary as a likely member of the AB~Dor YMG ($\sim$120~Myr) based on its kinematics and activity.
We found a near-IR spectral type of L0$^{+2}_{-1}$ for the companion along with subtle spectroscopic hints of 
low surface gravity (\citealt{Bowler:2012p23851}).  Based on the photometric distance
to the primary (50~$\pm$~10~pc), the 2$\farcs$4 projected separation corresponds to 120~$\pm$~20~AU.

In \citet{Bowler:2012p23851} we presented two epochs of relative astrometry for 
1RXS~J235133.3+312720~AB based on adaptive optics imaging with NIRC2 in 2011 
in $H$- and $K'$-bands.  In addition, we also obtained seeing-limited relative photometry from IRTF in $YJHKs$ filters.
Here we present new, more precise 1--3.8~$\mu$m relative photometry obtained with NIRC2 in 2013 (Table~7).
We infer a $Y$--$J$ color of 1.17~$\pm$~0.17~mag for 1RXS~J235133.3+312720~B
assuming a $Y$--$J$ color of 0.459~$\pm$~0.001~mag for the primary, which is the mean color for M2.0 dwarfs from \citet{Rayner:2009p19799}. 
Based on the MKO photometry of the primary from  \citet{Bowler:2012p23851}, we derive the following colors
for 1RXS~J235133.3+312720~B: ($J$--$H$)$_{\mathrm{MKO}}$=0.73~$\pm$~0.12~mag, ($H$--$K$)$_{\mathrm{MKO}}$ = 0.64~$\pm$~0.08~mag, ($J$--$K$)$_{\mathrm{MKO}}$ = 1.37~$\pm$~0.12~mag.
Finally, based on the typical $K_{\mathrm{MKO}}$--$L'$ color of 0.2~$\pm$~0.1~mag for M2 dwarfs from \citet{Golimowski:2004p15703},
we derive a $K_{\mathrm{MKO}}$--$L'$ color of 0.8~$\pm$~0.2~mag for the companion.
Compared to typical colors of M/L dwarfs (e.g., \citealt{Bowler:2012p23980}; \citealt{Golimowski:2004p15703}), our new photometry for 1RXS~J235133.3+312720~B 
corresponds to spectral types of $\approx$L0--L3, which is consistent with our published classification based on near-IR spectroscopy.

Altogether, our astrometry over 3 years shows tentative signs orbital motion.  The reduced $\chi^2$ values for constant 
and linear fits to the separation measurements are 1.58 and 1.38, respectively.  Reduced $\chi^2$ values for the 
PA are 3.91 and 0.47 for the constant and linear models, suggesting a slight change of
--0.076~$\pm$~0.017$^{\circ}$~yr$^{-1}$.  Additional astrometry in the future will be needed for confirmation.

\subsubsection{GJ~3629 B (2MASS~J10512059+3607255~B)}


\begin{figure*}
  \vskip -1.in
  \hskip -.1in
  \resizebox{7in}{!}{\includegraphics{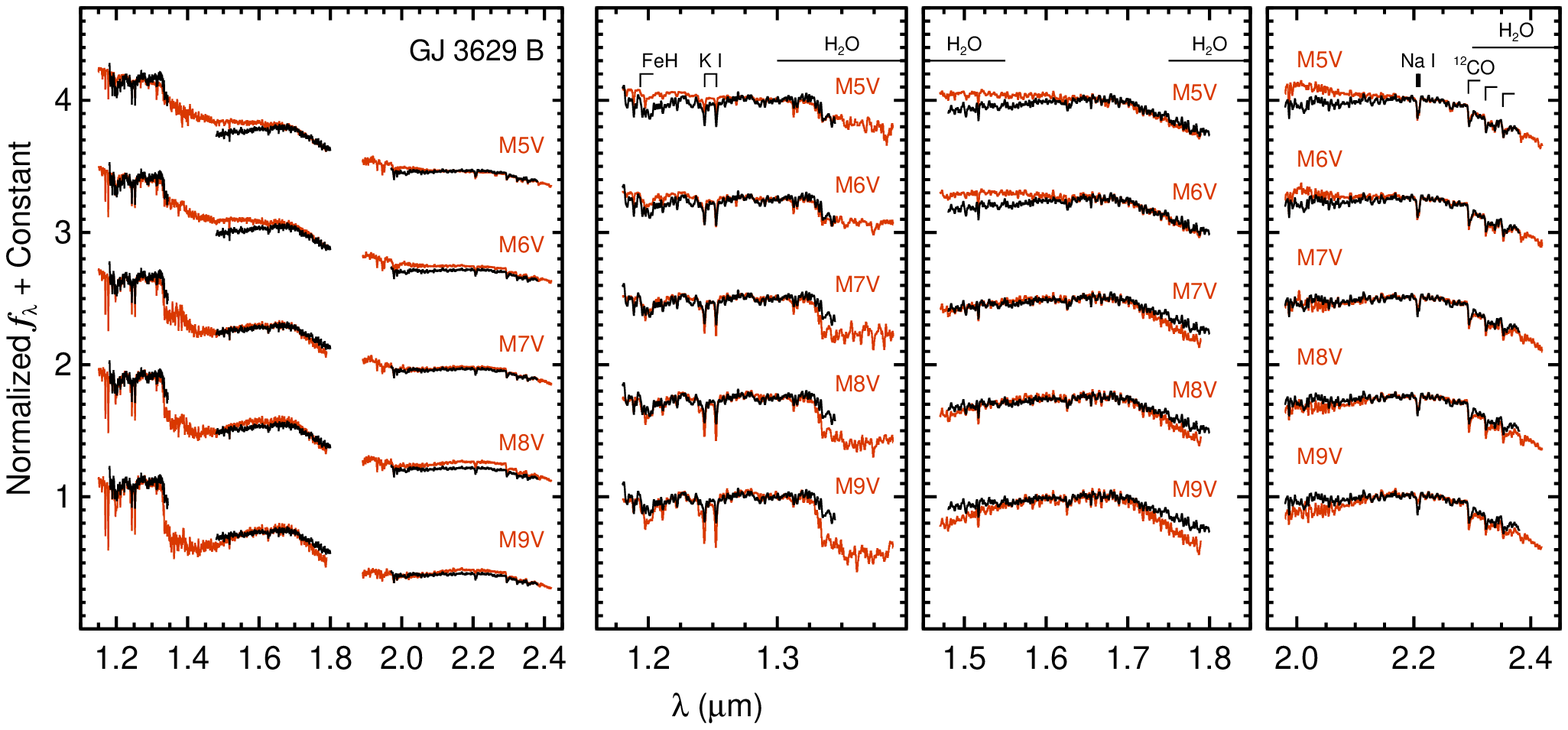}}
  \vskip -1in
  \caption{Our OSIRIS spectrum of GJ~3629~B compared with field objects from the IRTF SpeX Spectral Library 
  (\citealt{Cushing:2005p288}; \citealt{Rayner:2009p19799}).  Overall the 1.15--2.40~$\mu$m spectrum of 
  GJ~3629~B, which is largely driven by relative band-to-band flux calibration,
   best resembles M7 field template (left panel).  
  Compared to individual bandpasses, GJ~3629~B is similar to both M7 and M8 objects in $J$, $H$, and $K$ spectral regions.  
  Altogether we adopt a spectral type of M7.5~$\pm$~0.5 for GJ~3629~B.
   The IRTF library templates are Gl~51 (M5V), Gl~406 (M6V), Gl~644~C (M7V), Gl~752~B (M8V), and 
   DENIS-P J104814.7--395606.1 (M9V). All spectra have been smoothed to a common resolving power of $R$$\approx$2000. \label{fig:gj3629b_hgcomp}}
\end{figure*}


\begin{figure}
  \vskip -.5in
  \hskip -0.2in
  \resizebox{4.in}{!}{\includegraphics{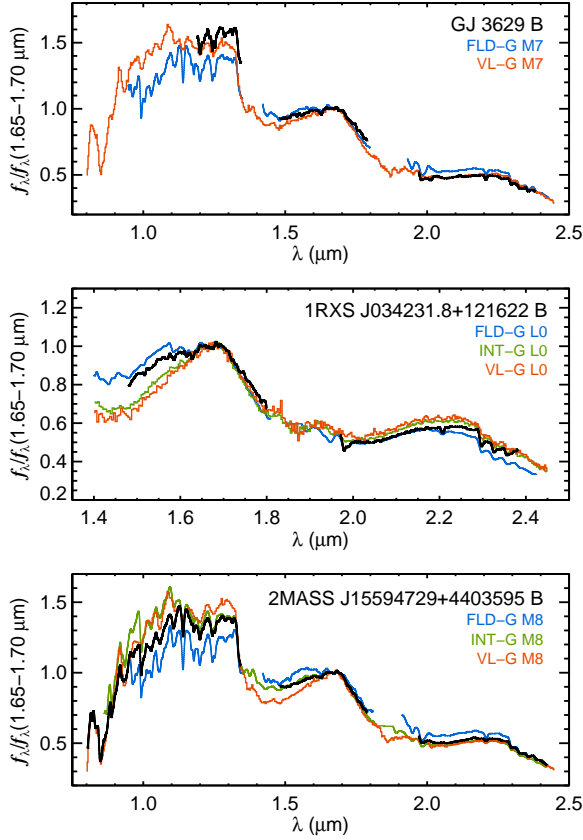}}
  \vskip -.6in
  \caption{Comparison of GJ~3629~B, 1RXS~J034231.8+121622~B, and 2MASS~J15594729+4403595~B
  with very-low gravity (VL-G), intermediate-gravity (INT-G), and field high-gravity (FLD-G) ultracool objects
  from \citet{Allers:2013p25314}.  GJ~3629~B does not appear to differ substantially from the FLD-G M7 object.  
  The $H$-band shape of 1RXS~J034231.8+121622~B somewhat less angular than the INT-G L0 template but is 
  noticeably more pronounced than the field object.  2MASS~J15594729+4403595~B agrees well with the INT-G M8 template. 
  The comparison objects from \citet{Allers:2013p25314} are 2MASSJ00034227-2822410 (FLD-G M7), 
  2MASS~J03350208+2342356 (VL-G M7), 2MASS~J17312974+2721233 (FLD-G L0), 2MASS~J15525906+2948485 (INT-G L0),  
  2MASS~J22134491--2136079 (VL-G L0), 2MASS~J08040580+615333 (FLD-G M8), and 2MASS~J00192626+4614078 (INT-G M8).
  TWA~27~A (VL-G M8) is from \citet{Looper:2007p3713}.  All spectra are smoothed to $R$$\sim$120 and normalized to the 1.65--1.70~$\mu$m region.
    \label{fig:palms4_bd_spec_comps}}
\end{figure}


\begin{figure}
 \vskip -.4in
  \hskip -.4in
  \resizebox{4in}{!}{\includegraphics{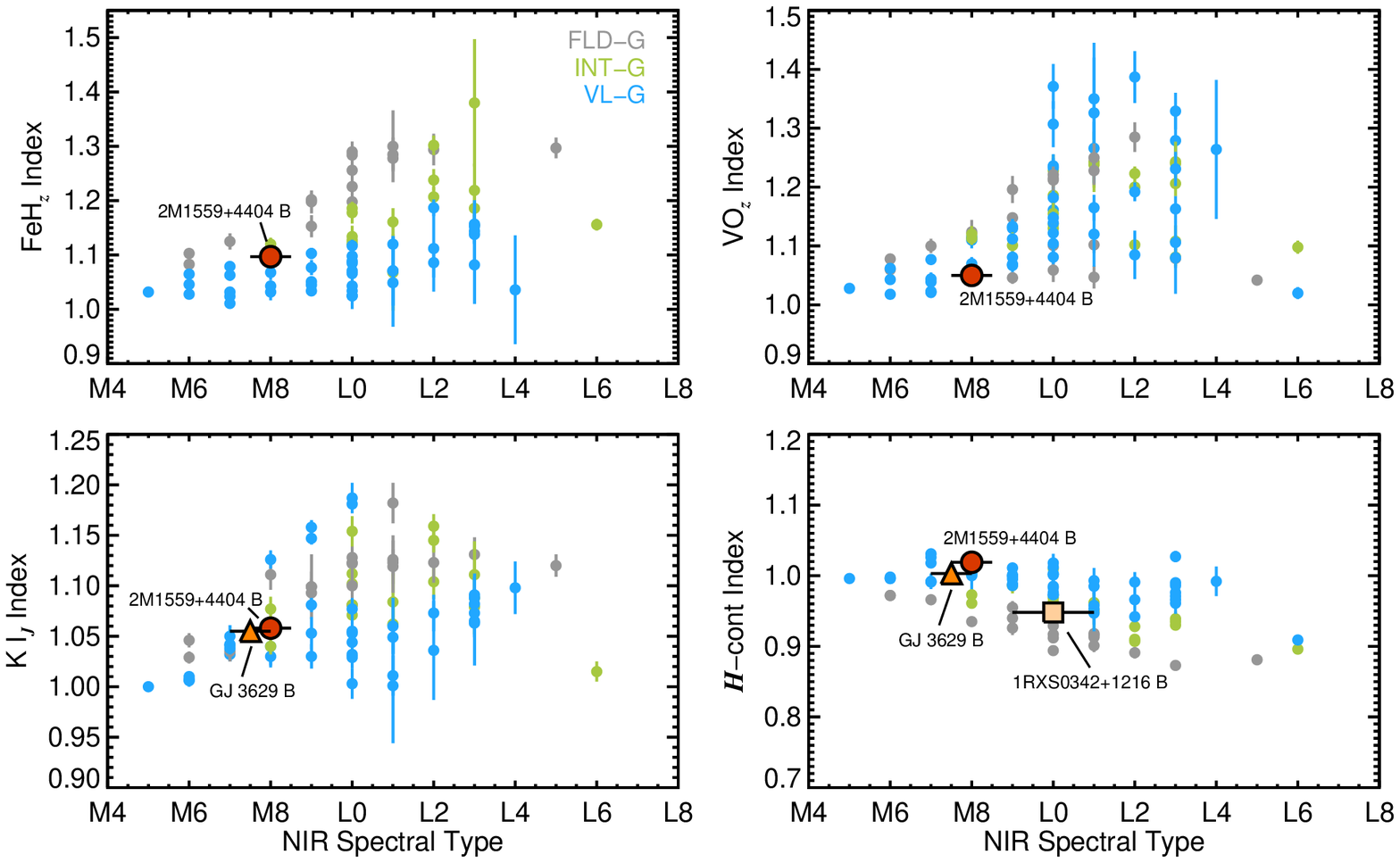}}
  \vskip -.3in
  \caption{Gravity indices as a function of NIR spectral type for  2MASS~J15594729+4403595~B, GJ~3629 B, and 1RXS J034231.8+121622 B
  following the \citet{Allers:2013p25314} classification scheme.  2MASS~J15594729+4403595~B is an INT-G brown dwarf; the other two
  companions show signs of low surface gravity but the limited spectral coverage prevents definitive assignments.  Objects with final gravity
  classes of field-gravity (FLD-G), intermediate-gravity (INT-G), and very low-gravity (VL-G) from \citet{Allers:2013p25314} are plotted in gray, 
  green, and blue for comparison.
    \label{fig:gravindices} } 
\end{figure}

GJ~3629~B is a modest-contrast companion located a mere 0$\farcs$2 from its active M3.0Ve host star GJ~3629~A (\citealt{Bowler:2012p23980}).
Based on its photometric distance of 22~$\pm$~3~pc and age of 25--300~Myr (\citealt{Shkolnik:2009p19565}), Bowler et al. 
inferred a mass of 46~$\pm$~16~\Mjup \ for the companion.  Recently \citet{Dittmann:2013tp} measured a parallactic
distance of 32.3~$\pm$~2.4~pc to the system.  This corresponds to a somewhat higher luminosity of 
log($L/L_{\odot}$) = --2.89~$\pm$~0.10~dex.  At the system age of 25--300~Myr, the \citet{Burrows:1997p2706}
evolutionary models imply a mass much closer to the hydrogen burning limit.  Assuming a log-normal luminosity distribution and
a linearly uniform age distribution, the resulting mass distribution from Monte Carlo simulations has a median value of 64~\Mjup.  
The 68.3\% confidence range about the median is 41--94~\Mjup, and the 95.4\% range spans 31--114~\Mjup.  
Altogether, the probability that GJ~3629 B is substellar ($<$75~\Mjup) is 62\%.  

Figure~\ref{fig:gj3629b_hgcomp} shows our resolved 1.15--2.4~$\mu$m Keck/OSIRIS spectrum of GJ~3629~B
compared to M5--M8 field stars from the IRTF SpeX Spectral Library (\citealt{Cushing:2005p288}; 
\citealt{Rayner:2009p19799}).  Overall the spectrum of GJ~3629~B is
typical of an ultracool M dwarf, exhibiting deep $\approx$1.4~$\mu$m and 1.9~$\mu$m steam bands, strong
2.2935~$\mu$m CO band heads, FeH absorption at $\approx$1.2~$\mu$m, and \ion{K}{1} and \ion{Na}{1}
doublets at 1.25 $\mu$m and 2.21~$\mu$m.  GJ~3629~B is an excellent match to the 
M7 template across the entire spectrum (left-most panel) and M7--M8 objects among 
individual bandpasses (right three panels).  Altogether we adopt a spectral type of M7.5~$\pm$~0.5.
Compared to a low- and high-gravity templates from \citet{Allers:2013p25314} 
in Figure~\ref{fig:palms4_bd_spec_comps}, there are no obvious signs of low surface gravity in GJ~3629~B, which
is prominently manifested as a angular $H$-band shape and shallow $J$-band alkali line strengths 
in young brown dwarfs (e.g., \citealt{McLean:2000p3926}; \citealt{Allers:2007p66}).
Our OSIRIS spectrum does not span the entire wavelength range to compute all gravity indices
using the \citet{Allers:2013p25314} scheme, but the \ion{K}{1}$_J$ and $H$-cont indices tentatively show
signs of youth (Table~6 and Figure~\ref{fig:gravindices}).

We also take the opportunity to update the physical properties of the primary with this new trigonometric distance.
Following the same methods as in \citet{Bowler:2012p23980}, the luminosity of the primary is 
log($L_\mathrm{Bol}/L_{\odot}$) = --1.54~$\pm$~0.08~dex and its mass is 0.3--0.5~\Msun \ using the
\citet{Baraffe:1998p160} evolutionary models.
Taking into account the updated component masses and projected separation (6.5~$\pm$~0.5~AU), 
the expected orbital period for this system is 30$^{+30}_{-13}$~yr assuming a projected-to-physical conversion
scale of 1.16$^{+0.81}_{-0.31}$ from \citet{Dupuy:2011p22603}.  
Our new astrometry taken with HiCIAO only a few months after
our last published epoch is consistent with our previously reported measurements.
Finally, we note that the system kinematics, $UVW$=\{--28.9~$\pm$~0.8, --15.8~$\pm$~1.0, --0.9~$\pm$~0.5\} km~s$^{-1}$, do not 
correspond to any known moving group.

\subsubsection{1RXS J034231.8+121622 B (2MASS~J03423180+1216225 B)}

1RXS J034231.8+121622 A is an active M4.0Ve star detected in the $ROSAT$ and $GALEX$ 
surveys (\citealt{Riaz:2006p20030}).
\citet{Shkolnik:2009p19565} first noted hints of youth (60--300~Myr) from its X-ray emission and ruled
out spectroscopic binarity from two epochs of high resolution ($R$$\sim$58,000) optical spectroscopy.

Recently, \citet{Dittmann:2013tp} measured a parallactic distance of 23.9~$\pm$ 1.1~pc to this star.
Its distance and radial velocity (35.4~$\pm$~0.4~km~s$^{-1}$) from \citet{Shkolnik:2012p24056} imply $UVW$
space velocities of \{--39.0~$\pm$~0.6, --13.4~$\pm$~0.8, --6.8~$\pm$~0.7~km~s$^{-1}$\}, which are similar to, though not formally 
consistent with, the Hyades moving group 
($UVW_\mathrm{Hyades}$ = \{--41.7, --19.3, --1.1\} $\pm$~0.4 km~s$^{-1}$; \citealt{Perryman:1998p25578}).  
1RXS~J034231.8+121622~A also shares a similar sky position with the Hyades  
and is only 24~pc from the cluster center (\citealt{Roser:2011p25028}), suggesting a possible association
with the larger but less coherent Hyades Supercluster or Stream (\citealt{Eggen:1958uq}).  However, the origin and relationship
of this kinematic overdensity with the Hyades cluster is not clear and is unlikely to be useful for age-dating
purposes (e.g., \citealt{Famaey:2005gy}).  So for this work we adopt the 60--300~Myr statistical age constraint from
\citet{Shkolnik:2009p19565} based on the X-ray luminosity and spectroscopic youth indicators of 1RXS~J034231.8+121622~A.

\citet{Janson:2012p23979} used Lucky imaging to resolve a close (0$\farcs$8) candidate companion, 
1RXS J034231.8+121622~B,
at two epochs in 2008.  However, they were not able to distinguish a background object from a comoving
companion from these data.  Their contrast measurement in $z$-band (5.20~$\pm$~0.27~mag)
imply a spectral type of $\gtrsim$L0 for the companion.

\begin{figure}
 \vskip -1.1in
  \hskip -0.7in
  \resizebox{4.3in}{!}{\includegraphics{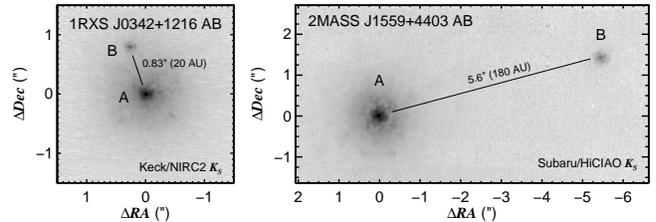}}
  \vskip -.9in
  \caption{NIRC2 and HiCIAO images of two new brown dwarf companions identified in this survey. 
  The inferred masses of 1RXS~J034231.8+121622~B (L0~$\pm$~1) and 2MASS~J15594729+4403595~B (M8.0~$\pm$~0.5) are
  35~$\pm$~8~\Mjup \ and 56~$\pm$~6~\Mjup, respectively.  North is up and East is left.
    \label{fig:bdimgs} } 
\end{figure}

\begin{figure}
  \begin{center}
  \resizebox{3.5in}{!}{\includegraphics{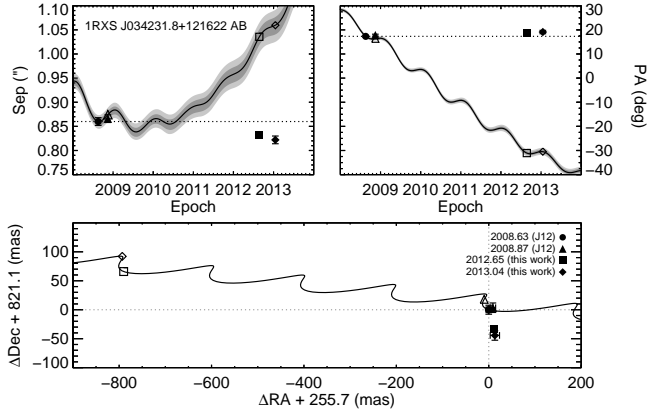}}
  \caption{Test for physical association for the companion to 1RXS J034231.8+121622.  The first two epochs in 2008 are from \citet{Janson:2012p23979}.  We
  confirm the companion is comoving and detect orbital motion with our AO imaging in 2012.   \label{fig:rxs0342_background} } 
\end{center}
\end{figure}

We imaged 1RXS J034231.8+121622 on 2012~August~23~UT and 2013~Jan~17~UT with NIRC2 in 
$Y$, $J$, $H$, $K_S$, and $L'$ bands (Table~7).
The companion was easily identified in all the data with contrasts between 4.3--3.6~mag (Figure~\ref{fig:bdimgs}).  
Figure~\ref{fig:rxs0342_background} shows our astrometry and that from Janson et al. 
between 2008 and 2013 compared to the expected track from a background object.
We confirm that 1RXS~J034231.8+121622~B is physically bound and detect slight orbital motion
in both PA and separation.  The reduced $\chi^2$ value for a constant fit in separation is 6.97
and for a linear fit is 0.98.  Similarly, for the PA, the constant fit gives 6.75 and 5.23.  Removing the 
2012.645 epoch $Y$-band PA point, in which the companion was only identified in three exposures, 
gives a reduced $\chi^2$ value of 3.67 and 0.46 for the constant and linear PA fits, respectively.  These imply orbital
motion of --8.1~$\pm$~1.5~mas~yr$^{-1}$ in separation and +0.33~$\pm$~0.09$^{\circ}$ yr$^{-1}$ in PA.
At a distance of 23.9~$\pm$~1.1~pc, the projected separation of the pair is 19.8~$\pm$~0.9~AU.

We use our measured contrasts to compute $J_\mathrm{MKO}$-, $H_\mathrm{MKO}$-, and $K_S$-band 
magnitudes for 1RXS~J034231.8+121622~B
based on photometry of the primary, which was first converted from the
2MASS to MKO filter system for the $J$ and $H$ filters with the relations from \citet{Leggett:2006p2674}.
Based on the typical $Y$--$J$ color of 0.524~$\pm$~0.01~mag for M4 stars from \citet{Rayner:2009p19799},
we derive a $Y$--$J$ color of 0.86~$\pm$~0.13~mag for 1RXS J034231.8+121622~B, which
suggests a photometric spectral type of M8~$\pm$~1.  

\begin{figure*}
  \vskip -1.in
  \hskip -.1in
  \resizebox{7in}{!}{\includegraphics{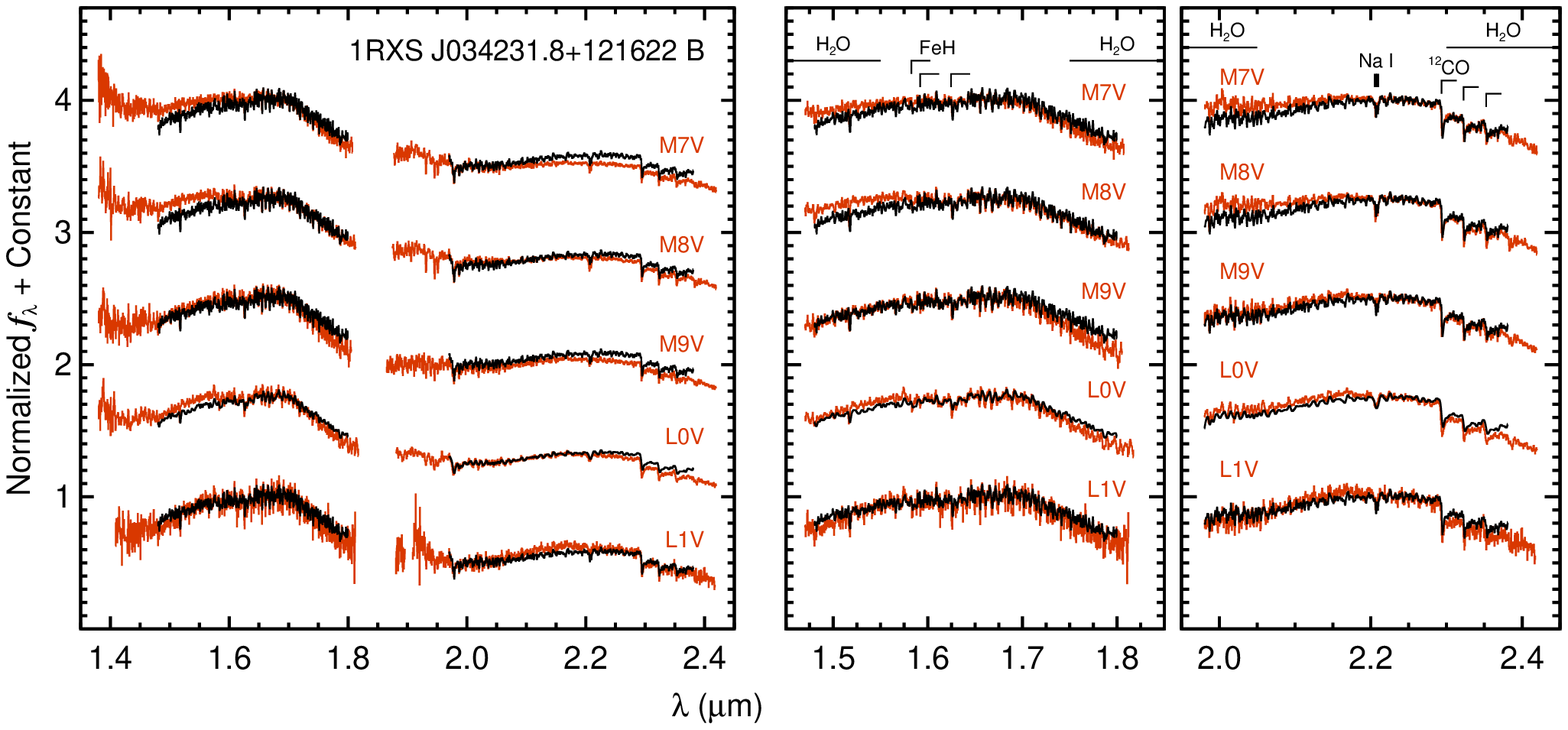}}
  \vskip -1in
  \caption{Our Keck/OSIRIS 1.4--2.4~$\mu$m spectrum of 1RXS~J034231.8+121622~B compared to field objects from 
  the IRTF SpeX Spectral Library.  We adopt a spectral type of L0~$\pm$~1 for 1RXS~J034231.8+121622~B.  M7--M9 templates
  are listed in Figure~\ref{fig:gj3629b_hgcomp}.  The L0 template is 2MASS~J17312974+2721233 from \citet{Allers:2013p25314}
  and the L1 template is 2MASS~J02081833+2542533 from \citet{Cushing:2005p288}.  All spectra have been smoothed to a common
  resolving power of $R$$\approx$2000 except for the L0 template, which has $R$$\approx$750.
    \label{fig:rxs0342b_hgcomp}}
\end{figure*}

Figure~\ref{fig:rxs0342b_hgcomp} shows our resolved Keck/OSIRIS $H$- and $K$-band spectra of 1RXS~J034231.8+121622~B
compared to field templates.  1RXS~J034231.8+121622~B is most similar to L0--L1 objects in the 1.4--2.4~$\mu$m region
and in $H$-band alone.  The $K$-band spectrum resembles field M9--L0 templates.  Altogether we adopt a spectral type
of L0~$\pm$~1.  Compared to younger L0 objects from \citet{Allers:2013p25314} in Figure~\ref{fig:palms4_bd_spec_comps},
the blue side of the $H$-band of 1RXS~J034231.8+121622~B appears somewhat shallower than the $H$-band spectral 
shape of the field object, but not as much as the intermediate- or very low-gravity brown dwarfs (Figure~\ref{fig:gravindices}).

Using this spectral type and the system distance, we compute an
$H$-band bolometric correction from \citet{Liu:2010p21195} and a bolometric luminosity of 
log~$L$/$L_{\odot}$ = --3.81~$\pm$~0.05~dex. Uncertainties in distance, spectral type, and photometry are
incorporated into our final error in an Monte Carlo fashion.
Based on the age of the system, the evolutionary models of \citet{Burrows:1997p2706}
imply a mass of 35~$\pm$~8~\Mjup.


\subsubsection{2MASS~J15594729+4403595~B}{\label{sec:2m1559}}

2MASS~J15594729+4403595~A is an M2.0 star exhibiting H$\alpha$ emission and
saturated X-ray emission (\citealt{Riaz:2006p20030}).
As part of an ongoing search for young low-mass members, 
Shkolnik et al. (in preparation) identify this active star 
from its $GALEX$ photometry (using \citealt{Shkolnik:2011p21923} criteria), 
which is similar to known young moving group members
in $NUV$--$W1$ color (11.11~$\pm$~0.04~mag; see \citealt{Rodriguez:2013fv}).
\citet{Malo:2013p24348} found that 2MASS~J15594729+4403595~A is a likely
member of the 120~Myr AB~Dor young moving group based on its sky position, 
proper motion, and high-energy emission.  Assuming group membership, they find a kinematic
distance of 33~$\pm$~4~pc to the primary and predict a radial velocity of --28.9~$\pm$~1.8 km~s $^{-1}$.

\begin{figure*}
  \vskip -.3in
  \hskip .6 in
  \resizebox{5.5in}{!}{\includegraphics{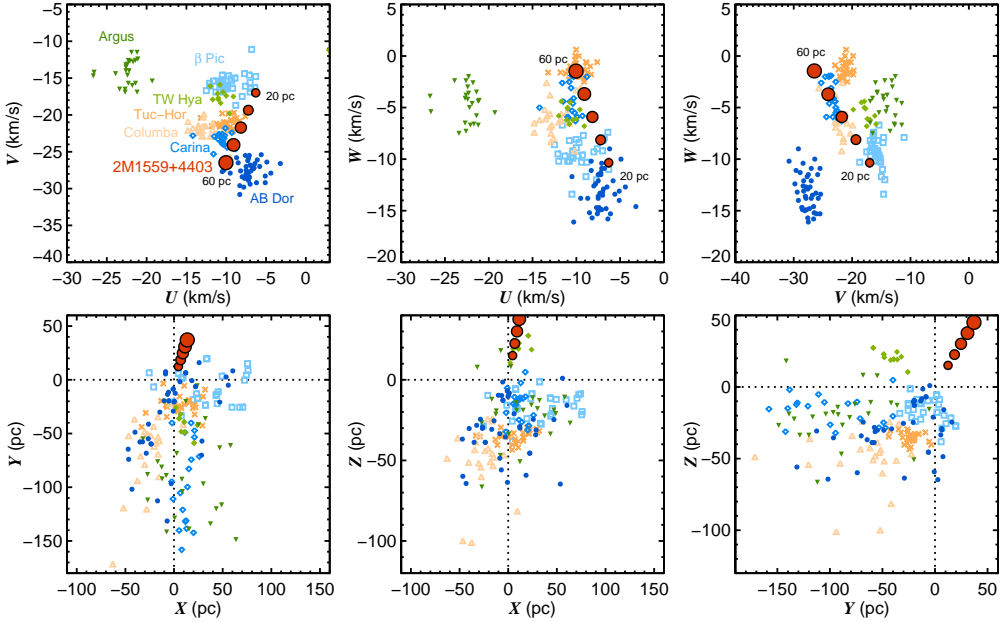}}
  \vskip -.3in
  \caption{$UVW$ galactic velocities and $XYZ$ space positions of 2MASS~J15594729+4403595~AB compared to nearby
    young moving groups from \citet{Torres:2008p20087}.  Although we lack a parallactic distance, the measured radial velocity 
    to the primary enables partial kinematic constraints.  2MASS~J15594729+4403595~AB is consistent with several moving
    groups in $UVW$ space for distances between $\sim$20--50~pc, but physically appears to be tens of 
    parsecs from these same groups.  Given this physical discrepancy, we find no convincing association with a known
    moving group. \label{fig:2m1559_uvwxyz} } 
\end{figure*}

Shkolnik et al. (in preparation) measure a radial velocity of --19.6~$\pm$~0.6 km~s $^{-1}$ for 2MASS J15594729+4403595
as part of their follow-up efforts to kinematically associate nearby young stars with moving groups.
Assuming the primary is not a single-lined spectroscopic binary, this velocity disagrees with the
prediction by Malo et al. for AB Dor.  In Fig~\ref{fig:2m1559_uvwxyz} we show the partial kinematic constraints 
for distances between 20--60~pc.  2MASS~J15594729+4403595 is consistent with $\beta$~Pic and Carina
at $\sim$20~pc and $\sim$50 pc.  However, the $XYZ$ positions disagree with all moving groups so we conclude
that it is probably not a member of these known groups, which prevents 
precise age-dating through coevality with a young cluster.  Additional radial velocities will help determine whether the primary
is RV-stable and this measurement represents the systemic RV.

\begin{figure}
  \begin{center}
  \resizebox{3.5in}{!}{\includegraphics{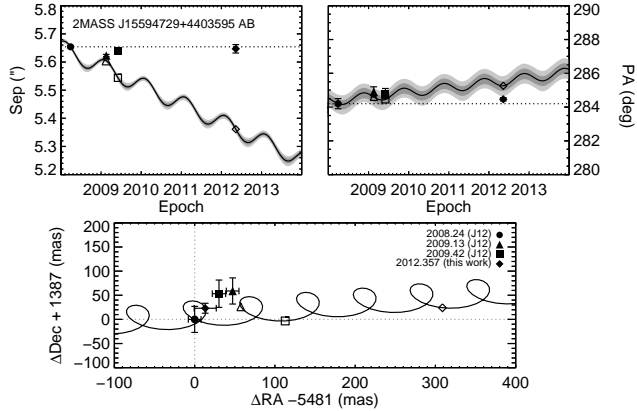}}
  \caption{Relative astrometry of 2MASS~J15594729+4403595 B compared to the expected background track of a stationary object (solid line).  The first three epochs in 2008 and 2009 are from \citet{Janson:2012p23979}.  We
  verify the companion is unambiguously comoving from our AO imaging in 2012.   \label{fig:2m1559_background} } 
\end{center}
\end{figure}

\citet{Janson:2012p23979} imaged the system three times between 2008 and 2009, identifying 
2MASS~J15594729+4403595~B at 5$\farcs$6 (187~$\pm$~23~AU) and confirming its physical association with the primary.
Their $i'$- and $z$-band contrasts imply a spectral type of $\approx$M8 for the companion.
We imaged the system in five filters at a single epoch in 2012 with HiCIAO (Figure~\ref{fig:bdimgs}).
Our astrometry listed in Table~7 are consistent with that of Janson et al. and
do not show signs of orbital motion (Figure~\ref{fig:2m1559_background}).

\begin{figure*}
  \vskip -1.in
  \hskip -.1in
  \resizebox{7in}{!}{\includegraphics{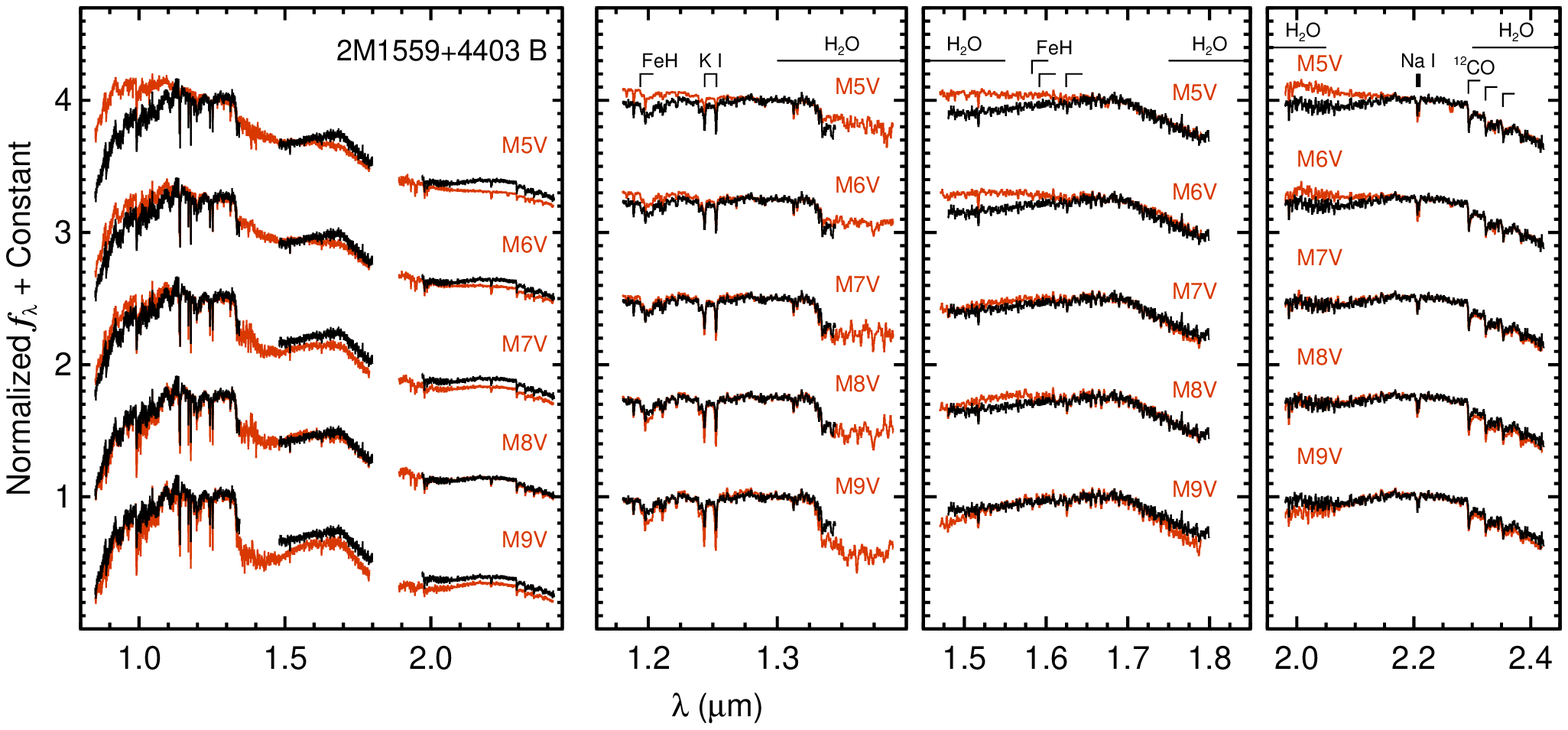}}
  \vskip -1in
  \caption{Our IRTF/SpeX SXD spectrum of 2MASS~J15594729+4403595~B compared to field templates from the IRTF
  SpeX Spectral Library. 2MASS~J15594729+4403595~B is a good match to the field M8 template across all bandpasses,
  though the $H$-band shape is noticeably more angular in 2MASS~J15594729+4403595~B.  We adopt a spectral type
  of M8.0~$\pm$~0.5.  The comparison objects are the same as in figure Figure~\ref{fig:gj3629b_hgcomp}.  All spectra have 
  been smoothed to a common resolving power of $R$$\approx$1200. \label{fig:twom1559b_hgcomp}}
\end{figure*}

We derive a $Y$--$J$ color of 0.80~$\pm$~0.05~mag for 2MASS~J15594729+4403595~B in a similar fashion as for 1RXS J034231.8+121622~B.
Compared to ultracool dwarfs in \citet{Rayner:2009p19799}, we infer a photometric spectral type of M7.5~$\pm$~1.
Our 0.8--2.45~$\mu$m IRTF/SpeX spectrum of 2MASS~J15594729+4403595~B is 
shown in Figure~\ref{fig:twom1559b_hgcomp}.
The best-fit spectral type across the entire spectrum is M8.  M8, M7, and M7--M8 templates provide the
best matches to field templates (Table~6).  
The gravity-insensitive index-based near-infrared classification schemes of 
\citet{Allers:2007p66} and \citet{Slesnick:2004jy} imply spectral types of M6.8~$\pm$~0.4, M7.6~$\pm$~1.1, and M7.6~$\pm$~0.4.
Altogether we adopt a spectral type of M8.0~$\pm$~0.5.
The shallow $J$-band alkali lines and angular $H$-band are immediately clear and point to low surface gravity,
which is supported by an ``INT-G'' gravity classification using the indices of \citet[see Table~6 
and Figure~\ref{fig:gravindices}]{Allers:2013p25314}.
Indeed, 2MASS~J15594729+4403595~B closely resembles the intermediate-gravity M8 object 
2MASS~J00192626+4614078 from \citet{Allers:2013p25314} in Figure~\ref{fig:palms4_bd_spec_comps}.

Although low-gravity features in young brown dwarfs are not yet fully calibrated with empirical benchmarks,
\citet{Allers:2013p25314} find that objects with intermediate-gravity spectra like 2MASS~J15594729+4403595~B are most 
closely linked to brown dwarfs with ages between $\sim$50--200~Myr.  Lacking a convincing association with a young moving group,
we adopt this spectroscopically-inferred age for the system.
The $H$-band photometric distance to the companion is 27~$\pm$~2~pc using the absolute magnitude-spectral type 
relation from \citet{Dupuy:2012p23924}.
An $H$-band bolometric correction from \citet{Liu:2010p21195} gives a bolometric luminosity
of --3.32~$\pm$~0.07~dex which together with the system age implies a 
mass of 43~$\pm$~9~\Mjup \ for 2MASS~J15594729+4403595~B based on the 
evolutionary models of \citet{Burrows:1997p2706}.  


\subsection{Stellar Binaries and Multiples from the PALMS Survey}

\begin{figure}
  \vskip -.3in
  \begin{center}
  \resizebox{3.5in}{!}{\includegraphics{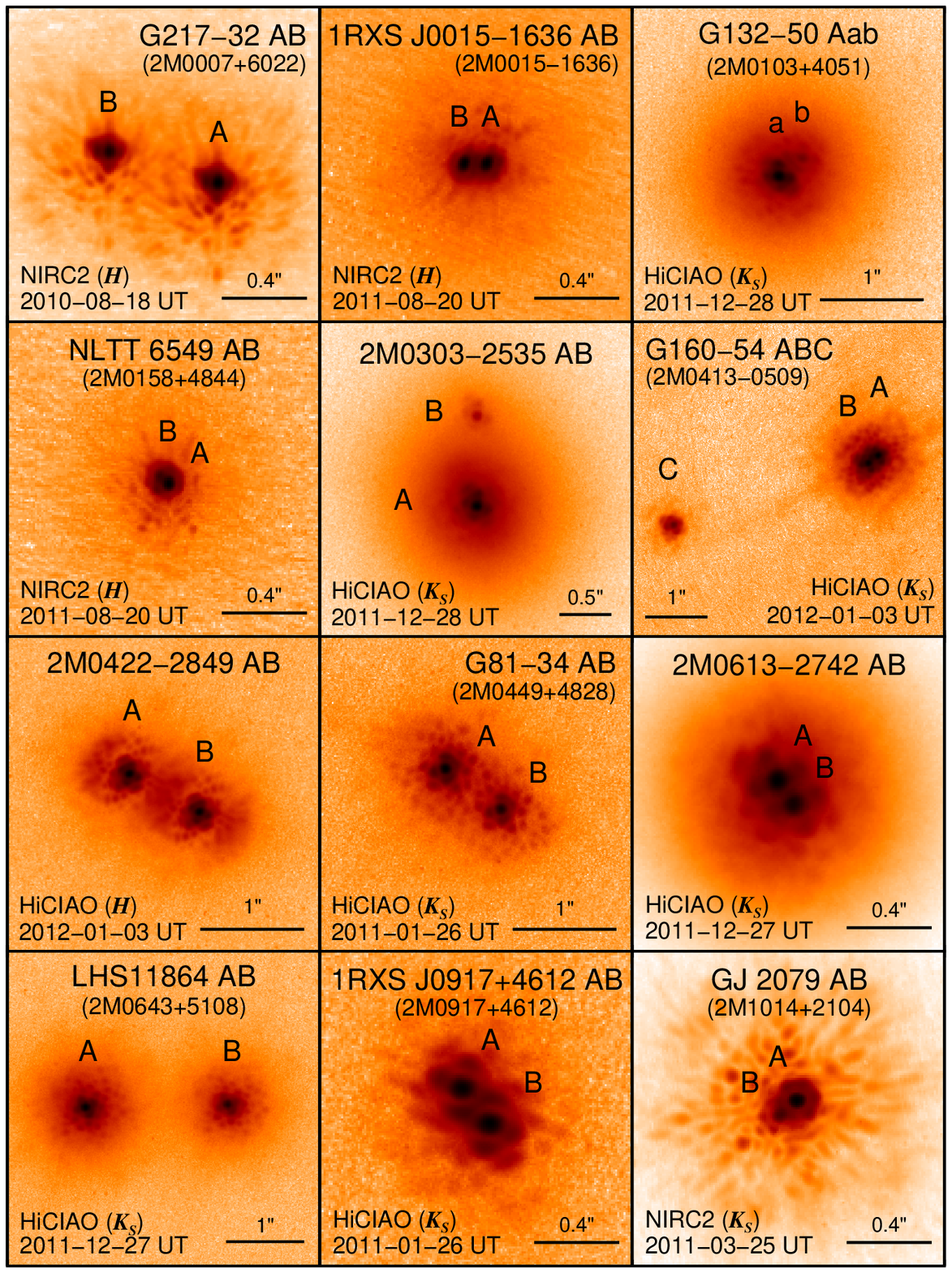}}
    \vskip -.4in
  \caption{Binary stars detected in our survey.  North is up and East is left.  \label{fig:binimgs_pg1} } 
\end{center}
\end{figure}

\begin{figure}
  \vskip -.3in
  \begin{center}
  \resizebox{3.5in}{!}{\includegraphics{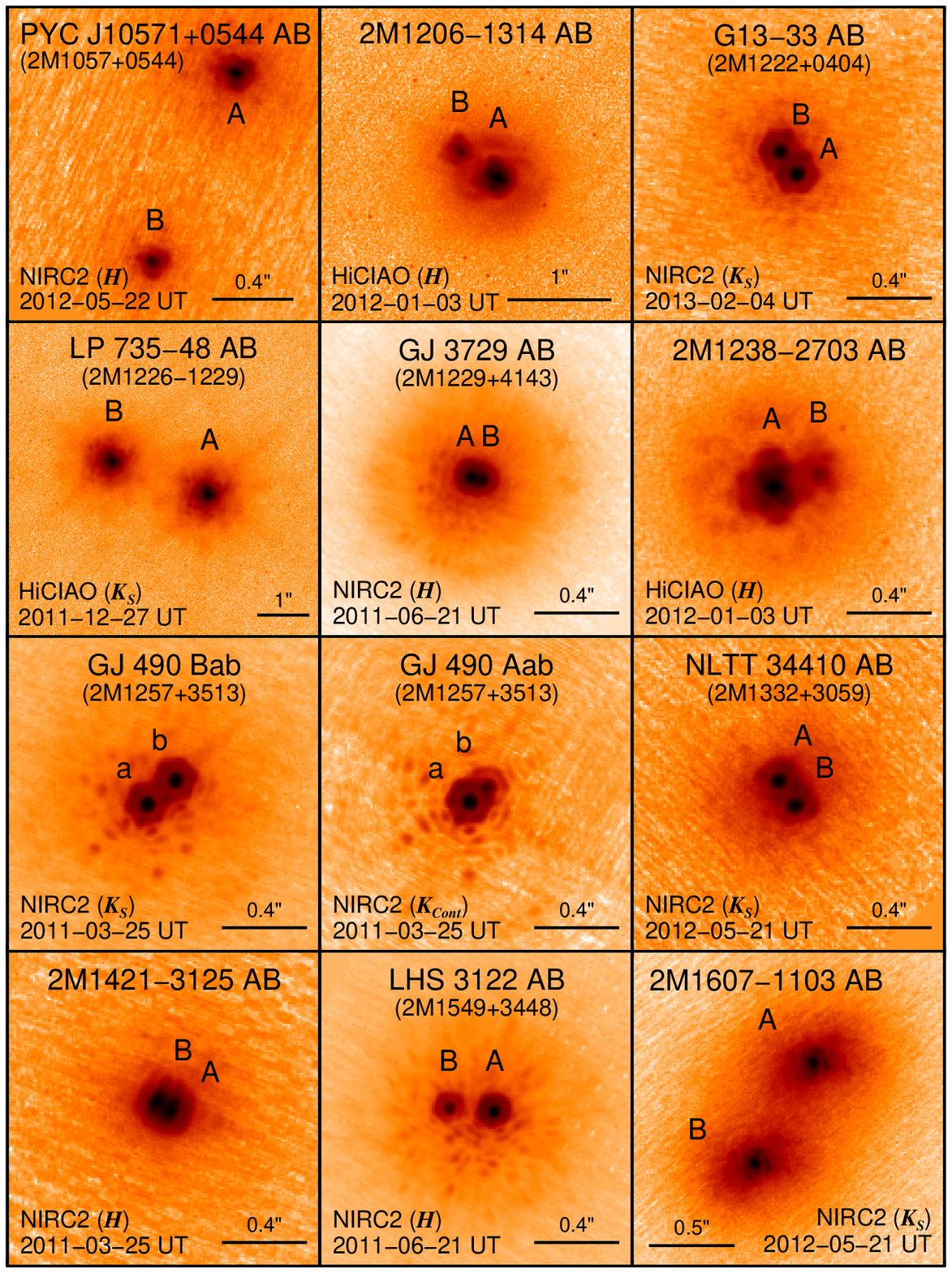}}
    \vskip -.4in
  \caption{Binary stars detected in our survey (continued).  North is up and East is left. \label{fig:binimgs_pg2} } 
\end{center}
\end{figure}

\begin{figure}
  \vskip -.3in
  \begin{center}
  \resizebox{3.5in}{!}{\includegraphics{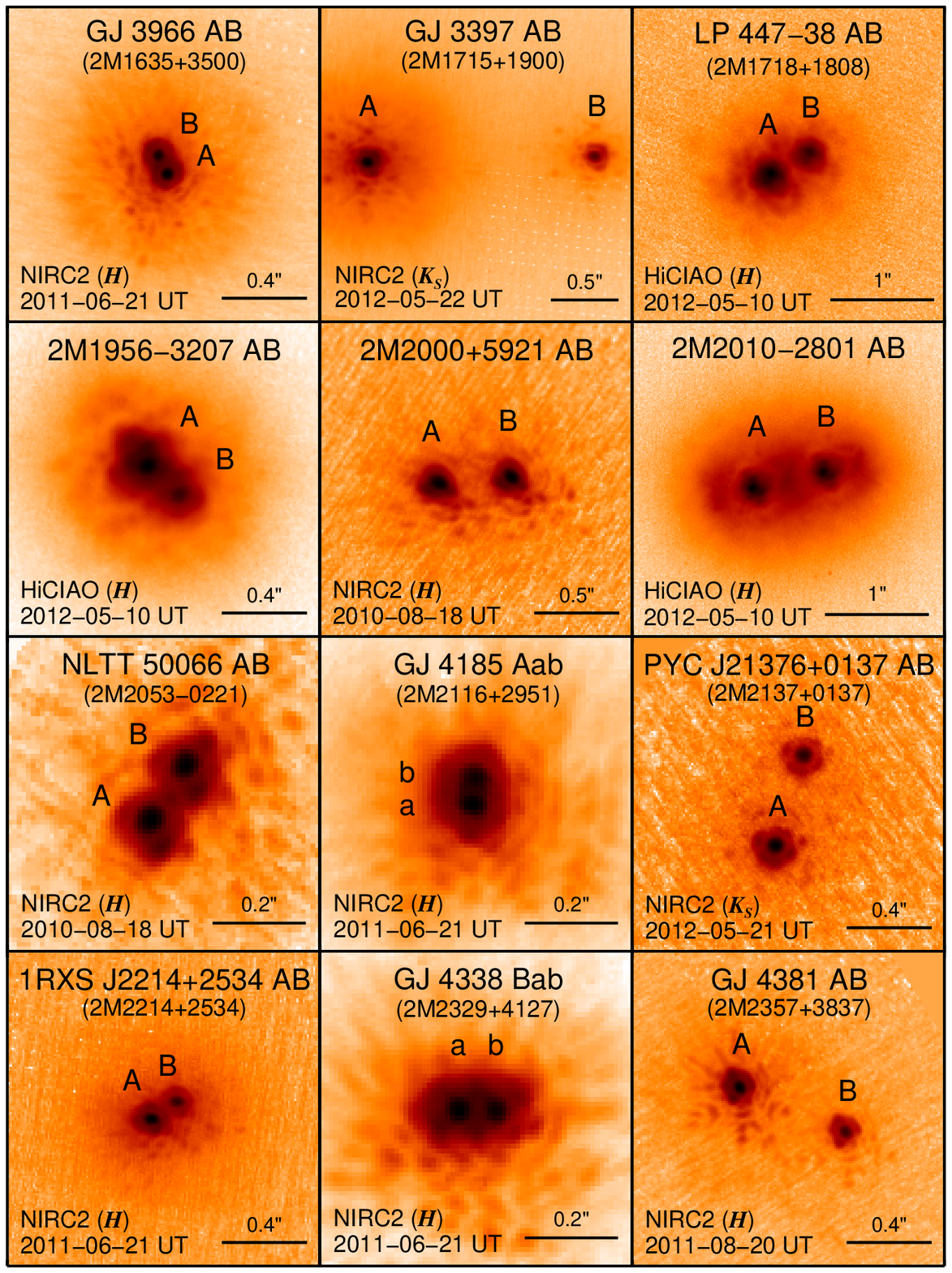}}
    \vskip -.4in
  \caption{Binary stars detected in our survey (continued).  North is up and East is left.  \label{fig:binimgs_pg3} } 
\end{center}
\end{figure}

Most of our targets have not been previously imaged with AO and, as expected, many 
were found to be close binaries (Table~7
and Figures~\ref{fig:binimgs_pg1}--\ref{fig:binimgs_pg3}).
Altogether 43 stars in our sample have stellar companions with projected separations less than 100~AU.
We resolve 38 systems into binaries with angular separations from 50~mas to several arcseconds;
17 of these are separated by $<$5~AU in projection, 29 are separated by $<$20~AU, and 
37 of these are separated by $<$100~AU. 
Among these, 27 are either new or spatially resolved for the first time in this work.  
An additional five targets not resolved
in our data were found to be close spectroscopic binaries either from the literature or from Shkolnik et al. (in preparation).  
One additional target not resolved in our survey, LP~449-1~AB, 
was identified as a 50~mas binary by \citet{Riedel:2014ce} with $HST$ Fine Guidance Sensor interferometry.

We intentionally vetted our initial target sample for previously known close binary systems
which were found in heterogeneous studies.  Since measuring the stellar 
companion mass function was not an original goal of this survey, 
we make no attempt to analyze the statistical properties of the multiples we uncovered.
Nevertheless, we find that at least 43 out of 122 of our targets have stellar companions 
within 100~AU, implying a minimum companion frequency of $>$35.4~$\pm$~4.3\%.
This agrees well with the established close companion fraction of 33~$\pm$~5\% 
(\citealt{Duchene:2013p25590}).

Several binaries were not seen in our first epoch of deep imaging and were only resolved 
in follow-up observations of wide planet candidates. 
These data are not incorporated into our statistical analysis so as to
maintain a homogeneous sample of single stars.
Note that the closest binaries with projected separations less than a few AU will yield 
dynamical masses on short timescales.

Seven targets in our sample form higher order hierarchical multiple systems.  Five of these are triple systems
(G~160-54~ABC, GJ~9652~Aab~+~GJ~9652~B, 2MASS~J19560294--3207186~AB~+~TYC~7443-1102-1, 
GJ~4185~Aab~+~GJ~4186~B, GJ~4338~Bab~+~GJ~4337~A) and two make up quadruple systems (G~132-50~Aab~+~G~132-51~Bab,
GJ~490~Aab+~GJ~490~Bab).


\subsection{Candidate Planets from the PALMS Survey}{\label{sec:candplanets}}

\begin{figure}
  \vskip -0.3in
  \hskip -.7in
  \resizebox{5in}{!}{\includegraphics{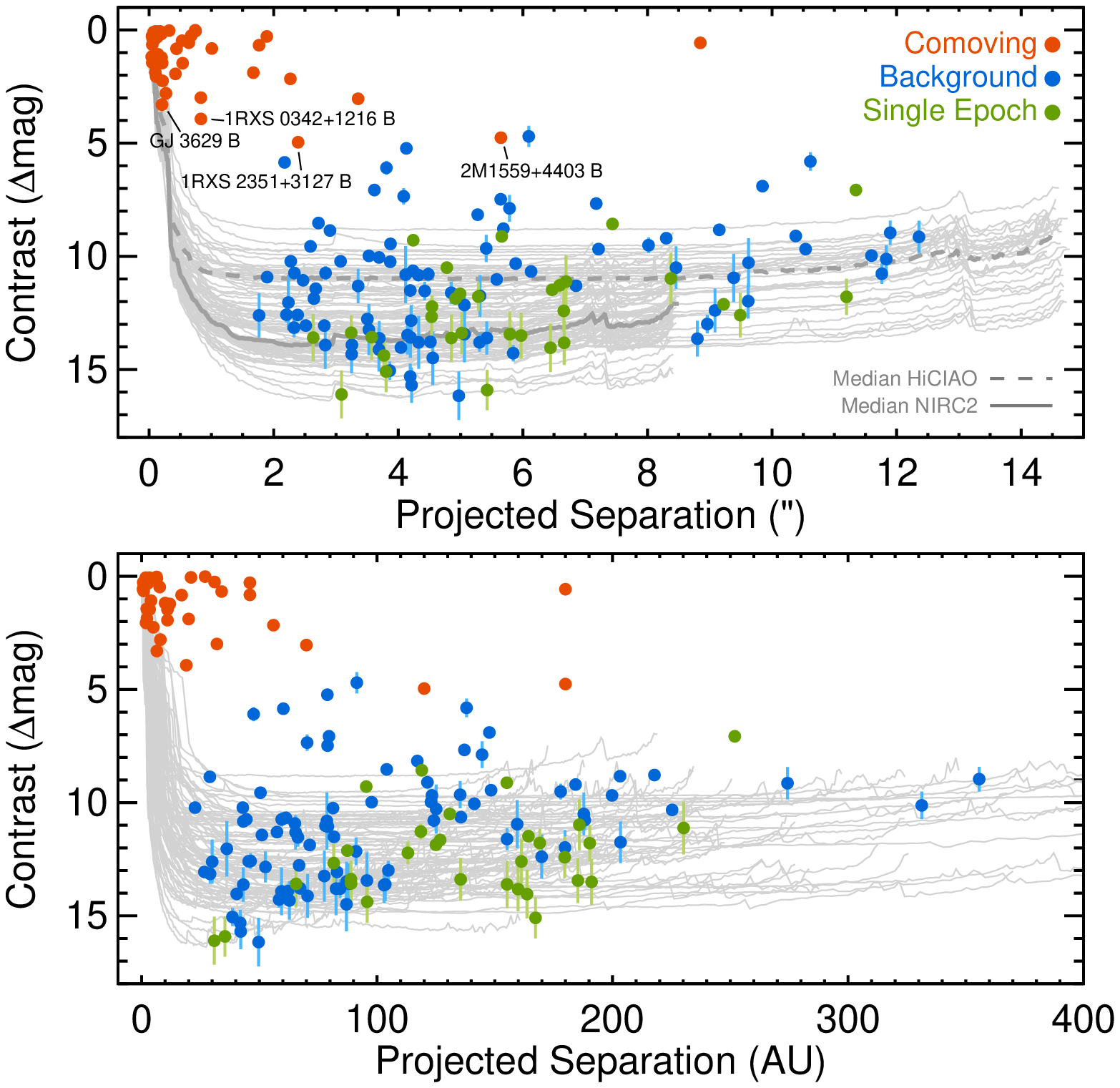}}
  \vskip -0in
  \caption{Point sources detected in our survey.  Comoving stellar and substellar companions are shown in red, background objects are plotted in blue, and objects with only a single epoch of astrometry are in green.  The individual and median NIRC2 and HiCIAO contrasts are overplotted in gray.  
In the top panel the measured angular separations and contrasts are displayed, while the projected separation is used in the bottom panel.  For clarity we have excluded several dozen single epoch candidates in the single crowded field surrounding GJ~9652~A.  \label{fig:candidates} } 
\end{figure}

Altogether 167 faint point sources were identified around 45 stars (singles and binaries) in our deep imaging (Figure~\ref{fig:candidates}).  
42 of these candidates are in the crowded low-galactic latitude field surrounding the single-line 
spectroscopic binary GJ~9652~A.  
Astrometry and relative photometry are listed in Table~8. 
Background stars are distinguished from comoving gravitationally bound companions using 
two or more epochs of follow-up imaging, with a prioritization for those at small projected separations under 100~AU.
In some cases candidates are visible in archival wide-field imaging 
surveys like the Digitized Sky Survey (first and second generations), the Sloan Digital Sky Survey (\citealt{Abazajian:2009p18572}), 
and 2MASS (\citealt{Skrutskie:2006p589}) and were rejected if their astrometry and/or colors 
were inconsistent with cool, comoving companions.  
Inevitably our second-epoch observations uncovered candidates not seen in our first epoch data, but 
most of these reside at wide separations beyond the main region of interest  ($\sim$10--100~AU).

For candidates with multiple epochs of astrometry we calculate
a reduced chi-squared value for a background scenario, $\chi^2_{\nu, BG}$, and a common proper motion scenario, $\chi^2_{\nu, CPM}$.
Here

\begin{displaymath}
\chi^2_\nu = \frac{1}{\nu} \sum_{i=1}^{N-1} \left( \frac{(\theta_{meas, i} - \theta_{pred, i})^2}{\sigma_{\theta, meas, i}^2 + \sigma_{\theta, pred, i}^2} + 
 \frac{(\rho_{meas, i} - \rho_{pred, i})^2}{\sigma_{\rho, meas, i}^2 + \sigma_{\rho, pred, i}^2} \right),
\end{displaymath}
where $\theta$, $\rho$, and $\sigma$ are the measured and predicted 
PA, separation, and their associated uncertainties at epoch $i$ for $N$ epochs of astrometry.
$\nu$ is the number of degrees of freedom, equal to 2$\times$$N$ -- 1. 
For the background case, the predicted measurements incorporate the proper motion and distance
to the target as listed in Table~2.  For the co-moving scenario, the predicted PA and
separation assume no orbital motion as expected for companions on wide orbits.

Table~9 summarizes our tests for common proper motion for candidates with at least two epochs of astrometry.  
93 candidates are consistent with background stars.  
The status of one candidate, 1RXS J124147.5+564506-CC1, is ambiguous.  The remaining 73 only have a single epoch of astrometry. 
We do not identify any planets in our sample.


\begin{figure*}
  \vskip -.6 in
  \hskip -0.5 in
  \resizebox{8in}{!}{\includegraphics{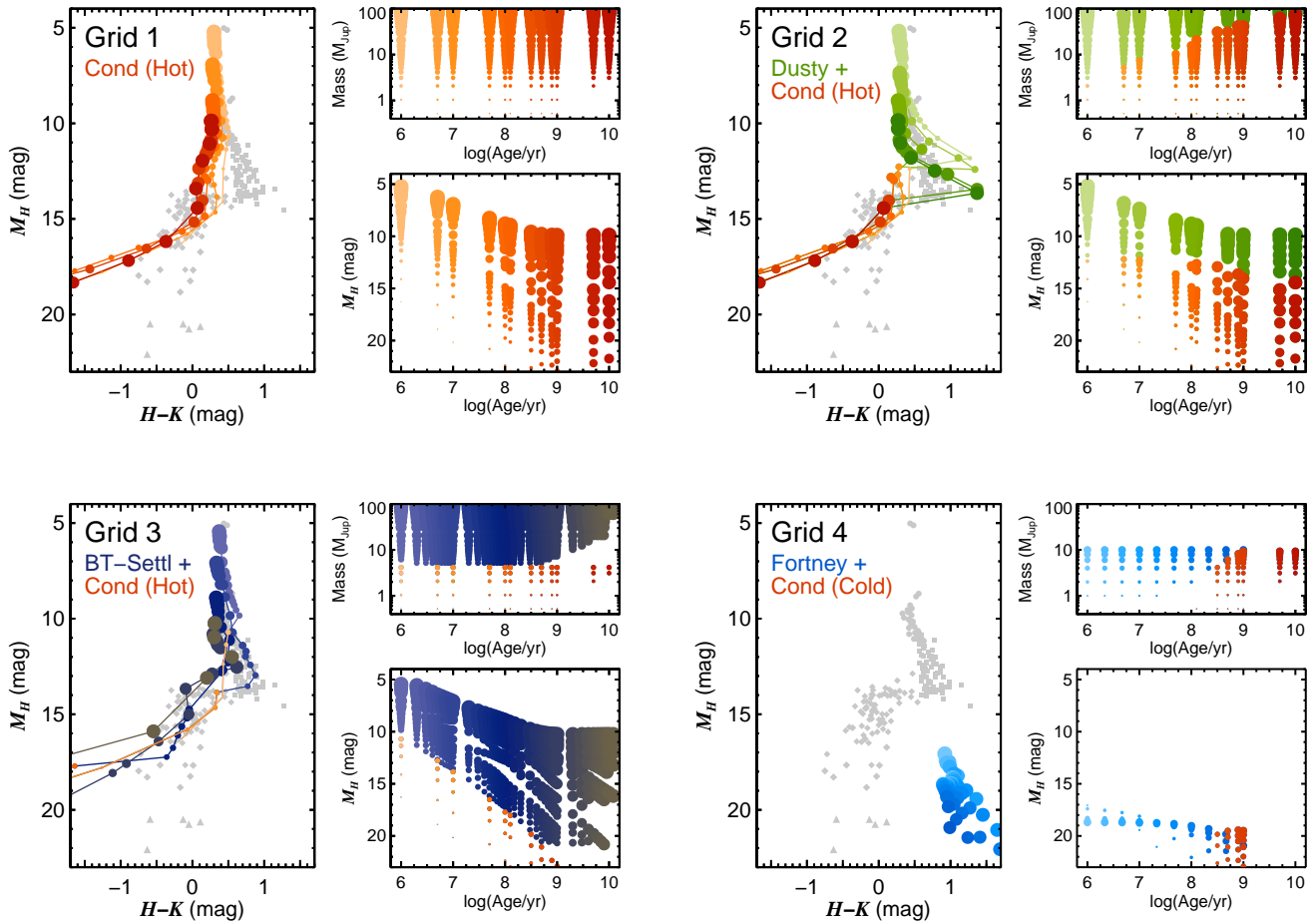}}
  \vskip -.2in
  \caption{Evolutionary model grids used in this work.  Three of the cases (Grids 1--3) are based on a hot-start formation scenario, 
  while the fourth (Grid~4) follows a cold-start prescription.  Each 3-panel set shows the predicted color-magnitude sequence in 
  $M_H$ vs. $H$--$K$ compared to the observed sequence of MLTY dwarfs (gray; from \citealt{Dupuy:2012p23924} and \citealt{Dupuy:2013ks}),
  the grid sampling in age and mass, and the evolution of $M_H$ with time.  Mass ranges of 0.5--100~\Mjup \ (0.5--10~\Mjup) are
  shown for hot-start (cold-start) cases with symbol sizes scaling with mass.  Grid~1 shows the Cond models of
  \citet{Baraffe:2003p587}, which poorly reproduce dusty L dwarfs and mid-to late-T dwarfs in color.  Grid~2 is a hybrid
  of Dusty models from \citet{Chabrier:2000p161} above 1500~K and Cond models at lower temperatures.  Grid~3 shows the 
  BT-Settl models from \citet{Allard:2011p23143} above 5~\Mjup \ and the Cond grid at lower masses, producing the best-fit
  to the M, L, and early-T sequence.  The cold-start scenario
  with slight (5$\times$ solar) metal-enrichment from \citet{Fortney:2008p8729} is shown in Grid~4  and is supplemented with
  Cond models at older ages.  The three hot-start cases predict similar evolution of absolute magnitude with planet mass and,
  overall, produce very similar statistical results in this study.
    \label{fig:evmods} } 
\end{figure*}


\begin{figure}
  \vskip -.4 in
  \hskip -1 in
  \resizebox{6in}{!}{\includegraphics{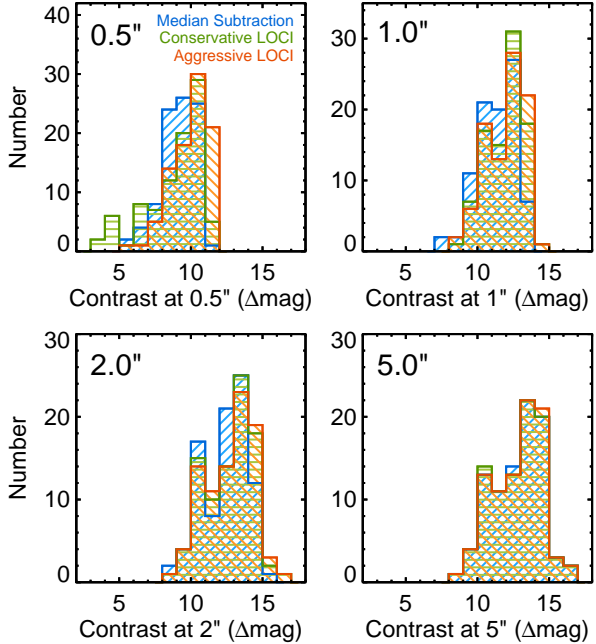}}
  \vskip -.5in
  \caption{Comparison of three methods of PSF subtraction between 0$\farcs$5--5$''$.  Scaled median subtraction (blue) systematically produces slightly worse (7-$\sigma$) contrasts compared to conservative (green) and aggressive (red) applications of LOCI.  At separations of 0$\farcs$5 near the IWA, the median subtraction outperforms conservative LOCI when sky rotation is small.  Beyond 3$''$ where the data are no longer contrast-limited, the median scaled subtraction method is used for all contrasts and the distributions are virtually identical.  \label{fig:locihists} } 
\end{figure}


\begin{figure}
  \vskip -.2in
  \hskip -.9in
  \resizebox{4.6in}{!}{\includegraphics{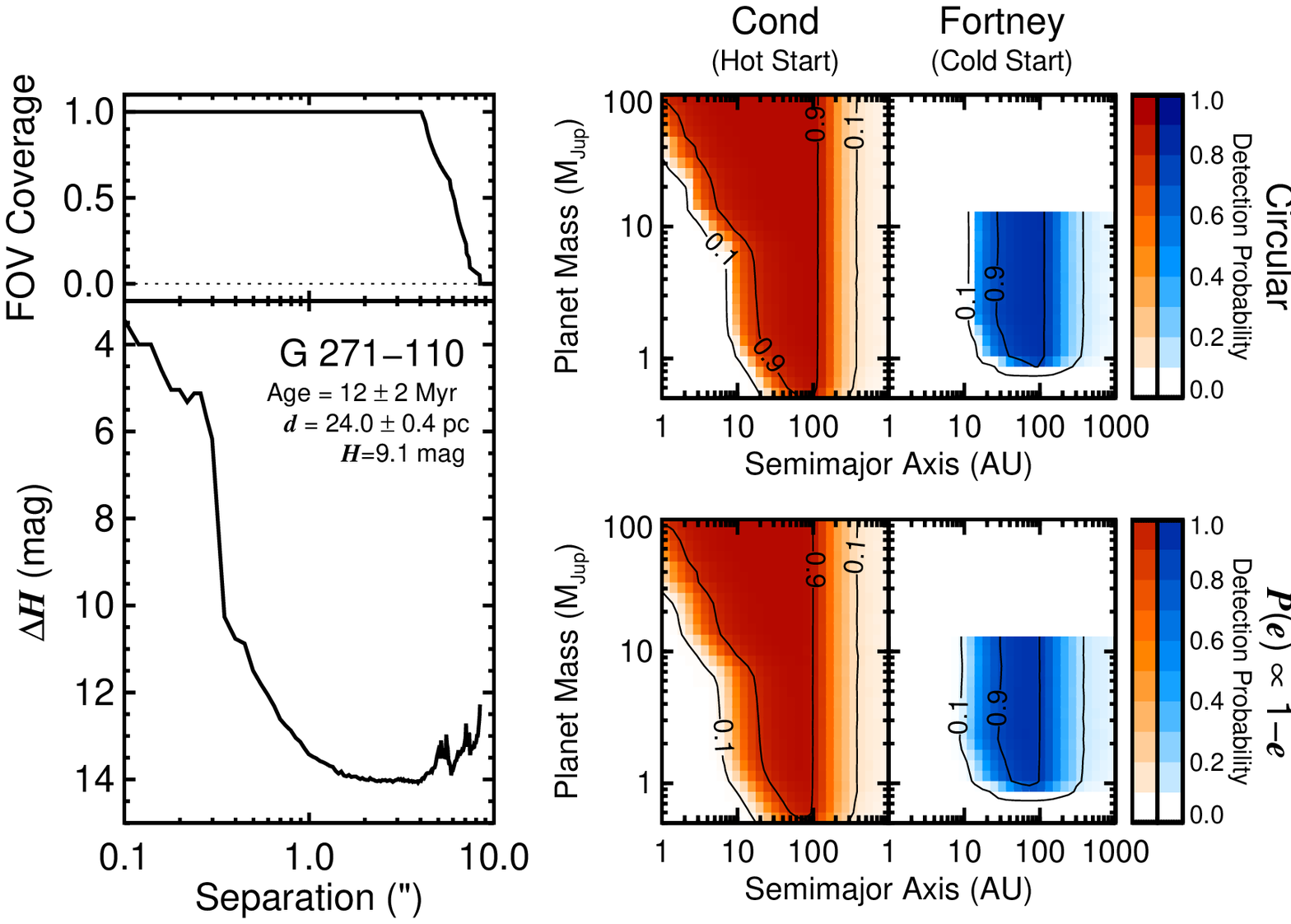}}
  \vskip -.4in
  \caption{Example of a sensitivity map for G~271-110.  Colors correspond to the fraction of simulated random orbits recovered
  for each \{mass, semi-major axis\} grid point based on two different evolutionary models: Cond (hot-start) and Fortney (cold-start).  
  The top panels show results for circular orbits, while the bottom panel follows a 1--$e$ eccentricity distribution.   \label{fig:plsensitivityex} } 
\end{figure}

\section{Statistical Analysis}{\label{sec:stats}}

Our null detection of planets provides powerful constraints on the outer architecture of planetary systems around low-mass stars.
Likewise, our four brown dwarf discoveries allow us to measure the frequency of brown dwarf companions to M dwarfs
over a variety of separations.
Since we are not uniformly sensitive to companions in mass and separation, a sensitivity map for each target
must be considered.  Similarly, our brown dwarf discoveries were made in projected separation and their distributions in
semi-major axis space must be inferred to derive accurate statistical constraints.
In the following analysis we assume two forms of eccentricity distributions to test their influence on our results:
circular orbits ($e$=0) and eccentricities following a 1--$e$ distribution, in which most planets have small or modest 
eccentricities (see Section~\ref{sec:masssensitivity}).  The latter case is motivated by the distribution of 
RV-detected planets (\citealt{Kipping:2013p25591}) and M dwarf binaries (\citealt{Duchene:2013p25590}).
Below we describe two approaches to derive statistical constraints over a range of companion masses and separations.

\subsection{Contrast Curve Selection Guidelines}

A common complication of large direct imaging surveys is that not all faint planet candidates can be followed up with
second-epoch astrometry.  Finite telescope time, different seeing conditions and AO correction, 
and varying field of view rotation can both prevent candidates found in first-epoch imaging from being recovered 
and reveal new fainter or wider point sources.
In this survey we found that 93 out of 167 faint point sources are stationary background stars.
Care must therefore be taken in our statistical treatment of  the remaining 74 candidates with unclear status.
Note, however, that only 8 of these are within projected separations of 100~AU around single stars in our sample
and most of them (42) come from a single low-galactic latitude target.

Following the recipe of \citet{Nielsen:2013jy}, we define selection guidelines for choosing contrast curves
to use in our statistical analysis.  These are considered on a case-by-case basis for each target in our survey:

\begin{enumerate}
  \item If no candidates are identified in a first-epoch observation and no subsequent deeper imaging is obtained, then 
            the contrast curve is used for our statistical analysis. 
  \begin{enumerate}
    \item If a later, deeper epoch of imaging uncovers candidates that are shown to be background, this deeper contrast curve is used.
    \item If a later, deeper epoch uncovers candidates that are \emph{not} shown to be background, then the initial 
              candidate-free first epoch contrast curve is used.
  \end{enumerate}
  \item If one or more candidates are identified at the first epoch and are shown to be background objects from subsequent imaging,
            and no other candidates are identified in the follow-up observation, then the deeper of the two contrasts is used.
  \begin{enumerate}
     \item If a second epoch reveals additional candidates that only have a single epoch of astrometry, then the first epoch contrast curve is used.
    \item If a second epoch fails to recover one or more candidates then this second epoch contrast curve is used.  This is analogous
              to (1b) but in reverse order.
  \end{enumerate}
   \item If only a single first-epoch observation is obtained and one or more candidates are identified then the floor
             of the contrast curve is defined to be 2-$\sigma$ above the brightest candidate with unknown status, where $\sigma$
             is the uncertainty in the relative contrast of that candidate.  Since we have no information about 
             whether single-epoch candidates
             are background or comoving, the raw contrast curve cannot be included in the statistical analysis.  Instead, we homogeneously
             remove all information about companions below the threshold of the brightest single-epoch candidate in the image.
\end{enumerate}

\subsection{Wide Stellar Binaries}

Wide stellar binaries beyond $\sim$100~AU can dramatically influence the outer regions of planetary systems by creating 
dynamically unstable zones where planets cannot exist on long timescales.  These wide binaries must therefore be taken into 
account in the statistical analysis of the survey.  Table~10 lists the multiplicity properties of the sample.  Altogether,
25 of our targets have companions beyond 100~AU.  

Our treatment of wide binaries follows that of \citet{Nielsen:2013jy} and is based on simulations 
by \citet{Holman:1999ti} of stability zones surrounding close-in planets with a wide stellar companion (S-type orbits)
and wide-separation circumbinary planets (P-type orbits).  Holmam \& Wiegert show that the region of stable
orbits is a strong function of both binary eccentricity and mass ratio.  The eccentricity distribution of
these (very) wide-separation binaries is unknown because of their long orbital periods.  
However, \citet{Abt:2006ux} showed that eccentricities become increasingly uniform 
(i.e., random) at long periods (10$^5$--10$^6$ days), with 
a mean eccentricity tending to 0.5.
We therefore adopt eccentricities of 0.5 for wide binary companions in our sample.  Assuming equal mass stars, 
the critical semi-major axis for stable S-type orbits from \citet{Holman:1999ti} is $\approx$10\% of the stellar semi-major axis.
For P-type orbits, the inner stability limit is $\sim$4 times the binary semi-major axis.

Wide binaries in our sample are complicated by projection effects and their unknown current orbital phase.  We therefore
adopt a median conversion factor of 1.14 from \citet{Dupuy:2011p22603} for the case of no discovery bias to transform
projected separations into semi-major axes.  For our statistical analysis, we then assume that the region 
between 10\% and 400\% of the binary semi-major axis is devoid of planets and does not contribute 
any information to our statistical analysis.  These allowable regions are listed in Table~10.  
As described above, these are conservative stability limits assuming a 
star-wide binary companion eccentricity of 0.5 and coplanar binary and planetary orbits.  Relaxing these constraints
would provide more room for dynamically-stable planets to reside.  This is further complicated by the uncertain 
conversion between projected and physical separation among our wide binaries.  
Indeed, \citet{Tokovinin:2006p22902} found the empirical limit for dynamical stability of triple star systems 
is near period ratios of 5 (that is, $P_3$/$P_1$$>$5), which may be a more realistic boundary.
Note that we do not exclude regions surrounding our brown dwarf discoveries for dynamical reasons 
since we had no \emph{a priori} knowledge of their existence.  In this way our sensitivity maps for these
targets contribute to the number of trials and the discoveries contribute to the number of detections in these regions
(see Section~\ref{sec:plfreq}).

\subsection{Mass Sensitivity}{\label{sec:masssensitivity}}

Converting contrast curves into sensitivity limits in planet mass and separation requires the use
of substellar evolutionary models.
These cooling curves in turn depend on assumptions about the way in which planets form.
``Hot-start" models (e.g., \citealt{Burrows:1997p2706}; \citealt{Saumon:2008p14070}) slowly radiate 
their initial gravitational potential energy over time
and therefore best represent formation via disk instability (e.g., \citealt{Boss:1997p18822}; \citealt{Mayer:2002p22604}).
On the other hand, ``cold start'' and ``warm start'' models 
(e.g., \citealt{Marley:2007p18269}; \citealt{Spiegel:2012p23707}; \citealt{Molliere:2012p25183}; \citealt{Bodenheimer:2013p25248}) 
follow a core accretion prescription, which assumes
significant loss of initial entropy at formation through punctuated energy dissipation associated with accretion events.
In addition to differences in initial conditions, though to a lesser degree, assumptions about the atmospheric properties
of giant planets can also influence both the rate at which planets cool and the evolution of their spectra (e.g., \citealt{Chabrier:2000p161}).

We adopt four sets of evolutionary models for this survey to reflect uncertainties in the formation
and atmospheric properties of giant planets.
Our choices are based on the accuracy of the models in reproducing the observed colors of brown dwarfs
and giant planets and on the sampling of the various publicly available grids in mass and age.
The properties of all four grids are summarized in Figure~\ref{fig:evmods}.
We selected solar-metallicity hot-start models incorporating three general prescriptions of 
photospheric dust: (1) the Cond models of \citet{Baraffe:2003p587}, in which dust is modeled as having
already formed and settled below the photosphere; (2) the Dusty models of \citet{Chabrier:2000p161},
which present an extreme view of photospheric dust formation and retention at all temperatures;
and (3) the BT-Settl isochrones from \citet{Allard:2011p23143}, which simulate the growth and sedimentation
of dust across the M/L/T transitions.
The Cond models are well sampled from ages of 1~Myr to 10~Gyr and masses of 0.5~\Mjup \ to 0.1~\Msun.
Dusty models produce better fits to the L dwarf color-magnitude sequence, but are too red 
below about 1500~K (Figure~\ref{fig:evmods}); we therefore supplement the Dusty grid with Cond models below that temperature,
resulting in a ``Dusty+Cond'' combination.  The BT-Settl models do a better job reproducing the M, L, and T
 sequence, but are not uniformly sampled at very low masses; we therefore supplement that grid with
 Cond models below 5~\Mjup.  
 For the cold start models we adopt the grid from \citet{Fortney:2008p8729},
 which assumes slight metal enrichment (5 times solar abundances), includes masses below
 13~\Mjup, and focuses on relatively young ages ($\lesssim$1~Gyr).  At older ages, all planetary-mass objects
 should have temperatures below $\sim$600~K, so we supplement the Fortney grid with Cond models
 in that region.
  
 Our strategy to infer planet detectability for each target in the \{planet mass, semi-major axis\} plane 
 is based on Monte Carlo realizations of simulated planets on random orbits.   For a given target and 
 semi-major axis $a$ we generate 
 10$^4$ orbits projected onto the sky with random ascending node position angles, 
 arguments of periastron, orbital inclinations (drawn from a sin~$i$ distribution), 
and periastron passage times.
We consider two possible eccentricity distributions, $e$=0 and $P(e)$~$\propto$~1--$e$,
to test whether adding modest eccentricities affect the results.
This choice of the eccentricity distribution is motivated by observations of modest-period 
(100--10000~day) M~dwarf binaries and extrasolar giant planets measured from RV surveys, 
which have similar distribution shapes that peak at small eccentricities and 
diminish roughly linearly to high values (\citealt{Duchene:2013p25590}; \citealt{Kipping:2013p25591}).

For a given companion mass $m$ we use the star's distance and age together with evolutionary models to assign an 
apparent magnitude to each simulated companion.  Gaussian age distributions are adopted
for stars that belong to young moving groups, while linearly uniform distributions are used for
the rest (see Table~2).  The uncertainty in the distance is also incorporated as a Gaussian distribution.
This allows us to then compare the apparent magnitudes and sky-projected separations of all orbits
for a given \{$m$, $a$\} to our contrast curves.  The fraction of simulated companions that fall above the curve 
(the ``detections'') is the 
overall sensitivity at that grid point.  Fractional FOV coverage is also incorporated by randomly assigning
``non-detections'' to planets with a probability equal to 1 minus the azimuthal coverage at that separation.

These simulations are repeated for all grid steps in mass (from 0.5--100~\Mjup) and physical separation 
(1--1000~AU), all four sets of evolutionary models, both circular orbits and eccentricity distributions
following $P(e)$~$\propto$~1--$e$, and our three methods of PSF subtraction.  
For our statistical analysis we adopt contrast curves from the aggressive version of the 
LOCI reduction because overall they produce the best contrasts, but the resulting mass sensitivities 
are similar for all cases.  Figure~\ref{fig:locihists} shows the distribution of contrasts for our three PSF
subtraction methods.  At 1$''$, the aggressive implementation of LOCI outperforms the scaled median subtraction
and our conservative version of LOCI by 0.5$^{+0.7}_{-0.3}$~mag and 0.12$^{+0.20}_{-0.17}$~mag,
respectively.  However, this gain in contrast is only marginal in planet mass: for the typical age of our sample 
($\approx$125~Myr), the Cond models of \citet{Baraffe:2003p587} predict an $H$-band brightness difference 
between a 9 and 10~\Mjup \ (4 and 5~\Mjup) planet of 0.50~mag (0.71~mag).  Our three hot-start model
prescriptions (Cond, Dusty+Cond, BT-Settl+Cond) 
produce similar sensitivity maps, so for the rest of this work we show representative results with Cond models. 

As an example, Figure~\ref{fig:plsensitivityex} shows Cond and Fortney sensitivity maps for G~271-110
based on the contrasts for this target.  As expected, our data are not sensitive to planetary companions
within $\sim$10~AU nor any companions beyond a few hundred AU because of the limited FOV coverage.
In this case most planets in the 10--100~AU range would have been detected.
In general, introducing non-zero eccentricities tends to slightly ``smear out'' the sensitivity plots, 
but the overall impact is small.

Finally, we note that our sensitivity maps are necessarily dependent on substellar cooling models, which
remain poorly constrained by observations.  In the few instances where they have been tested
through precise dynamical mass measurements of the benchmark brown dwarf systems HD~130948~BC and Gl~417~BC, 
\citet{Dupuy:2009p15627} and \citet{Dupuy:2014uj} found that low-mass evolutionary models 
systematically overpredict brown dwarf masses by $\approx$15--25\%.  A similar result was found
by \citet{Crepp:2012p23811} with the older HR~7672~AB system.   
This potential (and worrisome) uncalibrated systematic error in cooling models is much larger than any effects caused
by our choice of eccentricity distribution or PSF subtraction method.


\begin{figure}
  \vskip -.5in
  \hskip -4.5in
  \resizebox{8.5in}{!}{\includegraphics{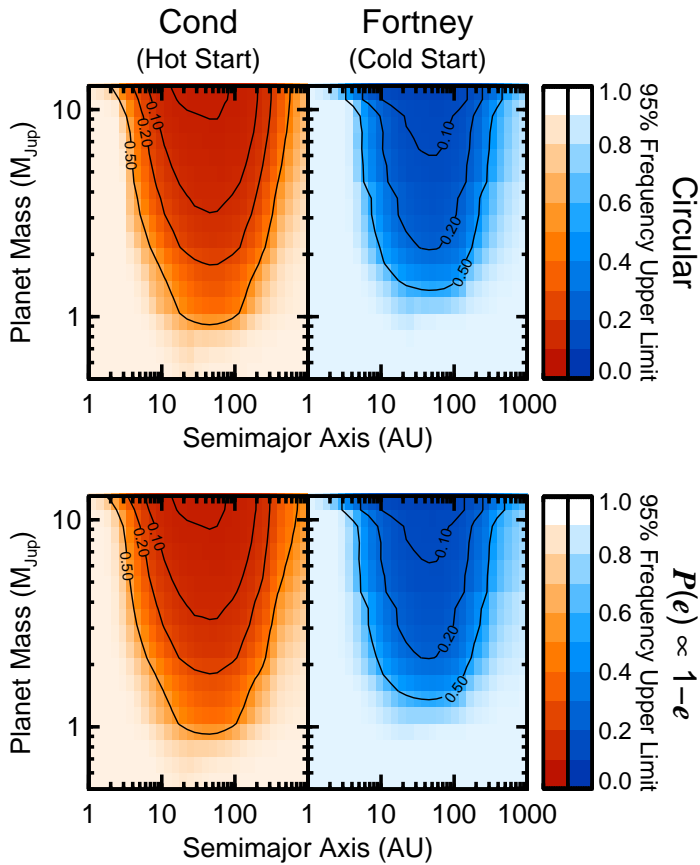}}
  \vskip -1.1in
  \caption{Upper limits on the frequency of gas giant planets.  Each grid point represents the 95\% confidence upper limit on the planet frequency.  The strongest
  constraints from our survey are for massive giant planets (5--13~\Mjup) between 10--100~AU.  Contours show the 5\%, 10\%, 20\%, and 50\% upper limits. \label{fig:uplimcont} } 
\end{figure}


\begin{figure*}
  \vskip -1in
  \resizebox{7in}{!}{\includegraphics{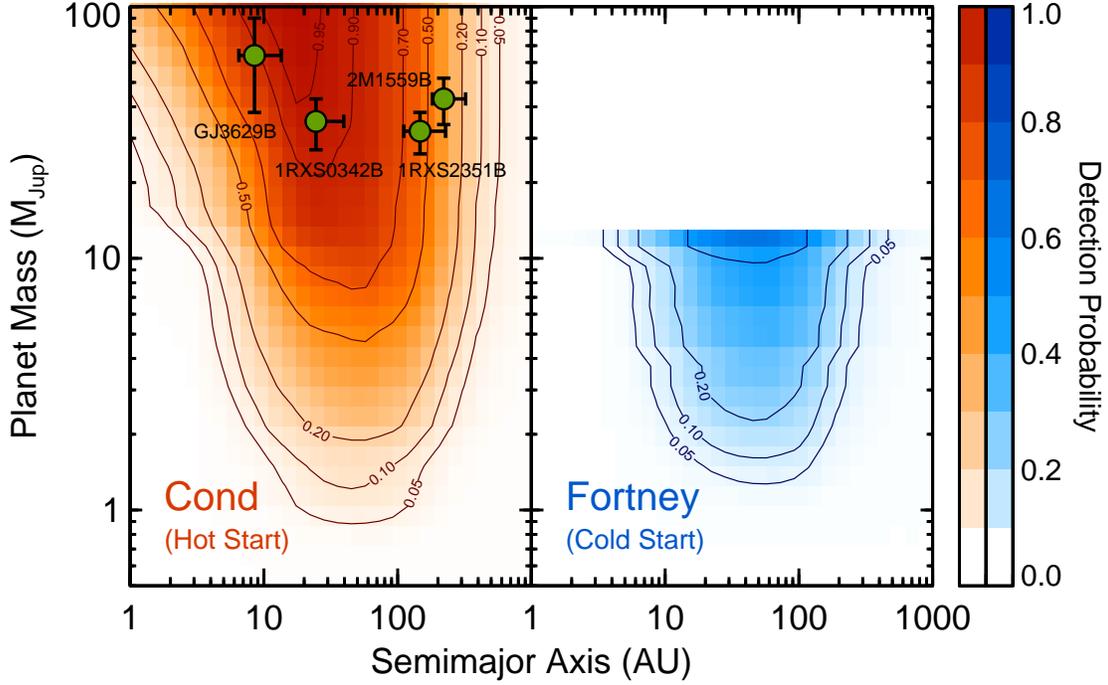}}
  \vskip -.4in
  \caption{Survey sensitivity map showing the fraction of targets sensitive to companions 
  between semi-major axes of 1--1000~AU and masses of 0.5--100~\Mjup.
  Our four brown dwarf discoveries are shown as green circles.  Contours show the 
  5\%, 10\%, 20\%, 50\%, 90\%, and 95\% levels for Cond (left) and Fortney (right)  evolutionary models.
     \label{fig:meansens} } 
\end{figure*}


\begin{figure}
  \vskip -.9in
  \hskip -.3in
  \resizebox{4in}{!}{\includegraphics{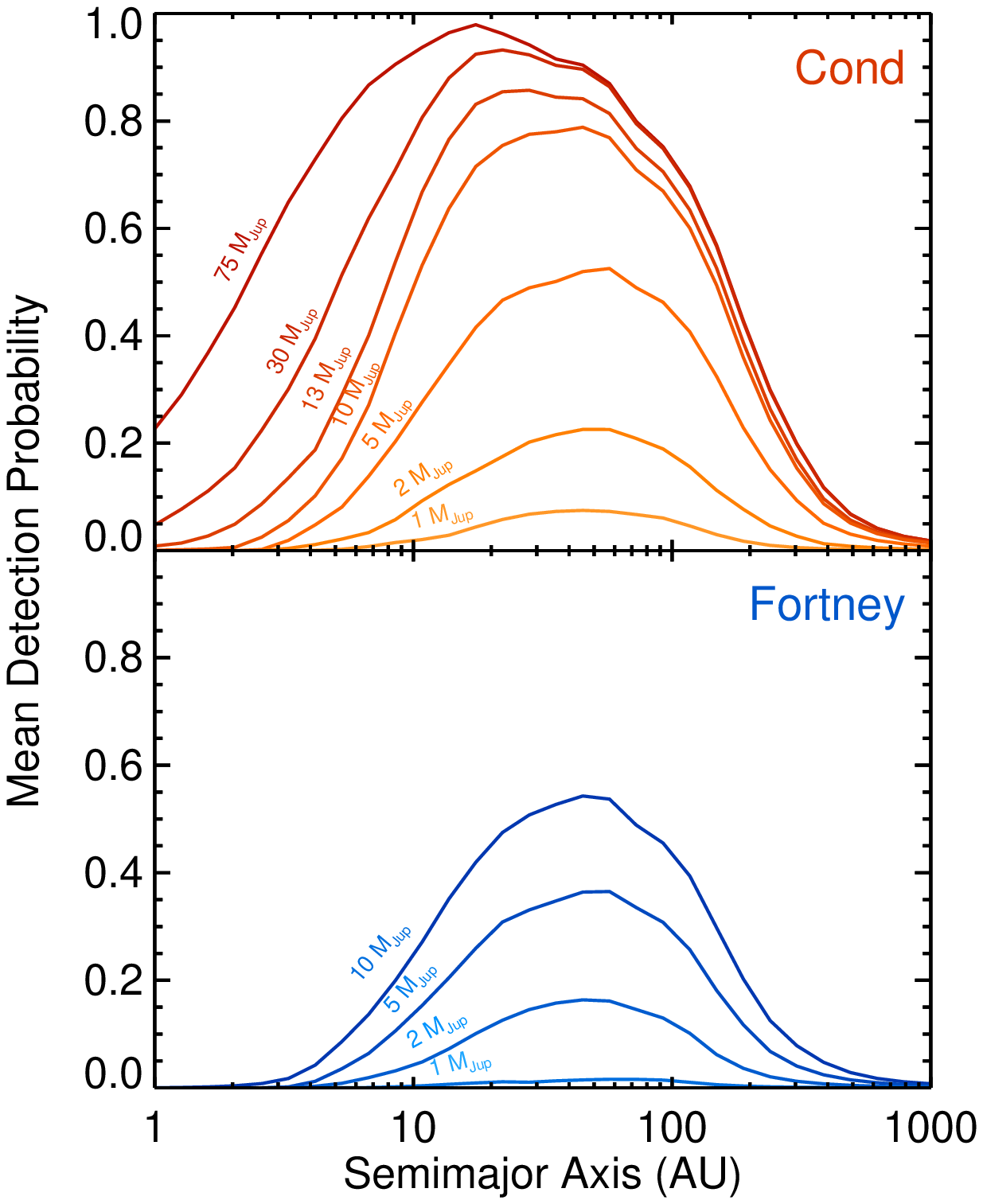}}
  \vskip -.7in
  \caption{Survey sensitivity map showing the mean detection probability as a function of semi-major axis for varying planet masses.  
     \label{fig:eurosens} } 
\end{figure}

\subsection{Giant Planet Frequency at a Given Planet Mass}{\label{sec:plfreq}}

Our first approach focuses on the following question: 
for a given planet mass and semi-major axis, 
what planet frequency is consistent with the non-detection from our survey?
Since we did not detect any planets, this analysis is concerned with
the (95\% confidence) upper limits on planet frequencies.
We use the sensitivity maps for each target (Section~\ref{sec:masssensitivity}) 
to compute the overall 95\% confidence upper limit at each \{$m$, $a$\} grid point.
For a given $m$ and $a$, the number of detections $N_\mathrm{det}$ is uniformly zero and the effective number of trials $N_\mathrm{trials}$
is simply the sum of the sensitivities at that grid point ($s(m, a)$) over all targets $N_\mathrm{tar}$:

\begin{equation}
N_\mathrm{trials} = \sum_{i=1}^{N_\mathrm{tar}}{s_i(m, a)},
\end{equation}

\noindent where $s$ is a number from 0 to 1 derived by the methods in Section~\ref{sec:masssensitivity}.
Since these constitute Bernoulli trials we can compute the probability distribution of the occurrence rate $f$ using the binomial distribution.
In a region 100\% sensitive to companions for all of our targets,
the number of trials would simply be equal to the number of targets, and the commonly used binomial distribution applies.  
On the other hand, for non-integer trials and successes the binomial coefficient can be generalized using Gamma functions.
The binomial distribution then becomes

\begin{equation}{\label{eqn:bin}}
P(f \mid n, k) = \frac{\Gamma (n+1)}{\Gamma(k+1) \Gamma(n-k+1)} f^k (1-f)^{n-k}(n+1), 
\end{equation}

\noindent where $n$ is the number of trials and $k$ is the number of successes.  The final $(n+1)$ factor is a 
normalization constant.\footnote{The meaning of a 
``trial'' and ``success'' becomes less intuitive with continuous rather than integer values.  However, noting that
$\Gamma(x+1)=x!$ for integer values of $x$, Equation~\ref{eqn:bin} reduces to its usual form when $k$ and $n$
are natural numbers.}
This is similar to the widely used method from \citet{Nielsen:2008p18024}, but here we use the more general 
binomial distribution instead of the Poisson distribution, which is only applicable for cases when
$N_\mathrm{trials}$ is large and $f$ is small.  For regions in \{$m$,$a$\} where the sensitivity to planets is low 
(small separations, large separations, and low masses), $N_\mathrm{trials}$ is small so the binomial distribution 
must be used to accurately measure upper limits.

Figure~\ref{fig:uplimcont} shows the results for the Cond and Fortney models with two assumptions about the planet eccentricity distributions.
Each colored grid point reflects the 95\% confidence upper limit on the planet frequency, and contours show the 50\%, 20\%, 10\%, and 5\% 
upper limits on planet frequency.
Table~11 summarizes the semi-major axes corresponding to these upper limits for each planet mass.
The best constraints are for high-mass planets between 10--100~AU, while the worst constraints are for small separations below $\sim$5~AU, large separations
beyond $\sim$500~AU, and planet masses below $\sim$1~\Mjup.
Assuming circular orbits and hot-start cooling models, we find that fewer than 10\% of single M dwarfs harbor 10~\Mjup \ (5~\Mjup) 
planets between 8.1--180~AU (13--130~AU). These results are insensitive to the choice of the hot-start model grid.
Naturally, cold start models produce poorer constraints; fewer than 10\% (20\%) of M dwarfs harbor 10~\Mjup \ (5~\Mjup) planets
between 21--96~AU (12--150~AU) using the Fortney models.


\begin{deluxetable*}{lcccc}
\tablenum{11}
\tabletypesize{\footnotesize}
\setlength{ \tabcolsep } {.12cm} 
\tablewidth{0pt}
\tablecolumns{5}
\tablecaption{Giant Planet Frequency Upper Limits for a Given Planet Mass (95\% Confidence)\label{tab:uplim1}}
\tablehead{
   \colhead{Mass} & \colhead{$\le$5\%} & \colhead{$\le$10\%} & \colhead{$\le$20\%} & \colhead{$\le$50\%}   \\  
   \colhead{(\Mjup)}     & \colhead{}         & \colhead{}                          & \colhead{}           & \colhead{}         
        }     
\startdata
\multicolumn{5}{c}{Cond (Circular Orbits)} \\

  13.0  &       13--85 AU   &   6.1--200 AU &   3.6--320 AU &   1.8--570 AU  \\
  10.0  &        36--61 AU  &   8.5--190 AU &   5.6--300 AU &   3.3--540 AU  \\
   7.0  &     $\cdots$   &   11--160 AU &   6.6--260 AU &   3.9--470 AU  \\
   5.0  &     $\cdots$   &   13--140 AU  &   7.3--240 AU &   4.0--440 AU  \\
   3.0  &     $\cdots$   &   $\cdots$  &   11--170 AU &   5.0--330 AU  \\
   2.0  &     $\cdots$   &   $\cdots$  &   18--120 AU  &   7.0--260 AU  \\
   1.0  &     $\cdots$   &   $\cdots$  &   $\cdots$  &    18--110 AU  \\

\multicolumn{5}{c}{Cond ($P(e)$ $\propto$ 1--$e$)} \\
  13.0  &       13--77 AU &    5.7--200 AU &     3.3--300 AU &      1.6--730 AU \\
  10.0  &       36--57 AU &    8.1--180 AU &     5.0--290 AU &      3.0--660 AU \\
   7.0   &    $\cdots$     &   11--150 AU &     6.2--260 AU &      3.4--520 AU \\
   5.0   &    $\cdots$     &    13--130 AU  &     6.7--240 AU &      3.6--460 AU \\
   3.0   &    $\cdots$     &    $\cdots$    &    10--160 AU &      4.6--320 AU \\
   2.0   &    $\cdots$     &  $\cdots$      &     17--110 AU &       6.3--260 AU \\
   1.0   &    $\cdots$     &  $\cdots$      &   $\cdots$     &       17--100 AU \\

\multicolumn{5}{c}{Fortney (Circular Orbits)} \\

  13.0 &   $\cdots$  &   10--150 AU &   5.6--240 AU &   3.3--450 AU \\
  10.0 &   $\cdots$  &   21--96 AU  &   9.7--180 AU &   5.2--340 AU \\
   7.0 &   $\cdots$  &   33--67 AU  &   11--160 AU &   5.4--300 AU \\
   5.0 &   $\cdots$  &   $\cdots$   &    12--150 AU &   5.7--290 AU \\
   3.0 &   $\cdots$  &   $\cdots$  &    17--110 AU &   7.4--220 AU \\
   2.0 &   $\cdots$  &   $\cdots$  &    33--63 AU   &   9.6--180 AU \\
   1.0 &   $\cdots$  &   $\cdots$  &   $\cdots$  &   $\cdots$ \\

\multicolumn{5}{c}{Fortney ($P(e)$ $\propto$ 1--$e$)} \\

  13.0   &    $\cdots$  &     10--140 AU &     5.3--250 AU  &    2.9--490 AU \\
  10.0   &    $\cdots$  &     20--85 AU  &     9.0--180 AU  &    4.8--330 AU \\
   7.0    &   $\cdots$  &     36--60 AU  &     10--150 AU  &    5.0--300 AU \\
   5.0    &   $\cdots$  &    $\cdots$  &     11--140 AU  &    5.2--290 AU \\
   3.0   &    $\cdots$  &   $\cdots$      &     17--98 AU   &    6.7--220 AU \\
   2.0   &    $\cdots$  &   $\cdots$      &    35--54 AU   &    8.7--170 AU \\
   1.0   &    $\cdots$  &   $\cdots$      &   $\cdots$       &    $\cdots$ 

\enddata
\end{deluxetable*}

\subsection{Giant Planet Frequency Over a Range of Planet Masses and Semi-major Axes}

Our second approach focuses on a related but slightly different question: what is the frequency of 
giant planets over a \emph{range} of planet masses and semi-major axes?
This can be addressed with our sensitivity maps and assumptions about the form of the underlying 
distributions of planet masses and semi-major axes.
For the following analysis we adopt logarithmically-flat distributions in mass from 0.5--100~\Mjup \ and 
semi-major axis from 1--1000~AU:
$dN/(d$log$a$~$d$log$m$) $\propto$ $m^\alpha$$a^\beta$, where $\alpha$=0.0 and $\beta$=0.0.
The choice of power law representations is partly motivated (but not defined) by planet populations at smaller separations ($<$10~AU), 
which have mass 
and period distributions that are well-reproduced with this functional form (e.g., \citealt{Cumming:2008p9188}; \citealt{Howard:2010p21412}).
Moreover, the logarithmically-flat forms are broadly consistent with the projected separation distribution and mass distribution for planets around M dwarfs 
found in microlensing surveys (\citealt{Gould:2010p23021}; \citealt{Cassan:2012p23478}).  This particular case
of a logarithmically-flat distribution in semi-major axis corresponds to ``\"Opik's law'', which is a good representation
of visual binaries in some circumstances (see \citealt{Duchene:2013p25590} for a summary).\footnote{Although 
it is a common practice in the analysis of direct imaging surveys to extrapolate power-law distributions from radial velocity-detected planets, 
it is not clear that extending the  
population of giant planets from within a few AU out to hundreds of AU is any more informative than the 
logarithmically uniform, scale-invariant Jeffrey's prior we have adopted.
In fact, it is conceivable that giant planets are better represented by other more complex
functional forms, like a power-law distribution in semi-major axis at small separations and 
a log-normal form at wide separations, especially if there are two modes of planet formation (e.g., \citealt{Boley:2009p19858}).}

In this case the number of trials for a given target is the average value over $a$ and $m$:

\begin{equation}{\label{eq:binom}}
N_\mathrm{trials} = \frac{\sum_{i=1}^{N_\mathrm{tar}} \sum_{j=1}^{N_a}  \sum_{k=1}^{N_m}   {s_i(m_k, a_j)}}{{N_a N_m}},
\end{equation}

\noindent where $N_a$ and $N_m$ are the number of grid points in $a$ and $m$ in the region of interest.
The number of detections is zero and, once again, the binomial distribution can be used to compute an upper limit
on the planet fraction at the desired level.

Table~12 summarizes the results for a various ranges of mass and semi-major axis 
for the Cond and Fortney models with circular and eccentric distributions.
Overall our survey is most sensitive to the 10--100~\Mjup \ range (Figures~\ref{fig:meansens} and \ref{fig:eurosens}), so we would expect the tightest
constraints in this region.
For masses between 1 and 13~\Mjup, semi-major axes between 10--100~AU,  
circular orbits, and a hot start formation, $N_\mathrm{trials}$ = 26.6, which translates into a 95\% frequency upper limit of $<$10.3\%.
That is, fewer than 10.3\% of M dwarfs harbor giant planets between 10--100~AU at the 95\% confidence level.
For the cold start models, $N_\mathrm{trials}$ is reduced to 16.1, and the upper limit is weakened to $<$16.0\%.
If we instead isolate the high-mass planet population of 5--13~\Mjup,  $N_\mathrm{trials}$ grows to 47.0 (27.7) and the upper limits tighten
to $<$6.0\% ($<$9.9\%) for the Cond (Fortney) cooling models.

As expected, exploring broader ranges of physical separation lowers $N_\mathrm{trials}$ and the constraints weaken
since we begin to sample regions with poor sensitivity, diluting each target's average sensitivity.  
For the Cond case with circular orbits between 1--10~AU, the upper limit 
over the entire planetary-mass range is $<$51\% and from 100--1000~AU it is $<$29\%.
Likewise, for the entire 1--1000~AU, 1--13~\Mjup \ range, the planet frequency is $<$20.0\%.
Adding modest eccentricities tends to dilute these statistics, but overall the effect is small.

\subsection{The Frequency of Brown Dwarf Companions to M Dwarfs}{\label{sec:bdfreq}}

Measuring the frequency of brown dwarfs over various ranges of $a$ involves the additional step of de-projecting the 
observed (sky-projected) separations onto the semi-major axis plane.  Like the above analysis, this involves assumptions
about the form of the semi-major axis distribution of substellar companions and their eccentricities (which
can be defined or parameterized and freely fit).
Here we adopt the same logarithmically-flat 
distribution in $a$ assuming both circular and mildly eccentric orbits (following 1--$e$).
Our approach is to simulate random sky-projected orbits at each step in a grid of semi-major axes, here 1--1000~AU.
The number of planets at each grid point is scaled according to the power law index used, resulting in a distribution of
projected separations at each step in $a$.  The cumulative distribution of projected separations
over the entire range of $a$ is then used to infer the original semi-major axis distribution  
based on the location a companion has been observed.  Uncertainties in the measured projected separation
due to errors in the target's distance and angular separation measurement are incorporated in a Monte Carlo
fashion.

\begin{figure}
  \vskip -.3in
  \hskip -.7in
  \resizebox{5.0in}{!}{\includegraphics{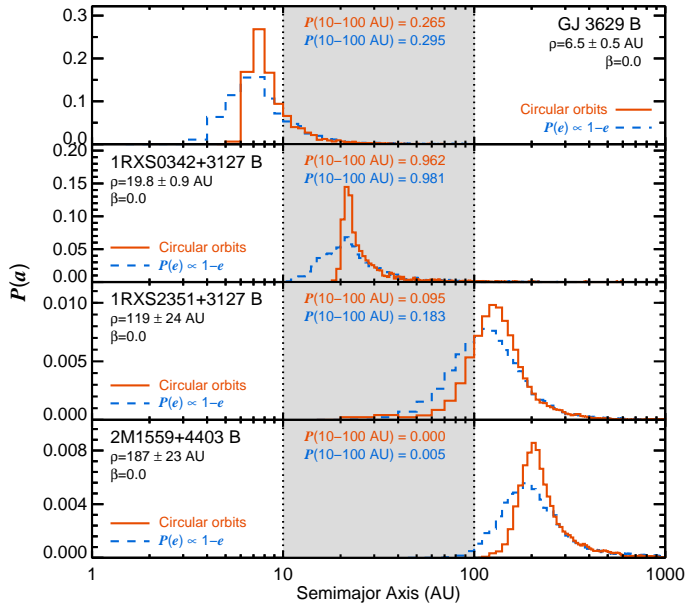}}
  \caption{Semi-major axis probability distributions for the four brown dwarfs discovered in our survey based on their observed sky-projected separations.  Two underlying eccentricity distributions are assumed: circular orbits and $P$($e$)~$\propto$~1-$e$.  The latter creates a broader shape since smaller semi-major axes can reproduce the observed projected separations.  The gray shaded region shows our method for computing $N_\mathrm{det}$, the number of detected companions.  In this example, $N_\mathrm{det}$ between 10--100~AU is the sum of all four probability distributions falling in that region (weighted by their likelihood of being substellar).  Here the  semi-major axis power-law index $\beta$ is flat (equal to 0.0) in log space.   \label{fig:sep2sma} } 
\end{figure}

The results of these simulations for our four brown dwarf discoveries are shown in 
Figure~\ref{fig:sep2sma}.  In general the eccentricity distributions ``smear out'' to smaller physical separations,
which is expected since planets can reach larger projected separations when they are on eccentric orbits.  
The inferred median $a$ and 68.3\% confidence range about the median for
GJ~3629 B, 1RXS J034231.8+121622 B, 1RXS J235133.3+312720~B, and 2MASS~J15594729+4403595~B is
8.5 (6.5--13.5)~AU, 24.5 (21.5, 39.5)~AU, 145 (105--235)~AU, and 225 (182--345)~AU, respectively (for circular orbits).
These \emph{a priori} semi-major axis distributions can then be used to compute the fraction that fall
within within a given range of $a$, or $N_\mathrm{det}$.  

If the mass of a brown dwarf companion is near the hydrogen-burning limit, or if its mass uncertainty is large enough,
 then it is possible to overestimate the inferred substellar occurrence rate since there is a chance that object might 
 be a low-mass star.  
To take this into account we weigh each of our four discoveries by
the probability they are substellar using the mass distributions we derived from their age and luminosity.
This mostly affects GJ~3629 B, which has a probability of 62\% of falling below the hydrogen-burning limit.
The corresponding probability for 2MASS~J15594729+4403595~B is 99.1\%, and is 100\% for both 
1RXS J034231.8+121622 B or 1RXS J235133.3+312720~B.
Once properly weighted by their substellar probabilities, the fractional detections within some range of semi-major axis
can be summed to determine $N_\mathrm{det}$.  Figure~\ref{fig:sep2sma} exemplifies this for the 10--100~AU region; 
for circular orbits, the total contribution from each companion is 
0.17, 0.988, 0.079, and 0.000, which sums to 1.23 ``detections.''

Applying the same analysis as in Section~\ref{sec:plfreq} to compute $N_\mathrm{trials}$ between \{13--75~\Mjup, 10--100~AU\} 
yields 66.8 ``trials,'' implying a substellar companion frequency of 2.8$^{+2.4}_{-1.5}$\%.
Similarly, we measure a frequency of 3.9$^{+4.8}_{-2.6}$\% for brown dwarfs between 1--10~AU.
Over the entire range of 1--1000~AU (encompassing all four weighted detections), we find a frequency of 11.1$^{+5.7}_{-4.3}$\%.
Results for all permutations of $a$ are listed in Table~12.


\section{Discussion}

The well-established correlation between stellar host mass and giant planet frequency
offers one of the strongest cases for core accretion  at small separations ($\lesssim$2.5~AU; 
\citealt{Johnson:2007p169}; \citealt{Johnson:2010p20950}).  
Since orbital period scales as $M_*^{-1/2}$, the timescale associated with planetesimal coagulation (a few Myr)
is faster for high-mass stars so more cores are able to form and accrete gaseous envelopes 
before protoplanetary disks disperse (e.g., \citealt{Laughlin:2004p19937}; \citealt{Kennedy:2008p18349}).  
In addition, there is now ample observational evidence that 
protoplanetary disk masses scale with stellar host mass, resulting in increased raw material for 
giant planet formation around high-mass stars compared to low-mass stars 
(\citealt{Andrews:2013ku}; \citet{Mohanty:2013p24943}).

Much less is known 
about the dependence of \emph{wide-separation} ($>$10~AU) giant planet frequency on stellar host mass.  
Like core accretion at small separations, disk instability predicts a positive trend
with primary mass assuming protoplanetary disk masses scale with protostellar mass (\citealt{Boss:2011p22047}).
A total of seven gas-giant planets have been directly imaged to date around three high mass 
(1.2--1.9~\Msun) young A-type stars and one G star 
(HR~8799, $\beta$~Pic, HD 95086, GJ~504; \
\citealt{Marois:2008p18841}, \citealt{Marois:2010p21591}, \citealt{Rameau:2013ds}, \citealt{Kuzuhara:2013p25347}).  
Around low-mass stars, companions near the deuterium-burning limit ($\approx$13~\Mjup)  
have been found at close separations within 100~AU (e.g., 2MASS~J01033563--5515561~C, \citealt{Delorme:2013p25184};
2MASS~J01225093--2439505~B, \citealt{Bowler:2013p25491}), and a growing population of planetary-mass 
objects on extreme orbits beyond 100~AU has been identified (e.g., GU~Psc~b, \citealt{Naud:2014jx}).  However, no companions
below 10~\Mjup \ have been imaged at $<$100~AU around stars between 0.1--1.0~\Msun\footnote{Interestingly, 
several planetary-mass companions are known around brown dwarfs (\citealt{Chauvin:2004p19400}; \citealt{Todorov:2010p20562}; 
\citealt{Liu:2012p24074}; \citealt{Han:2013by}), indicating an alternative formation mechanism of planetary-mass
companions around very low host masses.}, 
perhaps pointing to a correlation between stellar mass and giant planet occurrence rate (\citealt{Crepp:2011p22358}).

On the other hand, this apparent trend can also be explained by a selection bias since nearly all large direct imaging 
planet searches are focusing on high-mass stars.  For example, the NICI Planet-Finding Campaign (\citealt{Liu:2010p21647}),
Gemini Planet Imager Exoplanet Survey (\citealt{Macintosh:2014ke}), 
SEEDS (\citealt{Tamura:2006p21640}), 
LBTI Exozodi Exoplanet Common Hunt (\citealt{Skemer:2014um}), and 
the International Deep Planet Search (\citealt{Vigan:2012p24691}) 
concentrate on AFGK stars ($\approx$0.6--2~\Msun), so a paucity of imaged planets around low-mass stars
is not surprising.  

The only way to test whether giant planet frequency correlates with stellar host mass 
is to compare the statistical properties of long-period planets
in different stellar mass regimes.  
The largest imaging program targeting high-mass stars is the NICI Planet-Finding Campaign 
(\citealt{Liu:2010p21647}).  From their subsample of 70 young B- and A-type stars, 
\citet{Nielsen:2013jy} find that fewer than 20\% of 1.5--2.5~\Msun \ stars harbor $>$4~\Mjup \ planets between
59--460~AU.    
Other smaller surveys have mostly resulted in upper limits or, in some cases, weak constraints if the HR~8799 
and/or $\beta$~Pic systems are included (\citealt{Ehrenreich:2010p22443}; \citealt{Janson:2011p22503}; \citealt{Rameau:2013it}).  
For example, \citet{Vigan:2012p24691} targeted 
38 A stars and 4 F stars and arrived at a frequency of 4.3$^{+9.1}_{-1.3}$\% when  
$\beta$~Pic~b is excluded (\emph{a priori} knowledge of its existence can strongly bias the 
way the observations are conducted).  

Several large ($N>$50) direct imaging surveys have focused on young Sun-like stars.  
The analysis of 100 FGK stars by \citet{Nielsen:2010p20955}, which combined the
surveys of \citet{Masciadri:2005p18273}, \citet{Biller:2007p19401}, and \citet{Lafreniere:2007p17991}, 
is the largest study of wide-period planets around 0.6--1.2~\Msun \ host stars to date.
No planets were detected, yielding an upper limit of $<$20\% (at the 95\% confidence level) for the 
frequency of $>$4~\Mjup \ planets
between $\approx$40--470~AU.  More recently, \citet{Chauvin:2014tq} measured similar constraints of  
$<$15\% for $>$5\Mjup \ planets between 100--300~AU in their sample of 51 Sun-like stars.

We find an upper limit of $<$6.0\% in this survey of 78 single young M dwarfs, 
which is by far the most substantial program to date in the low-mass regime.
Taken together with similarly large surveys targeting A and FGK stars, 
\emph{there is currently no statistical evidence for a dependency of giant planet frequency 
with stellar host mass}.  
In the future, larger sample sizes will be needed to distinguish between 
small differences in the relative occurrence rates of long-period giant planets around A stars and M dwarfs.

\subsection{A Constant Substellar Companion Fraction with Host Mass}

The relative occurrence rates of brown dwarf companions as a function of stellar host mass also provides clues
about their formation.  
Large-scale hydrodynamical simulations of fragmenting molecular clouds 
by \citet{Bate:2009p20333} and \citet{Bate:2012p24259} produced brown dwarf companion frequencies with
no discernible dependency on the primary host star mass.  This seems to be consistent with observations:
\citet{Vigan:2012p24691} find a frequency of 2.8$^{+6.0}_{-0.9}$\% between 5--320~AU for massive A and F stars,
\citet{Metchev:2009p18431} find a frequency of 3.2$^{+3.1}_{-2.7}$\% (2-$\sigma$ limits) between 28--1590~AU around FGK stars,
and we infer a rate of 2.8$^{+2.4}_{-1.5}$\% (4.5$^{+3.1}_{-2.1}$\%) between 10--100~AU (10--200~AU) for M dwarfs.  Although the ranges
of semi-major axes being considered are different in these studies, they all point to comparable rates
of a few percent across all separations.  

\citet{Metchev:2009p18431} compared all published direct imaging searches for brown dwarf companions as of 2009 and found
a tentative trend between the frequency of brown dwarf companions and both stellar host mass and separation.  
Surveys targeting low-mass stars ($\approx$0.2--0.6~\Msun) at small separations ($\lesssim$150~AU) 
found a paucity of brown dwarfs compared to those focusing on more massive stars ($\gtrsim$0.7~\Msun) and
wide separations ($\gtrsim$150~AU).
However, our results do not support this correlation; our brown dwarf companion frequency of
a few percent is similar to the higher-mass, wide-separation surveys.  
As emphasized by Metchev \& Hillenbrand, most of these previous surveys did not correct for 
incompleteness in their observations, so the inferred substellar frequencies should be treated
with caution.  On the other hand, our deep observations probe the \emph{entire} substellar regime
and we correct for incompleteness in the regions in which we are not sensitive.
Incidentally, two \emph{additional} brown dwarfs were previously known at separations of $\approx$350--400~AU
around single stars in our sample (G~196-3 and NLTT~22741; \citealt{Rebolo:1998p19498}; \citealt{Reid:2006p22856}).
Neither were detected in our data so they were not included in our statistical results, but
together they imply that at least six out of 78 single M dwarfs in our sample host substellar companions,
a rate much higher than inferred from previous, less sensitive surveys targeting low-mass stars
listed in Metchev \& Hillenbrand.
Our results are supported by the $HST$ multiplicity survey by \citet{Dieterich:2012p24214}, which found
a multiplicity rate of 2.3$^{+5.0}_{-0.7}$\% for L0--T9 companions to field M dwarfs.\footnote{Note that old low-mass stars
have effective temperatures reaching early-L spectral types, so this frequency is slightly different from the substellar
companion fraction.}

An ongoing debate over whether gas giants can form via direct gravitational collapse of a massive protoplanetary disk 
has consumed much of the discussion about planet formation for the past decade, especially after the discovery of the
 HR~8799 planets (e.g., \citealt{Boss:2007p18983}; \citealt{Durisen:2007p8061}; \citealt{Boley:2009p19858};
\citealt{DodsonRobinson:2009p19734}; \citealt{Nero:2009p20063}; \citealt{Kratter:2010p20098}).
Simulations show that protoplanetary disks can collapse when conditions
 are both cool enough and disk surface densities are high enough.
 The region between a few tens to a few hundreds of AU occupies this  ``sweet spot'' and is the most likely place 
 for giant planets to form from this mechanism (e.g., \citealt{Stamatellos:2009p14111}; 
 \citealt{Vorobyov:2010p22077}; \citealt{Boss:2011p22047}).
 Our constraints on the frequency of giant planets in this region for the most common type of star imply that, overall,
 disk instability is not an efficient mechanism for producing gas giants around low-mass stars. 

\section{Summary and Conclusions}

We have carried out a deep direct imaging search for giant planets 
around nearby ($\lesssim$40~pc) young ($\lesssim$300~Myr) low-mass stars with Keck and Subaru.  
Out of 122 targets, 44 are resolved
into close visual binaries with separations ranging from $\approx$0$\farcs$05--2$''$; 27 of these
are new or spatially resolved for the first time.
Because known binaries were removed prior to the start of this survey, we infer a minimum
stellar companion frequency of $>$35.4~$\pm$~4.3\% within 100~AU.
38\% of our sample are confirmed or likely members of young moving groups spanning ages of 8--620~Myr and
57\% of our targets have measured parallaxes.
Below we summarize results of our deep imaging search for planets around the 78 single M dwarfs in our sample:
\begin{enumerate}
\item Four comoving brown dwarfs with masses between 30--70~\Mjup \ and 
projected separations of 6--190~AU were discovered in our survey: 
1RXS J235133.3+312720~B (\citealt{Bowler:2012p23851}), 
GJ~3629~B (\citealt{Bowler:2012p23980}), 1RXS~J034231.8+121622~B,  and 2MASS~J15594729+4403595~B.
1RXS J235133.3+312720 is a likely member of the $\approx$120~Myr 
AB~Dor moving group.

\item Taking into account our detection limits, we measure a brown dwarf companion fraction of 
2.8$^{+2.4}_{-1.5}$\% (4.5$^{+3.1}_{-2.1}$\%) between 10--100~AU  (10--200~AU) 
around single M dwarfs.  These results are consistent with the brown dwarf 
occurrence rate found around high- and intermediate-mass primaries,
which is also in general agreement with
hydrodynamical simulations of turbulent fragmentation by \citet{Bate:2009p20333}.

\item No planets were confirmed in our survey.  Among 102 candidates detected around 38 single stars in our deep imaging, 
60 are shown to be stationary background stars.  The status of the remaining 42 candidates with only a single epoch
of astrometry is unclear, but only  
8 of these are located at projected separations less than 100~AU.  

\item  Our null detection of planets implies that $<$10.3\% ($<$6.0\%) of single M dwarfs harbor 1--13~\Mjup \ (5--13~\Mjup) planets 
between 10--100~AU assuming hot start evolutionary models and logarithmically-uniform distributions in planet mass and semi-major axis.

\item The dearth of massive planets at tens to hundreds of AU around the most common type of star in our galaxy implies that, overall, 
disk instability is not a common mechanism of giant plant formation.  

\item Finally, comparing the largest direct imaging planet searches in three mass regimes (A, FGK, and M stars),
there is currently no statistical evidence for a correlation between stellar host mass and giant planet frequency
at large separations ($>$10~AU).  
 
\end{enumerate}

In the future, much larger samples of several hundred stars in each stellar mass bin will be needed to discriminate 
differences  in the relative frequencies of giant planets at $\gtrsim$10~AU.  
We caution that for large homogeneous analyses incorporating our contrast curves and those of any other surveys,
not all planet candidates have been rejected as background stars and so targets and contrast curves must be 
carefully selected on a case-by-case basis.  Ultimately, large statistical comparisons with the current generation of instruments on 
8--10~meter class telescopes will set the stage--- and statistical baseline--- for 
the next generation of thirty-meter telescopes to image true Jupiter analogs in the 3--10~AU region.

\appendix{}

\section{HiCIAO Distortion Correction}{\label{sec:append1}}

The HiCIAO optical distortion, plate scale, and orientation were measured using 
$H$-band images of the globular cluster M5 obtained on the nights of 2011 January 27 UT and 2012 May 10 UT.
We targeted a $\approx$20$''$~$\times$~20$''$ region near the center of the cluster covering the same 
dense stellar field as in \citet{Cameron:2009p21658}.
The HiCIAO camera lens was changed in April 2011, so we generated two distortion maps: one for 
our January 2011 observing run, and one for our December 2011/January 2012 and May 2012 runs.
NIRC2 images of the same field obtained on 2006 February 07 UT using the wide camera (A. Kraus, private communication) 
were used as an absolute reference frame.
We first corrected the NIRC2 optical distortions using the solution created by B. Cameron (2007, private communication). 
Stars were then identified in the images from both instruments  with the DAOPHOT photometry package (\citealt{Stetson:1987p10559}).
Finally, the AMOEBA downhill simplex algorithm was used to fit for relative $x$/$y$ offsets (in pixel coordinates) 
between the two systems, a relative magnification scale for HiCIAO, an overall rotation of HiCIAO, 
and 18 coefficients comprising a 2-dimensional, third-order polynomial fit following \citet{Anderson:2003p23978}.
A total of 297 and 344 stars are used for our 2011 and 2012 calibration measurements, respectively.

\begin{figure}
  \vskip -.5in
  \hskip 1.2in
  \resizebox{5in}{!}{\includegraphics{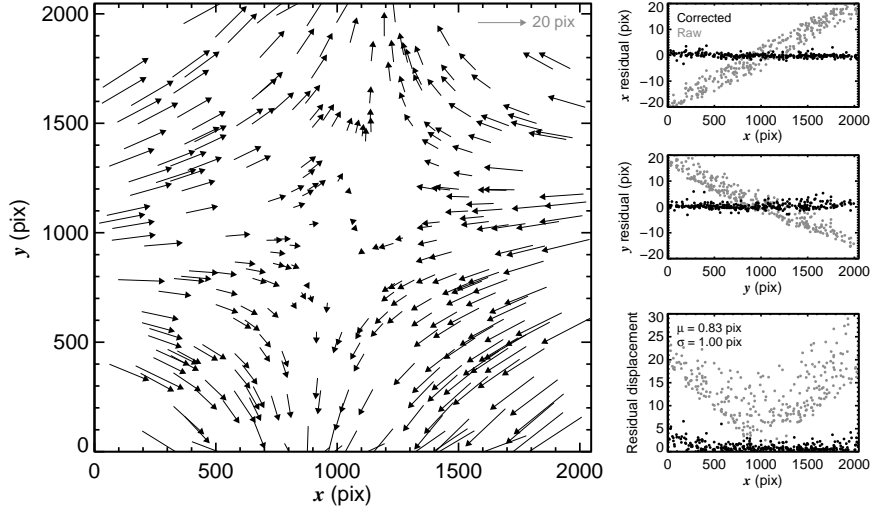}}
  \vskip -.3in
  \caption{Our HiCIAO distortion solution from January 2011.  Arrow bases and heads indicate
  the measured and corrected stellar positions, respectively, in our images of M5 after applying 
  third-order polynomial polynomial fit in $x$ and $y$.  For visual purposes all arrow lengths have been increased
  by a factor of 10.  The average residual
  displacement after correction is 0.8~pix (7.8~mas in $H$ band).    \label{distsol_jan11} } 
\end{figure}

\begin{figure}
  \vskip -.5in
  \hskip 1.2in
  \resizebox{5in}{!}{\includegraphics{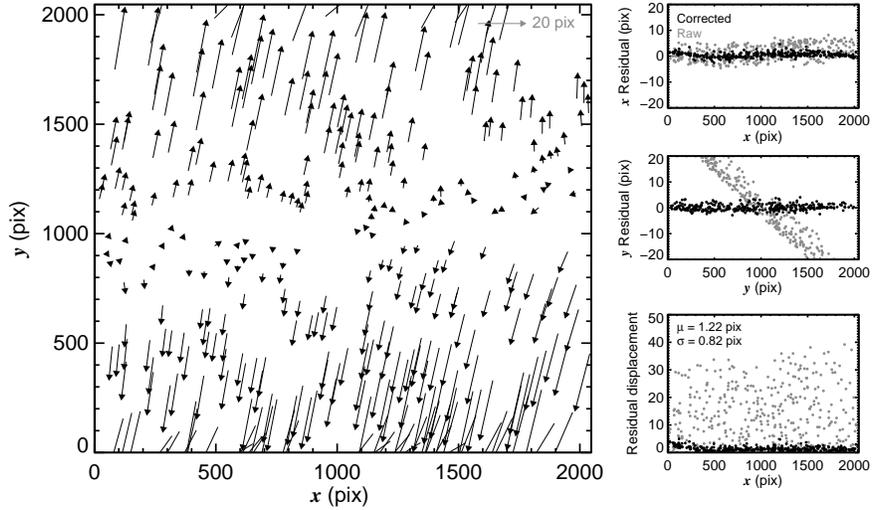}}
  \vskip -.3in
  \caption{Our HiCIAO distortion solution from May 2012.  In April 2011 the HiCIAO camera lens
  was replaced, creating a substantially different distortion map compared to the January 2011 one shown in Figure~\ref{distsol_jan11}.
  After correction, the typical residual displacement is 1.2~pix (11.7~mas in $H$ band).  Arrow lengths have been increased by a factor of 10
  for visual purposes.  \label{distsol_may12}}
\end{figure}

The best-fit distortion solutions are shown in Figures~\ref{distsol_jan11} and \ref{distsol_may12}.  The upgraded camera lens
created a significant qualitative difference in the optical distortion, with most of the optical aberrations occurring in the $y$-direction
along the detector columns with the new lens in place.  The uncorrected optical distortion produces significant positional offsets 
of up to 30 pixels near the edges of the detector at both epochs.  After applying our solution, the average total residual 
displacement between the HiCIAO 
and NIRC2 positions is 0.8~pix and 1.2~pix for the 2011 and 2012 calibration datasets, respectively, showing little dependence
on spatial position across the entire 2048$\times$2048 pixel array.  We therefore adopt 1~pix as a typical systematic positional uncertainty caused
by optical distortions ($\sigma_{d}$) for our HiCIAO observations.
The best-fit solutions give magnification scales of 4.103 and 4.100 times smaller than the NIRC2 wide camera for the
2011 and 2012 data.  
We also solved for HiCIAO distortion solutions at each epoch using the same field and instrument setup, 
except with the coronagraph slide in place to test its influence on the astrometry.  The results are virtually
identical to the solutions without the coronagraph in place, giving magnification scales of 4.104 and 4.101 at each epoch.  
Because of these similarities, we assume 
identical magnification factors of 4.102~$\pm$~0.002.  Based on the NIRC2 wide camera plate scale of 39.884~$\pm$~0.039 
mas pix$^{-1}$ measured by \citet{Pravdo:2006p22531}, this implies a HiCIAO $H$-band plate scale of 9.723~$\pm$~0.011 
mas pix$^{-1}$.   The HiCIAO detector appears to be aligned very closely with celestial north;
the best-fit solutions imply rotations of --0$\fdg$01 (--0$\fdg$09) and +0$\fdg$03 (+0$\fdg$03) for the 2011
and 2012 datasets without (with) the coronagraph (positive is East from North).  We conservatively adopt
a detector orientation of 0$\fdg$0~$\pm$0$\fdg$1.

Our observations of the 5$\farcs$6 pair 2MASS~J15594729+4403595~AB in the $Y$, $J$, $H$, and $K_S$ filters
at the same position on the detector show that the HiCIAO plate scale varies significantly with wavelength.  
The separation in $H$-band is 580.8~$\pm$~1.4~pix,
or 5647~$\pm$~15~mas using our plate scale measurement, which is in excellent agreement with the value of  
5638~$\pm$0$\farcs$004 measured by \citet{Janson:2012p23979}  
(no orbital motion is expected given the system's $\sim$190~AU separation).
On the other hand, the separations in $Y$, $J$, and $K_S$ bands are 575.5~$\pm$~1.4~pix, 579.2~$\pm$~1.4~pix, and 583.7~$\pm$~1.4~pix,
respectively, implying plate scales of 9.81~$\pm$~0.04 mas~pix$^{-1}$, 9.75~$\pm$~0.04 mas~pix$^{-1}$, 9.67~$\pm$~0.03 mas~pix$^{-1}$.
We adopt these wavelength-dependent plate scales for our astrometry.
On the other hand, the PAs are consistent within 0$\fdg$01, so we do not make any corrections to that.


\section{NIRC2 600 mas Coronagraph Attenuation}{\label{sec:nirc2_transmission}}

NIRC2 has several circular, partly transmissive occulting spots located on a clear slide in its first focal plane.
We made use the 600 mas diameter spot for our survey.  To calibrate the transmission, we observed
the young, 2$\farcs$9 separation M1.5+M4.0 binary NLTT 32659 (\citealt{Shkolnik:2009p19565}) on 2012 May 21 UT with and without 
 the primary under the spot.  The observations and resulting flux ratios are
listed in Table~\ref{tab:cortrans}.  Aperture photometry using an extraction radius of 5~pix yields
a transmission of 7.51~$\pm$~0.14~mag (0.099~$\pm$~0.013\%) in $H$ and 6.65~$\pm$~0.10~mag (0.22~$\pm$~0.02\%) in $K_S$.
Incidentally, our PA and separation measurements are in excellent agreement with \citet{Shkolnik:2012p24056}.


\begin{deluxetable*}{lcccccccc}
\tabletypesize{\scriptsize}
\tablewidth{0pt}
\tablecolumns{9}
\tablecaption{NIRC2 Coronagraph Calibration Measurements of NLTT 32659 AB\label{tab:cortrans}}
\tablehead{
        \colhead{Date}   &   \colhead{Coronagraph}  &  \colhead{Filter}  &  \colhead{$N$ $\times$ Coadds $\times$}   &  \colhead{FWHM}   &    \colhead{Strehl}  & \colhead{Separation}    &    \colhead{PA}   &   \colhead{$\Delta$mag} \\
        \colhead{(UT)}   &   \colhead{}                           &  \colhead{}            &    \colhead{Exp. Time (s)}                                 &  \colhead{(mas)}    &    \colhead{}             & \colhead{(")}                   &    \colhead{($^{\circ}$)}   &   \colhead{}
        }   
\startdata
 2012 May 21   &  None &  $K_S$ &  28~$\times$~10~$\times$~0.15  &  61~$\pm$~7  &  0.16~$\pm$~0.08  &   2.899~$\pm$~0.002        &  88.54~$\pm$~0.03   &   2.03 $\pm$ 0.04  \\
 2012 May 21   &  600 mas &  $K_S$ &  13~$\times$~1~$\times$~3.0        &  62~$\pm$~4  &  0.11~$\pm$~0.03  &   2.898~$\pm$~0.002        &  88.64~$\pm$~0.03   &   4.60 $\pm$ 0.09  \\
2012 May 21   &  None &  $H$  &  17~$\times$~1~$\times$~0.15     &  50~$\pm$~4  &  0.10~$\pm$~0.04  &   2.897~$\pm$~0.003        &  88.59~$\pm$~0.06   &   2.18 $\pm$ 0.07  \\
2012 May 21   &  600 mas      &  $H$  &  12~$\times$~1~$\times$~2.5        &  53~$\pm$~4  &  0.08~$\pm$~0.02  &   2.902~$\pm$~0.004        &  88.59~$\pm$~0.07   &   5.30 $\pm$ 0.11  
\enddata
\end{deluxetable*}
\clearpage

\section{Notes on Individual Objects}{\label{sec:indnotes}}

\textit{GJ 3030 AB (2MASS~J00215781+4912379)}.  
GJ~3030~AB was first identified as a visual binary in the Washington Double Star Catalog 
(WDS; \citealt{Mason:2001p23846}) and later by \citet{McCarthy:2004p18279} in their coronagraphic
search for brown dwarf companions, although no astrometry is provided in the latter.  
The WDS catalog lists a companion to the M2.4 primary with a contrast of 2.9~mag in the optical
at a separation of 2$\farcs$5--2$\farcs$9 and a position angle of 290--291$^{\circ}$ 
from two epochs in 1995 and 1998.
We confirm the physical separation of the pair and detect modest orbital motion.
Unresolved light curves of GJ~3030~AB from the HATNet survey (\citealt{Hartman:2011p22788}) reveal a 
photometric period of 6.166~$\pm$~0.014 days likely corresponding to the rotation period of the primary.

\textit{NLTT 1875 (2MASS~J00350487+5953079)}.
\citet{Shkolnik:2012p24056} proposed this M4.3e star as a candidate kinematic member of IC~2391 based on their measured
radial velocity (--1.0~$\pm$~0.1~km~s$^{-1}$) and its photometric distance of 26~$\pm$~6~pc.  Recently, \citet{Dittmann:2013tp}
presented a trigonometric parallax of 38.3~$\pm$~2.2~pc  to NLTT~1875.  At this revised distance, the star's $U$, $V$, and $W$ 
space velocities are \{--34~$\pm$~2, --21.8~$\pm$~1.2, --3.1~$\pm$~0.3~km~s$^{-1}$\}, respectively, which do not correspond
to any known young moving groups.  Moreover, it does not appear overluminous compared to normal main sequence stars 
on the $M_V$ vs. $V$--$K_S$ diagram.  We therefore adopt a wider age range of 100--500~Myr for this star.

\textit{G 271-110 (2MASS~J01365529--0647363)}.
This active M4 star is a very wide (14,600~AU) companion to EX Cet (\citealt{AlonsoFloriano:2011uz}), 
a young G5 star with an $Hipparcos$ distance of 24.0~$\pm$~0.4~pc.
One faint ($\Delta K$=13.8~$\pm$~1.0~mag) point source was identified at a separation of 
6$\farcs$662~$\pm$~0$\farcs$003 (160~AU) and a PA of 23.18~$\pm$~0.20$^{\circ}$ from this star in our first epoch of imaging, 
but we were unable to recover it in several follow-up attempts because conditions were worse 
or integration times were insufficient.

\textit{1RXS J022735.8+471021 (2MASS~J02273726+4710045)}.
Based on the radial velocity of --6.0~$\pm$~0.7 from \citet{Shkolnik:2012p24056} and the parallactic distance of 
27.4~$\pm$~1.7~km~s$^{-1}$ from \citet{Dittmann:2013tp}, the $UVW$ space velocities for the M4.6 star 1RXS J022735.8+471021
are \{--7.9~$\pm$~1.2, --23.9~$\pm$~1.6, --14.7~$\pm$~1.4~km~s$^{-1}$\}.  These agree well with the AB~Dor moving
group (e.g., \citealt{Torres:2008p20087}), so we assign 1RXS J022735.8+471021 as a probable member of this moving group.

\textit{2MASS~J03033668--2535329~AB}.
\citet{Makarov:2005p25256} first noted this M0 star as a likely binary from significant differences between 
$Hipparcos$ and Tycho proper motions.
This 0$\farcs$83 binary ($\Delta$$K_S$=2.99~$\pm$~0.06~mag) was later resolved as part of the the Astralux Lucky imaging 
survey by \citet{Bergfors:2010p23053} and \citet{Janson:2012p23979}.
Their astrometry from 2008.88 ($\rho$=0$\farcs$834~$\pm$~0$\farcs$005, $\theta$=7.6~$\pm$~0.3$^{\circ}$) and 
2010.08 ($\rho$=0$\farcs$834~$\pm$~0$\farcs$005, $\theta$=3.5~$\pm$~0.3$^{\circ}$) together with our measurements from 2011
reveal  a constant separation but a PA changing by $\approx$3$^{\circ}$~yr$^{-1}$.  

\textit{2MASS~J04220833--2849053 AB}.
This star is a 0$\farcs$74 equal-flux K7Ve binary system. 
\citet{Torres:2006p19650} found strong  H$\alpha$ emission (EW=12~\AA) and 
Li$\lambda$6708 absorption ($EW$=70~m\AA).  
Based on the stars' $V$--$I$ color of 1.2~mag from UCAC4, 
the Li depletion implies an age consistent with the Pleiades (\citealt{Torres:2008p20087}).
We therefore adopt a conservative age range of 50--200~Myr for this system.

\textit{2MASS~J04472312--2750358 and 2MASS~J04472266--2750295}.
2MASS J04472312--2750358 is the M2Ve secondary companion to the bright M0V star
2MASS~J04472266--2750295 separated by 8$\farcs$8.  
The stars share similar radial velocities and proper motions (\citealt{Torres:2006p19650}),
and imaging dating to the early twentieth century shows some orbital motion (\citealt{Mason:2001p23846}).
The system was detected by $ROSAT$ and both components are detected in $GALEX$.
The primary shows no H$\alpha$ emission but \citet{Torres:2006p19650} found the companion is in emission,
suggesting an upper age limit of $\sim$1.2~Gyr.  We therefore adopt the lower limit of 400 Myr from
\citet{Shkolnik:2009p19565} and an upper limit of 1.2~Gyr for the system. 

\textit{L~449-1 AB (2MASS~J05172292--3521545)}. 
This nearby (11.9~pc; \citealt{Riedel:2014ce}) active pair of mid-M dwarfs was first noted by \citet{Scholz:2005p22579}.  
\citet{Riedel:2014ce} identify a close stellar companion to the M4.0e primary at a separation of 47~mas
from $HST$ Fine Guidance Sensor interferometry from 2008.  We did not resolve the companion in
our HiCIAO observations from 2011.
Interestingly, deep VLT/SINFONI observations from 1.4--2.5~$\mu$m by \citet{Janson:2008p19399} revealed
a candidate marginally-resolved ($\approx$50~mas) low-contrast companion, though they attribute it to a PSF artifact.
\citet{Riedel:2014ce} find no evidence the system is particularly young and tentatively associate it with
the UMa moving group based on its kinematics.

 \textit{AP Col (2MASS~J06045215--3433360)}.
\citet{Scholz:2005p22579} first drew attention to this active, optically variable M4.5 star because of its strong X-ray emission and
proximity to the Sun.  \citet{Riedel:2011p22580} measured a parallactic distance of 8.4~pc and kinematically associate 
it with the young ($\sim$40--50~Myr) Argus or IC~2391 moving groups.  The origin and relationship of these
two groups remains ambiguous (see Section 4.1 of \citealt{Riedel:2011p22580} for a detailed discussion), but because of its
proximity to Earth compared to typical IC~2391 members ($\sim$150~pc), we adopt Argus as the physical 
association. Deep adaptive optics imaging of AP~Col by \citet{Quanz:2012p25296} did not reveal
any planetary companions down to contrasts of $\Delta$$L'$$\sim$11~mag at 0$\farcs$5, corresponding to
planetary masses near 1~\Mjup.  
Our HiCIAO observations in $H$ band reach a sensitivity of 13.5~mag at 1$''$ and we identify a single wide candidate
companion at 7.1$''$ (60~AU).  Our follow-up astrometry at Keck shows it is a background star.

\textit{1RXS J091744.5+461229 AB (2MASS~J09174473+4612246)}.
This M2.5 star was resolved into a 0$\farcs$25 binary by \citet{Janson:2012p23979},
who also confirmed the physical nature of the pair from two epochs of astrometry in 2008 and 2009.
We detect modest orbital motion with our new astrometry.  
A rotational period of 0.562~days for the unresolved system was measured in the HATNet survey (\citealt{Hartman:2011p22788}).

\textit{GJ 354.1 B (2MASS~J09324827+2659443)}.
This star is a widely-separated (72$''$, $\approx$1300~AU)
M5.5 companion to the young K0 star DX~Leo (\citealt{Gaidos:1998p19641}; \citealt{Montes:2001p19646}; \citealt{Lowrance:2005p18287}).
The primary has a long history of potential kinematic matches to YMGs:
\citet{Gaidos:2000p25311} list it as a candidate member of the Pleiades;
\citet{Montes:2001p19646} and \citet{Maldonado:2010p25313} broadly associate it with the Local Association;
\citet{Gaidos:1998p19641} and \citet{Fuhrmann:2004p17709} link it with the Her-Lyr group 
(though this is refuted by \citealt{LopezSantiago:2006p18285});
\citet{Nakajima:2012p23430} find Tuc-Hor to be the best match; and 
\citet{Brandt:2014hc} link it with Columba.
While the $UVW$ kinematics of DX~Leo are in good agreement with members of the
Carina, Tuc-Hor, and Columba YMGs, its $XYZ$ space positions do not entirely agree 
with a single group.  Because of its close kinematic agreement with Carina members, we adopt that
association and age ($\approx$30~Myr) for GJ~354.1~B, though a complete kinematic traceback analysis
is needed to confirm this.

\textit{PYC J09362+3731 AB (2MASS~J09361593+3731456; HIP~47133)}.
This star is an equal-mass M0.5 SB2 system identified by 
\citet{Schlieder:2012p23477} and \citet{Schlieder:2012p25080} as a likely member of the 
$\beta$~Pic moving group based on its $UVW$ kinematics.  However, \citet{Malo:2013p24348}
note that the spatial position of PYC J09362+3731 AB disagrees with established members by $\sim$40~pc.
This casts doubt on the membership of PYC J09362+3731 AB, especially since the activity 
detected by $ROSAT$ and $GALEX$ could be a result of tidal interactions rather than youth.  
We therefore assume it is a
member of the field for this study and adopt a conservative age range of 10~Myr--10~Gyr.

\textit{NLTT 22741 A (2MASS~J09510459+3558098)}.
LP~261-75~A is an active M4.5e star with an L6.5 companion separated by 12$''$ (\citealt{Reid:2006p22856}).  
At a distance of 33~pc (\citealt{Bowler:2013p25491}; \citealt{Dittmann:2013tp}; F. Vrba, in preparation), this corresponds to
$\approx$360~AU in projected separation. Combining its distance and radial velocity from \citet{Shkolnik:2012p24056} gives
$UVW$ space velocities of \{--14.1~$\pm$~0.7, --24.3~$\pm$~1.5, --1.1~$\pm$~0.7\} km~s$^{-1}$.  The $U$ and $V$ velocities are consistent 
with the Columba association, but differ by $\approx$5~km~s$^{-1}$ in $W$.  Note, however, that the NIR spectrum of NLTT~22741~B
(L4.5~$\pm$~1.0 spectral type) from \citet{Bowler:2013p25491} does not have the angular $H$-band features expected for a young ($\lesssim$100~Myr) brown dwarf.
Lacking a likely young moving group match,
we therefore adopt the age estimate of 100--200~Myr from \citet{Reid:2006p22856} for this system.  

\textit{GJ 2079 AB (2MASS~J10141918+2104297)}.
This star (also known as DK Leo, HIP~50156) is an active M0.5~$\pm$ 0.5 star with a 
parallactic distance of 23.1~$\pm$~0.1~pc (\citealt{Perryman:1997p534}; \citealt{vanLeeuwen:2007p12454}).
\citet{Makarov:2005p25256} and \citet{Frankowski:2007p22906} found evidence for a close astrometric companion
based on differences between $Hipparcos$ and $Tycho$-$2$ proper motions.
Similarly, \citet{Shkolnik:2012p24056} identified GJ~2079 as an SB1 from variable radial velocity measurements 
spanning a decade.
We resolved the likely culprit with AO imaging at Keck: a tight ($\sim$90~mas) companion with a $K_S$-band contrast of 1.8~mag.
The system is unresolved in our 2011 December 28 UT Subaru data, but two epochs at Keck (obtained before and after our HiCIAO data
on 2011 March 25 UT and 2013 February 4 UT) 
separated by $\sim$2~yr show substantial orbital motion.
GJ~2079 was also imaged by the Subaru SEEDS program on 2011 December 24 UT--- just a week before our own non-detection 
with HiCIAO reported here--- 
and the companion was not resolved; \citet{Brandt:2014hc} report an upper limit of $\sim$20~mas, suggesting 
GJ~2079~B had moved too close to the primary to resolve at that epoch.

\citet{Schlieder:2012p23477} identify GJ 2079 as a probable member of the $\beta$~Pic YMG, but 
\citet{Shkolnik:2012p24056} suggest the Carina YMG is more likely based on their more recent radial
velocity measurement, the lack of Li absorption, and weak H$\alpha$ emission (see note~$k$ in their Table~6).
Similarly, \citet{Malo:2013p24348} propose GJ~2079 is a member of the Columba group regardless of its (varying) radial velocities.
Since GJ~2079 is a close binary, continued monitoring is clearly needed to derive a systemic RV before reassessing its kinematic
membership to YMGs.  For this work we follow Shkolnik et al. in adopting GJ~2079~AB as a member of the Carina YMG with
an age of $\sim$30~Myr.

\textit{2MASS J11240434+3808108}.
This M4.5 star has a known M8.5 companion located at 8$\farcs$3 ($\approx$170~AU given its photometric
distance of $\approx$20~pc), 2MASS~J11240487+3808054 (\citealt{Close:2003p20000}; \citealt{Cruz:2003p67}).
In addition to their common proper motion, the radial velocity of the companion (--14~$\pm$~3 km~s$^{-1}$) measured by \citet{Reiners:2009p20785}
agrees with that of the primary (--11.5~$\pm$~0.5 km~s$^{-1}$) from \citet{Shkolnik:2012p24056}.
\citet{Shkolnik:2009p19565} assign an age range of 40--300~Myr for the primary from its high X-ray emission and lack of 
spectroscopic indicators of youth, while \citet{Shkolnik:2012p24056} tentatively assign it to the Ursa Major moving group 
($\sim$500~$\pm$~100~Myr; \citealt{King:2003p22818}) from its kinematics.
\citet{Burgasser:2004p574} obtained a low-resolution near-infrared spectrum of the companion, which does not show
obvious signs of low gravity in the form of an angular $H$-band shape, supporting an age $\gtrsim$100~Myr (e.g., \citealt{Allers:2013p25314}).
The HATNet survey measured a fast rotation period of 0.475 days for the primary (\citealt{Hartman:2011p22788}); unfortunately,
rotation periods become unreliable age indicators for stars that are fully convective (\citealt{Irwin:2011p22865}).

The 2MASS $H$-band spectrophotometric distance to the companion 2MASS~J11240487+3808054 is 20.3~$\pm$~1.3~pc
using the relations from \citet{Dupuy:2012p23924}.  (This error incorporates the rms spread from Dupuy \& Liu, a
spectral type uncertainty of 0.5 subclasses, and the photometric uncertainty.)
At this distance, the $UVW$ space velocities of the system (\{14.8~$\pm$~0.7, 2.8~$\pm$~0.3, --6.7~$\pm$~0.5\}~km~sec$^{-1}$) 
are an excellent kinematic match with Ursa Major (see
Table 2 of \citealt{AmmlerVonEiff:2009p22831}). We therefore adopt an age of  500~$\pm$~100~Myr for this system.
The corresponding luminosity of the companion is log~$L_\mathrm{Bol}/L_{\odot}$=--3.35~$\pm$~0.06~dex,
which translates into a mass of 81~$\pm$~5~\Mjup \ using evolutionary models from \citet{Burrows:1997p2706}.
This is very near the substellar boundary; however, the probability that the mass is below 75~\Mjup \ is only 15\%.
Regardless, it was not detected in our high-contrast imaging due to its large angular extent so does not enter into our statistical analysis.

\textit{TWA 30 A (2MASS~J11321831--3019518)}.
This young M5 star was identified as a new member of the TWA moving group by \citet{Looper:2010p20583}.
It exhibits \ion{Li}{1}~$\lambda$6708 absorption and forbidden optical line emission, probably a result of outflow activity.  
\citet{Looper:2010p21597} identified a very wide ($\sim$3400~AU) companion, which shows similar forbidden
lines emission.  Although TWA~30~B is much fainter (5~mag in $K$ band), its earlier spectral type (M4)
and variable reddening suggests it harbors an edge-on disk.  
Our NIRC2 data show that TWA~30~A is single down to $\approx$0$\farcs$06 (2.5~AU),  and deep imaging did not reveal
any substellar candidates. 
Note that TWA~30~A is strongly variable in the NIR (\citealt{Looper:2010p21597}).  This affects 
the conversion of relative contrast curves to absolute contrasts and companion mass sensitivities.  For this work
we have adopted the 2MASS $K_S$-band magnitude for the primary, which may not accurately represent the 
apparent brightness of TWA~30~A during our deep ADI observation.

\textit{2MASS J12062214--1314559 AB}.  
This M3.5 system was first resolved from a single epoch of imaging 
by \citet{Janson:2012p23979}  into a $\approx$0$\farcs$4 binary with a $z$-band contrast of 2.2~mag. 
We confirm the physical nature of the pair and detect orbital motion relative to the astrometry from
Janson et al. at epoch 2010.11 ($\rho$=0$\farcs$420~$\pm$~0$\farcs$003, $\theta$=65.9~$\pm$~0.3$^{\circ}$).
\citet{Riaz:2006p20030} identified 2MASS~J12062214--1314559~AB as a chromospherically 
active star ($EW$(H$\alpha$)=--4.9~\AA).  

\textit{LHS~2613 (2MASS~J12424996+4153469)}.
This single, X-ray active M4.0 dwarf has been identified by Shkolnik et al. (in preparation) as a possible nearby young star.
It's parallactic distance of 10.6~$\pm$~1.3~pc (\citealt{vanAltena:1995p25079}) combined with its measured 
RV of --4.0~$\pm$~0.2 km~s$^{-1}$ (Shkolnik et al., in preparation)
imply $UVW$ kinematics of \{--23~$\pm$~3, --13.7~$\pm$~1.6, --5.2~$\pm$~0.3\} km~s$^{-1}$, which 
agree well with the Argus YMG.  The large uncertainty in $U$ is mostly influenced by the error in the distance to the system.
Association with the $\sim$40~Myr Argus group agrees with the star's 
placement on the color-magnitude diagram; with $V$=12.4~mag (\citealt{Zacharias:2013p24823})
and $M_V$=12.3~mag, LHS~2613 lies $\sim$0.5--1~mag above the main sequence given its $V$--$K_S$ color of 5.16 
(see, e.g., Figure~4 from \citealt{Riedel:2014ce}).  This is further bolstered by an 86\% membership probability 
by the BANYAN~II web tool from \citet{Gagne:2014gp}\footnote{http://www.astro.umontreal.ca/~gagne/banyanII.php}.  
If confirmed with a more precise distance, 
LHS~2613 will be among the nearest pre-main sequence stars.

\textit{GJ 1167 A (2MASS~J13093495+2859065)}.
A 194$''$ companion to the M3.5 star GJ~1167~A at a PA of 23$^{\circ}$ (LP~322-835; GJ~1167~B) is listed in the 
Washington Double Star catalog (\citealt{Mason:2001p23846}).  However,
the proper motion of GJ~1167~B ($\mu_\alpha$cos$\delta$=--232~$\pm$~7 mas~yr$^{-1}$, $\mu_\delta$=--160~$\pm$~5 mas~yr$^{-1}$; 
\citealt{Monet:2003p17612}) disagrees with GJ~1167~A ($\mu_\alpha$cos$\delta$=--338~$\pm$~8 mas~yr$^{-1}$, 
$\mu_\delta$=--211~$\pm$~8 mas~yr$^{-1}$; \citealt{Zacharias:2013p24823}), 
so the pair are unlikely to be physically related.

\textit{G 227-22 (2MASS J18021660+6415445)}.
The parallactic distance of 8.5~$\pm$~0.3~pc to G~227-22 from \citet{Dittmann:2013tp} combined with the radial velocity measurement
of --1.2~$\pm$~0.2~km~s$^{-1}$ from \citet{Shkolnik:2012p24056} yield $UVW$ space velocities of 
\{15.0~$\pm$~0.6, 4.2~$\pm$~0.3, --8.3~$\pm$~0.4\} km~s$^{-1}$.  These are in excellent agreement with the Ursa Major moving group
(\citealt{AmmlerVonEiff:2009p22831}), so we assign G~227-22 as a likely member of that association and adopt the group age of
500~$\pm$~100~Myr for this star.

\textit{2MASS~J20003177+5921289  AB}.  This near equal-flux M4.1-type 0$\farcs$3 binary was first identified by 
\citet{Janson:2012p23979} from imaging in 2008 and 2009.  Our 2010 data show continued  
outward orbital motion by $\approx$20~mas in separation and $\approx$5$^{\circ}$ in PA.

\textit{NLTT 48651 (2MASS J20043077--2342018)}.
NLTT 48651 is a single M4.5 dwarf detected by $ROSAT$ and $GALEX$ and identified by Shkolnik et al. (in preparation)
as a possible nearby young star.  The RV of --7.5~$\pm$~0.7 km~s$^{-1}$ measured by Shkolnik et al. enable partial
constraints on the star's kinematics, which agree well with AB~Dor moving group members at a distance of $\sim$18~pc.
Indeed, the BANYAN~II web tool suggests an AB Dor membership probability of 93\%, so we adopt the AB~Dor age
of 120~$\pm$~10 Myr for this star.  A parallax will be needed for unambiguous confirmation of group membership.

\textit{2MASS J20284361--1128307}.
This X-ray active M3.5 dwarf has mostly gone unnoticed in the literature.  \citet{Riaz:2006p20030} measured
moderately strong H$\alpha$ emission (6.3~\AA) and, more recently, \citet{Riedel:2014ce} presented
a trigonometric distance of 18.8~$\pm$~0.6~pc.  Although lacking a radial velocity, Riedel et al. argue that
this star is a probable member of the Argus association based on its position on the HR diagram and 
partially-constrained kinematics.  Assuming membership to Argus, they predict a radial velocity of
--25.4~km~s$^{-1}$.  Shkolnik et al. (in preparation) measure a radial velocity of --25.2~$\pm$~0.3 km~s$^{-1}$ 
implying $UVW$ kinematics of \{--24.4~$\pm$~0.4, --17.7~$\pm$~0.5, --3.8~$\pm$~0.7\} km~s$^{-1}$.
These are in good agreement with the Argus moving group, so we consider 2MASS J20284361--1128307 
a likely member.

\textit{NLTT 50066 AB (2MASS~J20531465--0221218)}.
This M3.0 equal flux binary ($\Delta$$H$=0.1~mag) 
was first resolved by  \citet{Janson:2012p23979}.  The pair has undergone significant
orbital evolution since the Janson et al. first epoch in 2008.  Its parallactic distance
of 37.9~$\pm$~5.7~pc (\citealt{Shkolnik:2012p24056}) implies a physical separation of $\sim$3--5~AU.
With an expected orbital period of $\sim$10--20~yr, 
astrometric monitoring should continue in order to yield a dynamical mass.

\textit{G 68-46 (2MASS~J23512227+2344207)}.
Lacking a parallax for G~68-46, previous studies have tentatively associated this active M4.0e star with 
the $\beta$~Pic (\citealt{Malo:2013p24348}) and Cha-Near (\citealt{Shkolnik:2012p24056}) moving groups.  
However, based on the trigonometric distance of 21.0~$\pm$~1.3~pc
from \citet{Dittmann:2013tp} and radial velocity of --2.1~$\pm$~0.5 km~s$^{-1}$ from \citet{Shkolnik:2012p24056},
we find that the $UVW$ kinematics of G 68-46 (\{--19.4~$\pm$~1.5, --16.4~$\pm$~1.1, --10.5~$\pm$~1.0\} km~s$^{-1}$) do
not match those of any nearby young moving groups.  We therefore adopt the age estimate of 35--300~Myr 
from \citet{Shkolnik:2009p19565}.

\acknowledgments

We are grateful to our anonymous referee for helpful comments,
Katelyn Allers for the low gravity spectral templates used in this work,
Adam Kraus and Trent Dupuy for assistance with some of the observations, 
John Johnson for constructive comments on this paper, and Kimberly Aller for measuring gravity indices.
It is a pleasure to thank the telescope operators and support astronomers Jun Hashimoto, Alan Hatakeyama, Ryo Kandori, 
Tomoyuki Kudo, Nobahiko Kusakabe, and Joshua Williams at Subaru Telescope and
Joel Aycock, Randy Campbell, Al Conrad, Heather Hershley, Marc Kassis, Jim Lyke, Jason McIlroy, 
Barbara Schaefer, Terry Stickel, Hien Tran, and Cynthia Wilburn at Keck Observatory for their support with the observations.
B.P.B. and M.C.L. have been supported by NASA grant NNX11AC31G and NSF grant AST09-09222.
We utilized data products from the Two Micron All Sky Survey, which is a joint project of the University of Massachusetts and the Infrared Processing and Analysis Center/California Institute of Technology, funded by the National Aeronautics and Space Administration and the National Science Foundation.
 NASA's Astrophysics Data System Bibliographic Services together with the VizieR catalogue access tool and SIMBAD database 
operated at CDS, Strasbourg, France, were invaluable resources for this work.
This research has made use of the Washington Double Star Catalog maintained at the U.S. Naval Observatory.
Finally, mahalo nui loa to the kama`\={a}ina of Hawai`i for their support of Keck and the Mauna Kea observatories.
We are grateful to conduct observations from this mountain.

\facility{{\it Facilities}: \facility{Keck:II (NIRC2)}, \facility{Subaru (HiCIAO)}, \facility{Keck:II (OSIRIS)}, \facility{IRTF (SpeX)}}

\newpage

\bibliographystyle{apj}
\bibliography{palms4_revised.bbl}


\begin{figure}
  \vskip -.3in
  \hskip .7in
  \resizebox{6in}{!}{\includegraphics{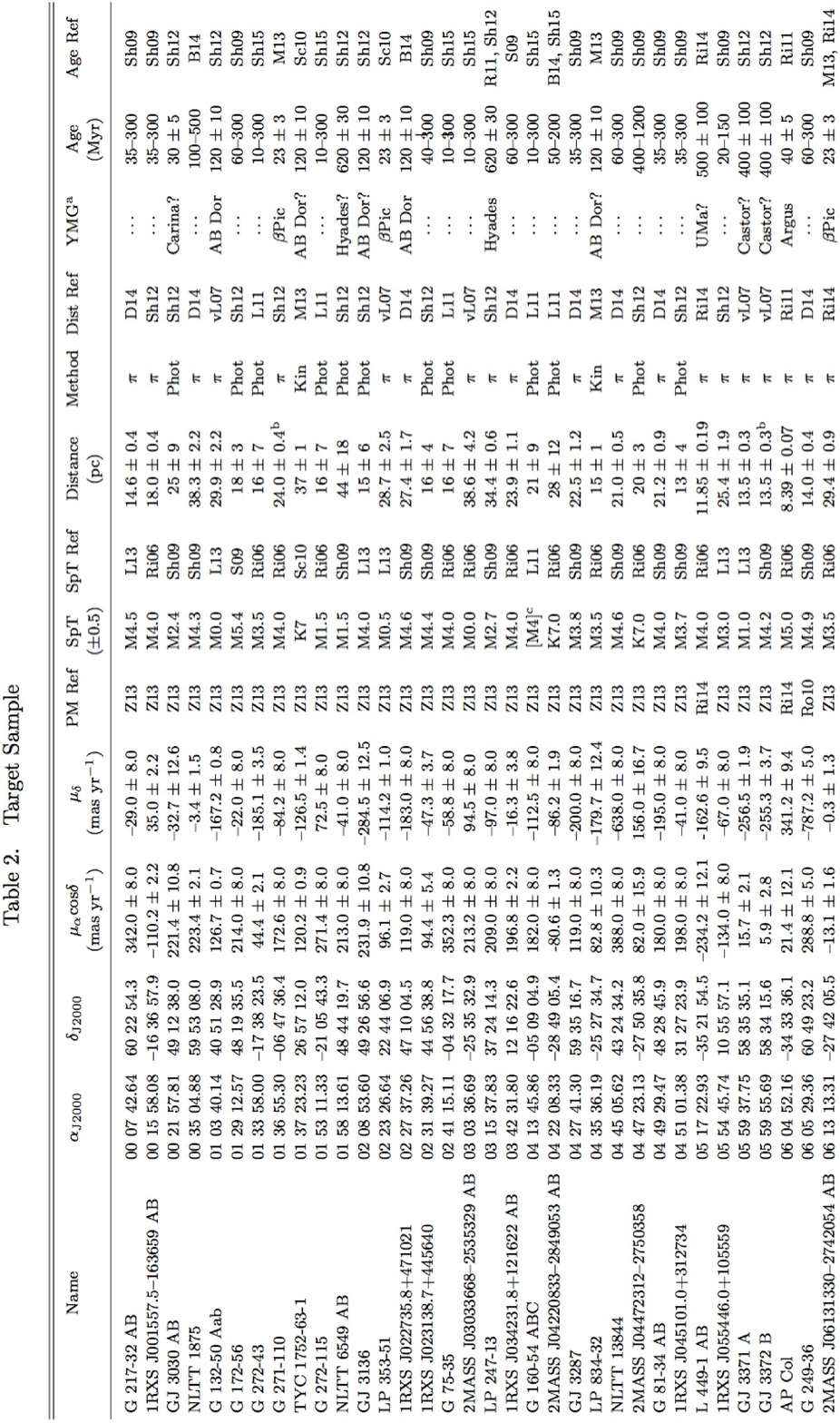}}
  \vskip 2in
  \caption{Table 2  \label{test}}
\end{figure}
\clearpage


\begin{figure}
  \vskip -.3in
  \hskip .7in
  \resizebox{6in}{!}{\includegraphics{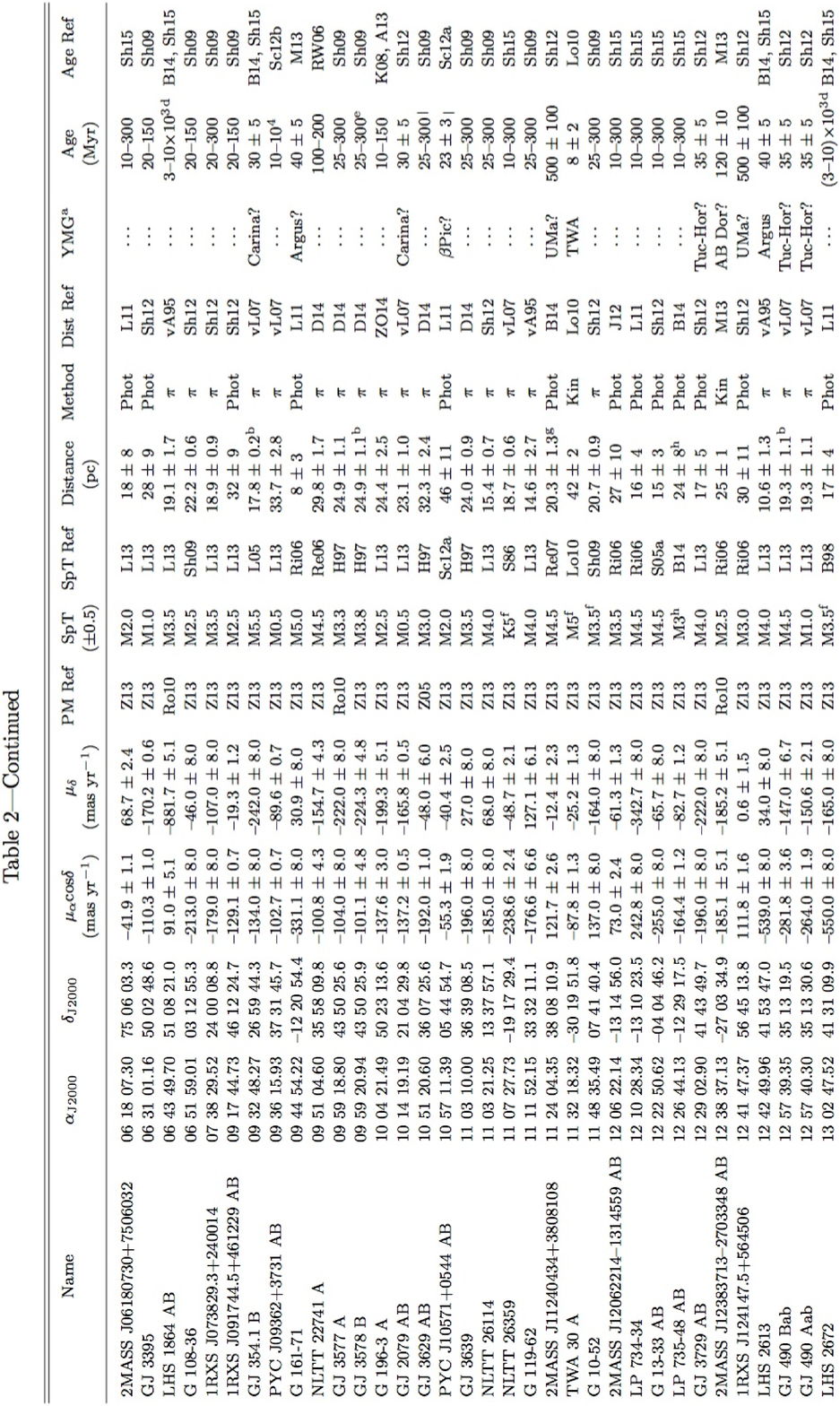}}
  \vskip 2in
  \caption{Table 2  \label{test}}
\end{figure}
\clearpage


\begin{figure}
  \vskip -.3in
  \hskip .7in
  \resizebox{6in}{!}{\includegraphics{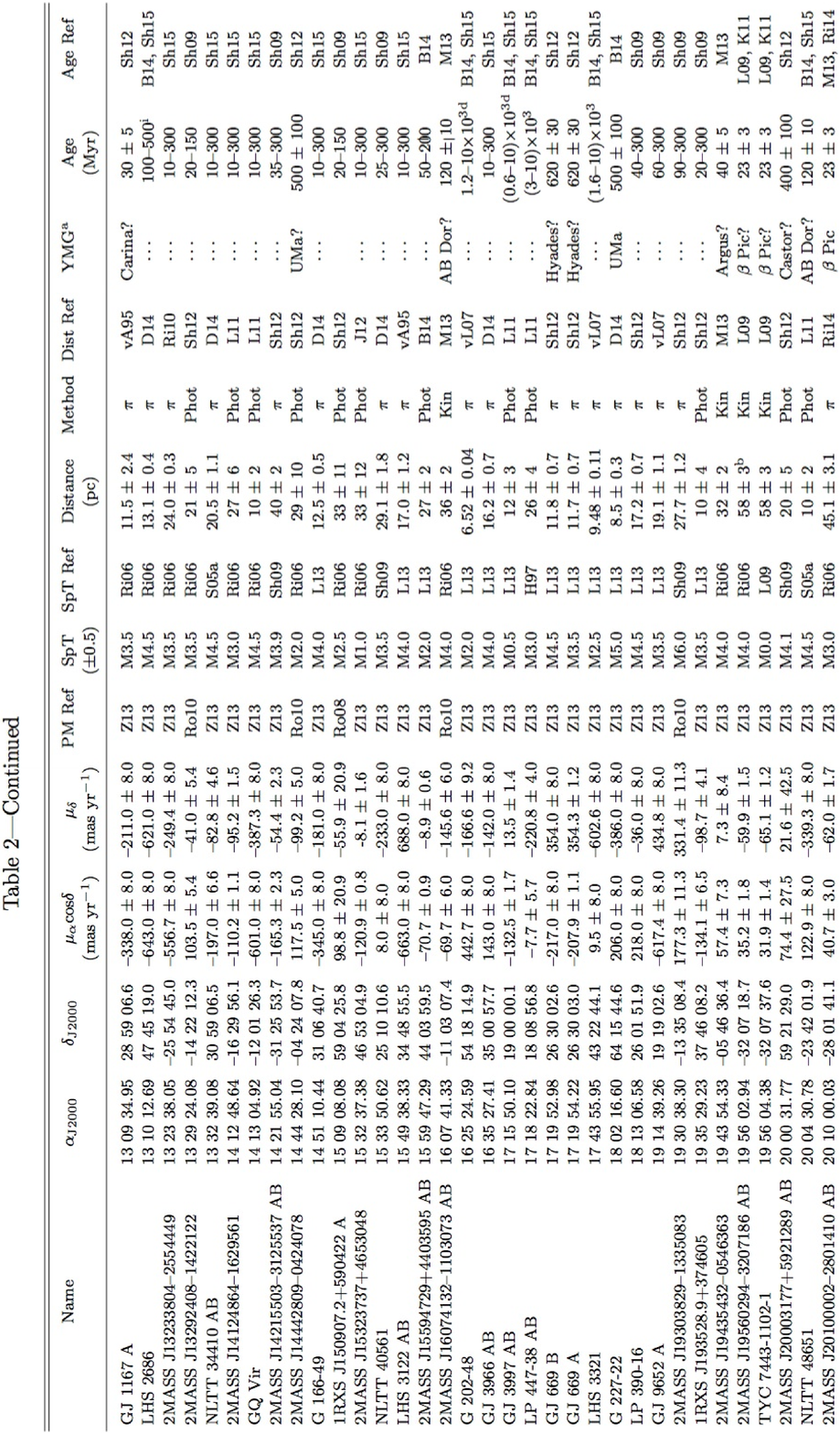}}
  \vskip 2in
  \caption{Table 2  \label{test}}
\end{figure}
\clearpage


\begin{figure}
  \vskip -.3in
  \hskip .7in
  \resizebox{6in}{!}{\includegraphics{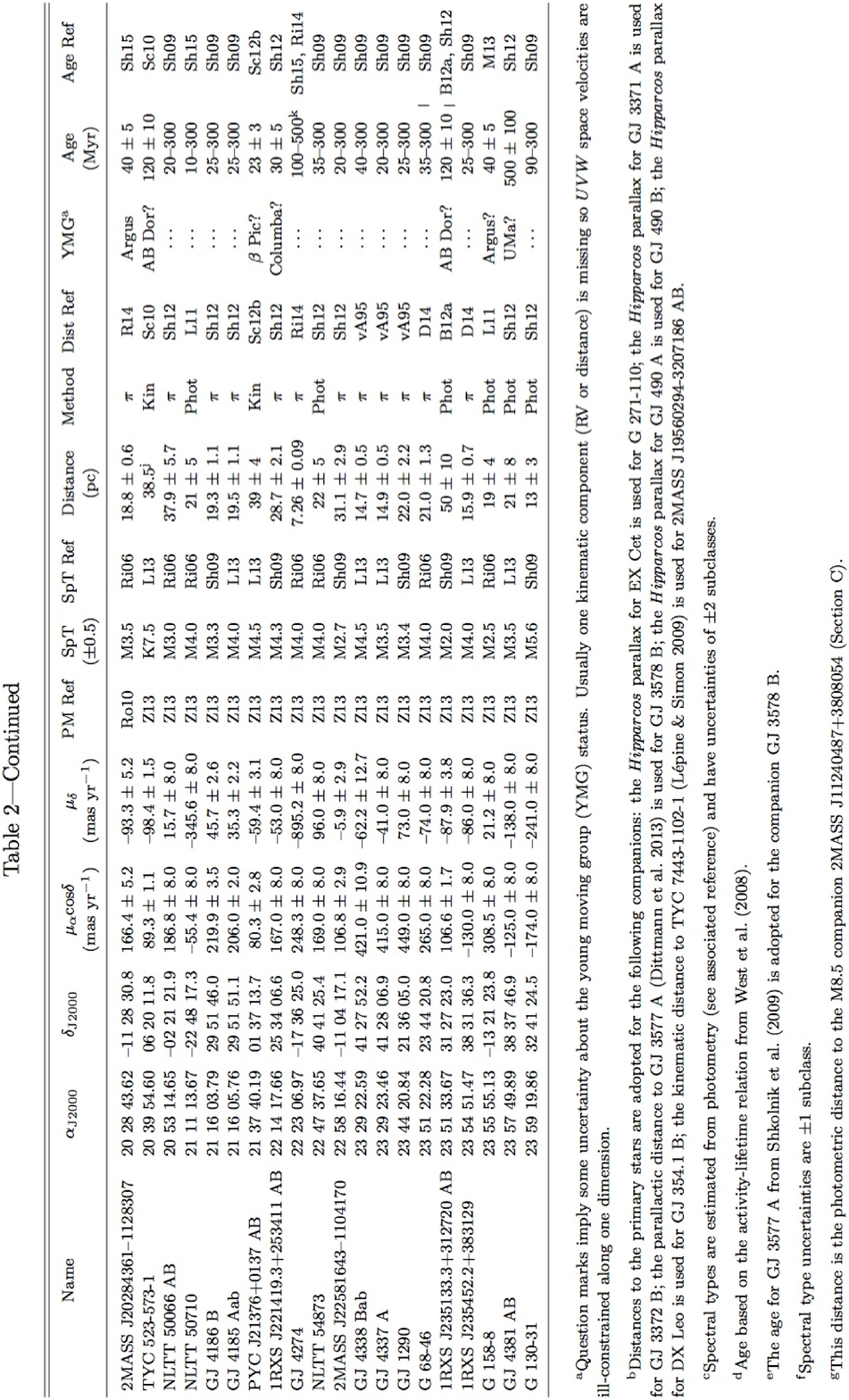}}
  \vskip 2in
  \caption{Table 2  \label{test}}
\end{figure}
\clearpage


\begin{figure}
  \vskip 0in
  \hskip .7in
  \resizebox{3.in}{!}{\includegraphics{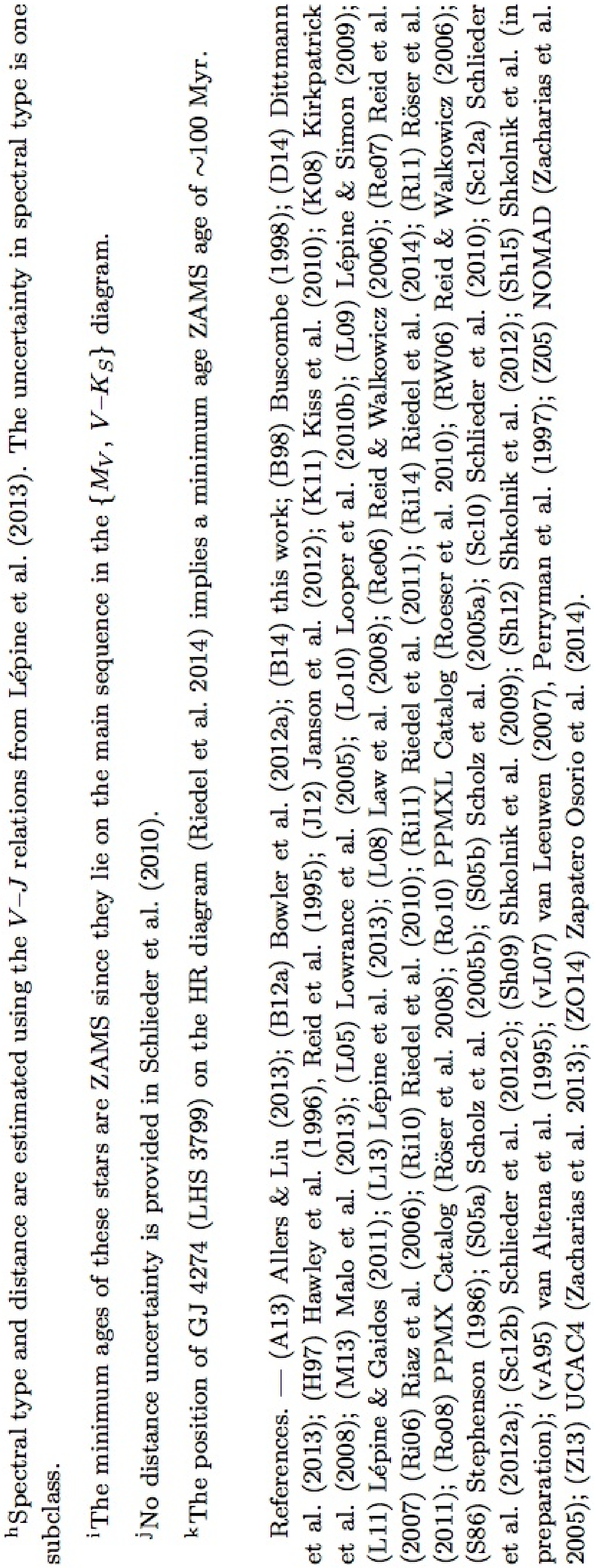}}
  \vskip 2in
  \caption{Table 2  \label{test}}
\end{figure}
\clearpage


\begin{figure}
  \vskip .8in
  \hskip .1in
  \resizebox{7.2in}{!}{\includegraphics{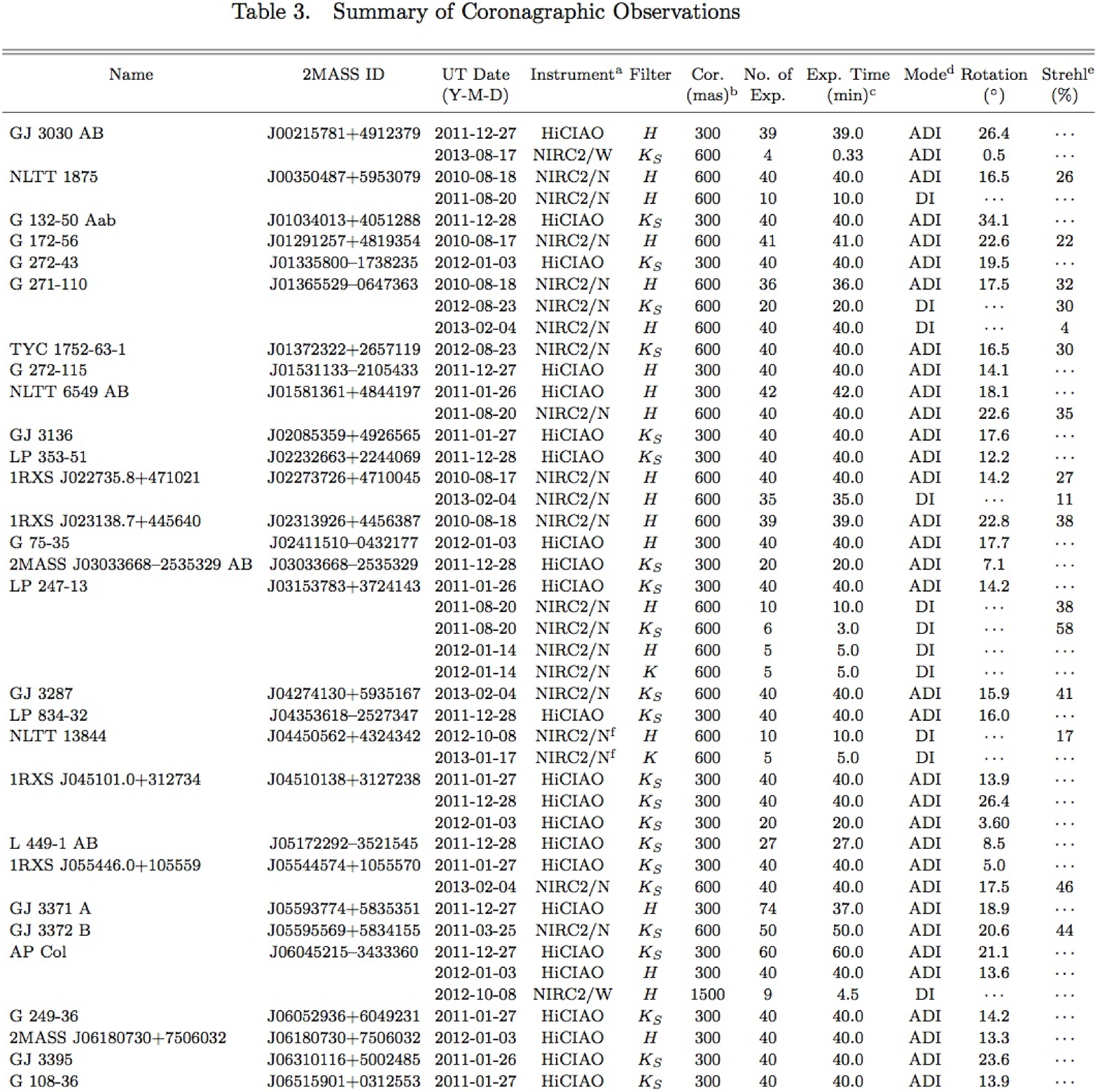}}
  \vskip 2in
  \caption{Table 3  \label{test}}
\end{figure}
\clearpage


\begin{figure}
  \vskip .8in
  \hskip .1in
  \resizebox{7.2in}{!}{\includegraphics{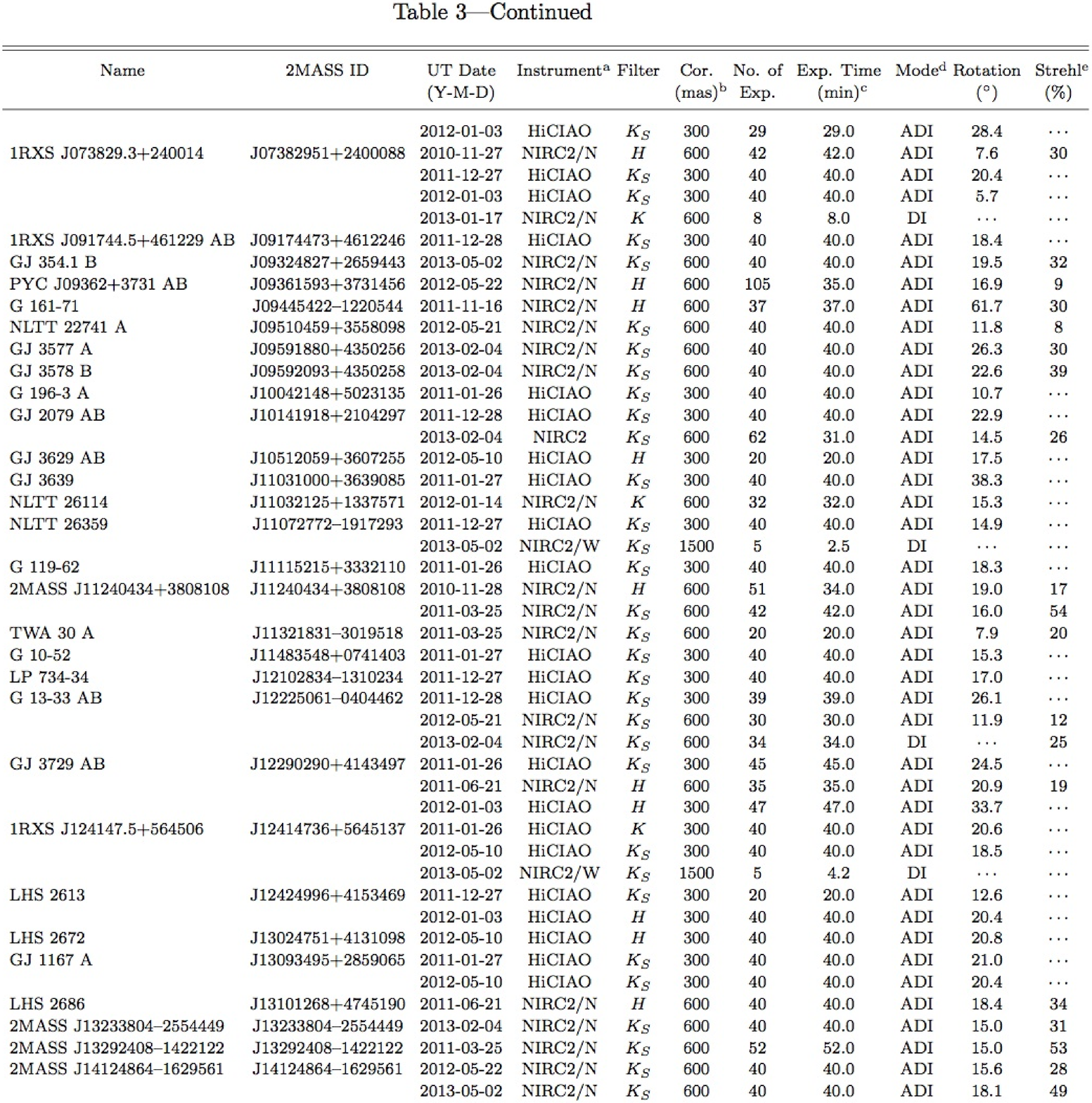}}
  \vskip 2in
  \caption{Table 3  \label{test}}
\end{figure}
\clearpage


\begin{figure}
  \vskip .8in
  \hskip .1in
  \resizebox{7.2in}{!}{\includegraphics{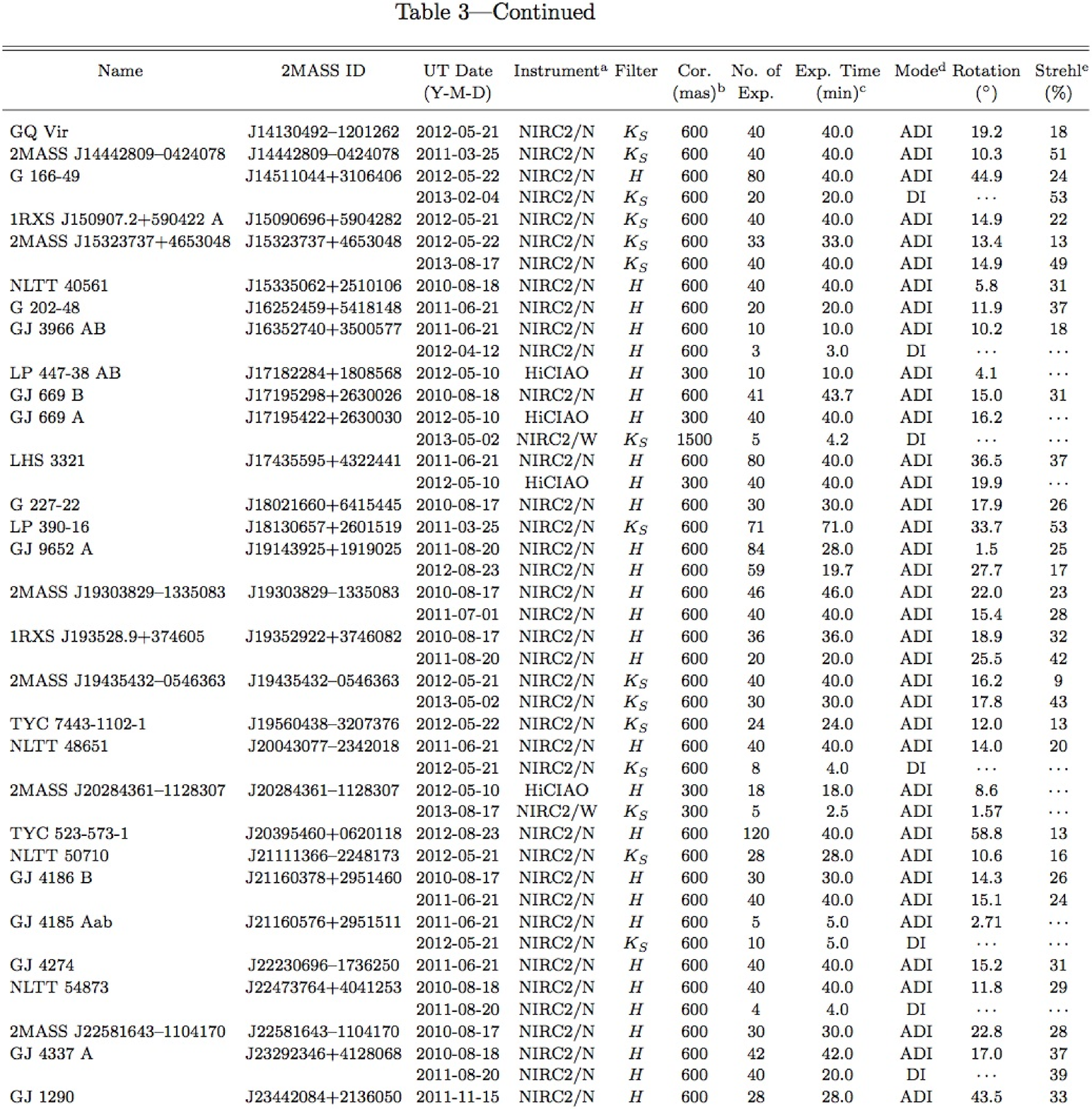}}
  \vskip 2in
  \caption{Table 3  \label{test}}
\end{figure}
\clearpage


\begin{figure}
  \vskip .8in
  \hskip .1in
  \resizebox{7.2in}{!}{\includegraphics{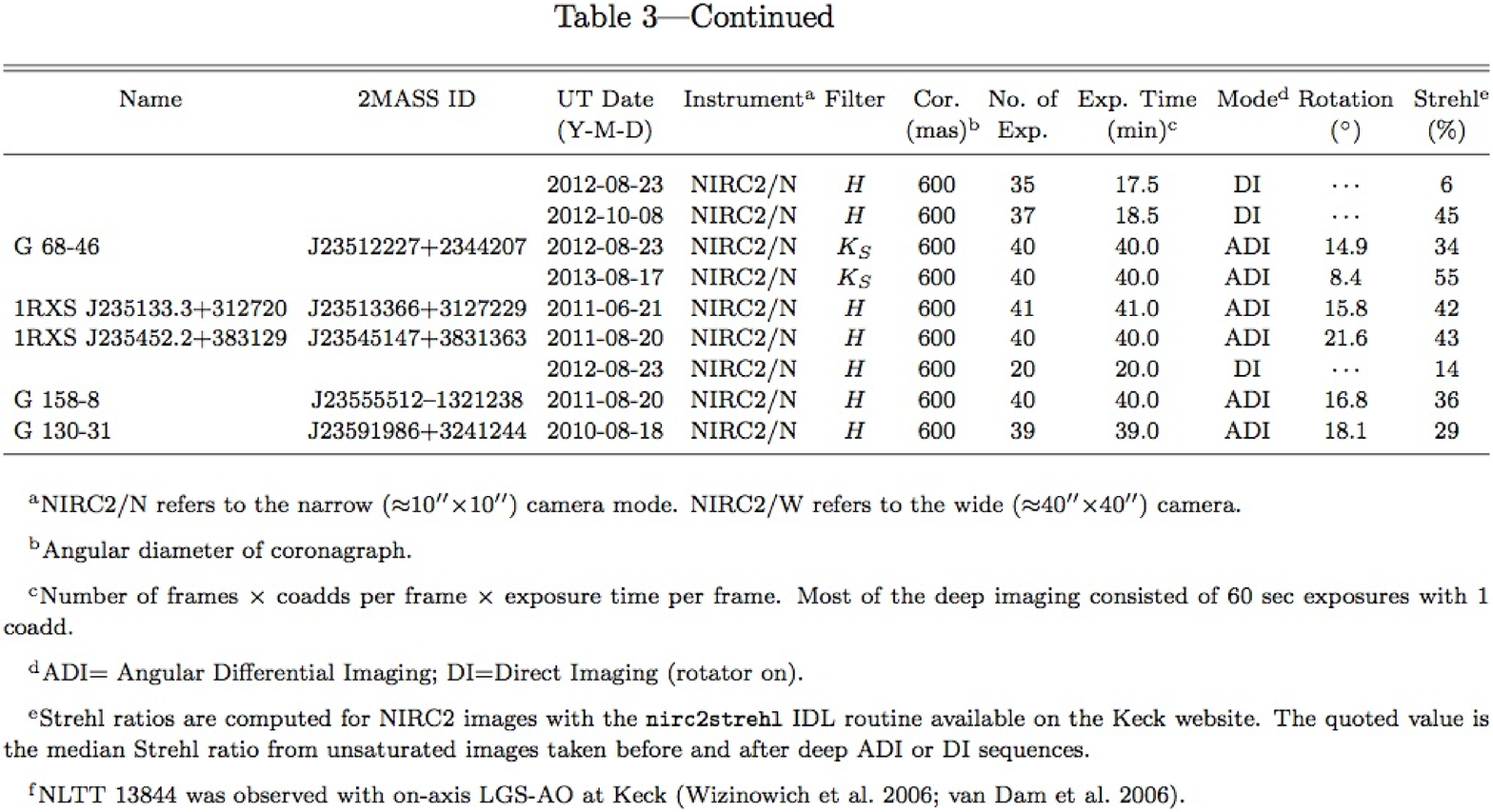}}
  \vskip 2in
  \caption{Table 3  \label{test}}
\end{figure}
\clearpage


\begin{figure}
  \vskip -.1in
  \hskip .1in
  \resizebox{7.in}{!}{\includegraphics{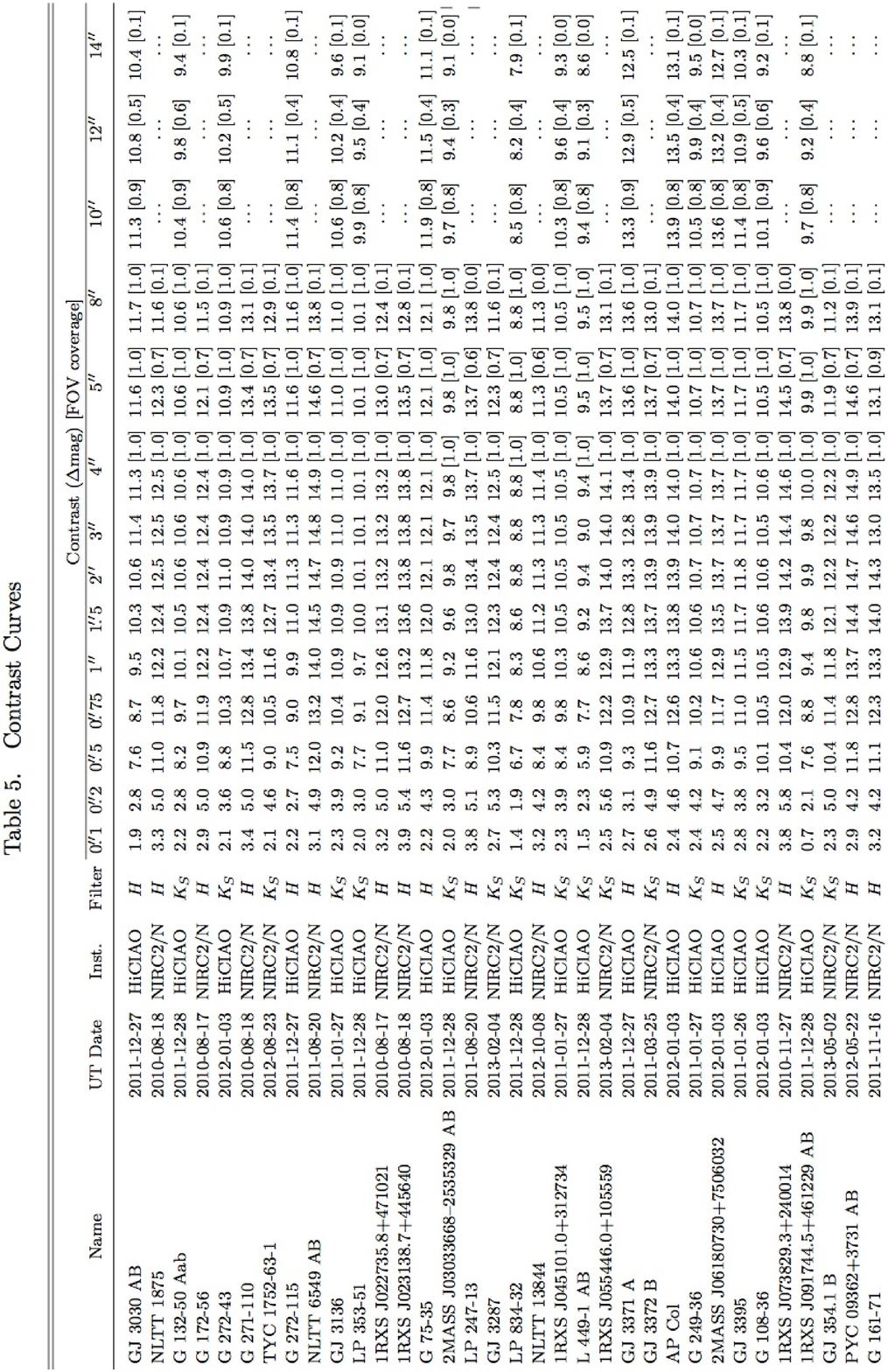}}
  \vskip 2in
  \caption{Table 5  \label{test}}
\end{figure}
\clearpage


\begin{figure}
  \vskip -.3in
  \hskip .2in
  \resizebox{7.in}{!}{\includegraphics{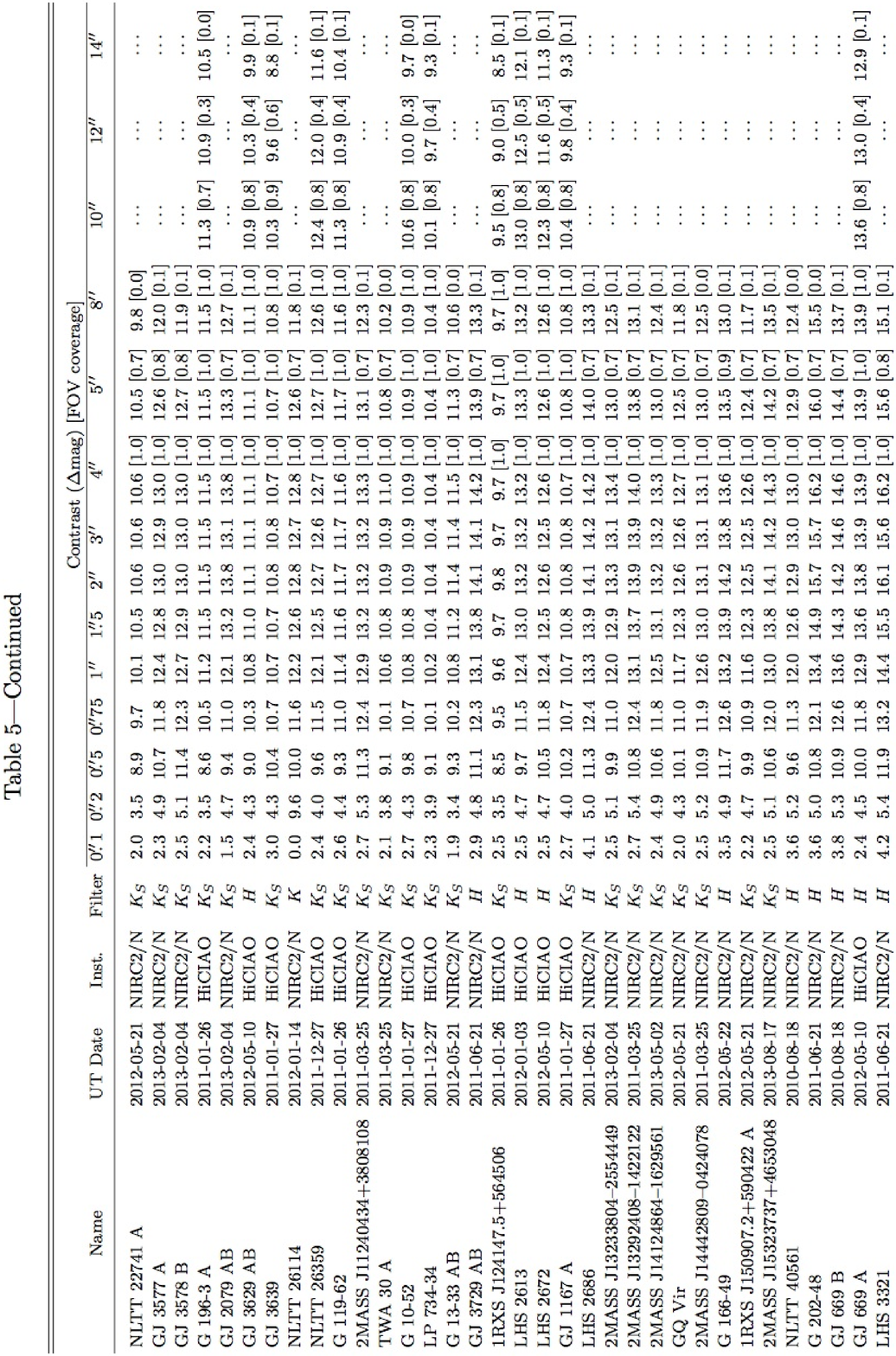}}
  \vskip 2in
  \caption{Table 5  \label{test}}
\end{figure}
\clearpage


\begin{figure}
  \vskip -.3in
  \hskip .2in
  \resizebox{7.in}{!}{\includegraphics{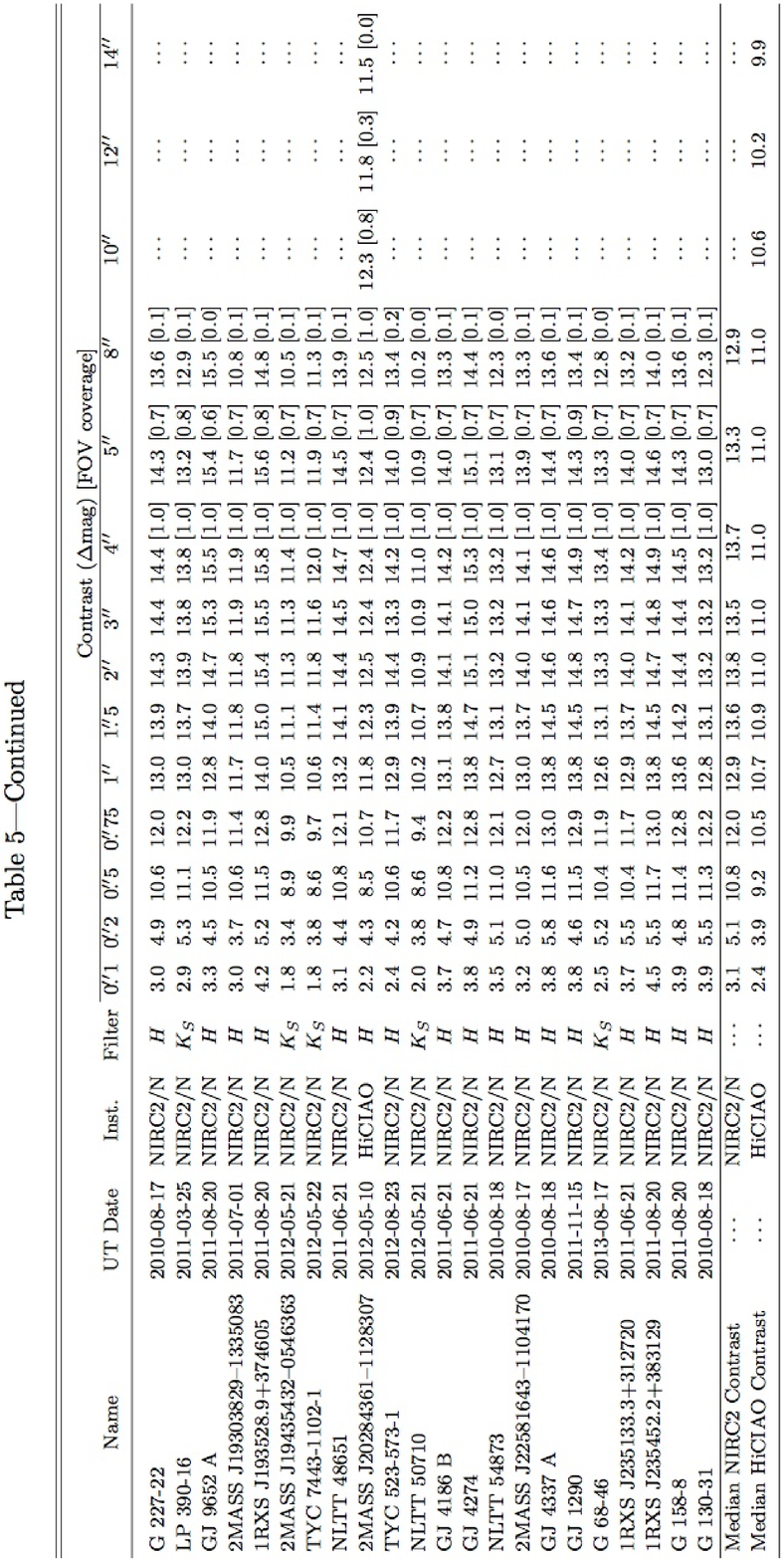}}
  \vskip 2in
  \caption{Table 5  \label{test}}
\end{figure}
\clearpage


\begin{figure}
  \vskip .1in
  \hskip 1.3in
  \resizebox{2.5in}{!}{\includegraphics{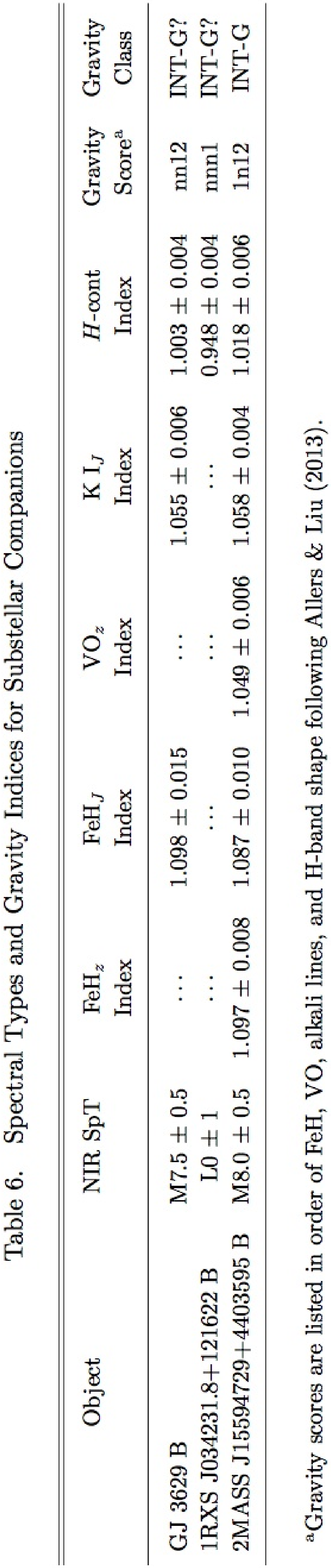}}
  \vskip 2in
  \caption{Table 6  \label{test}}
\end{figure}
\clearpage


\begin{figure}
  \vskip -.2in
  \hskip .3in
  \resizebox{6.5in}{!}{\includegraphics{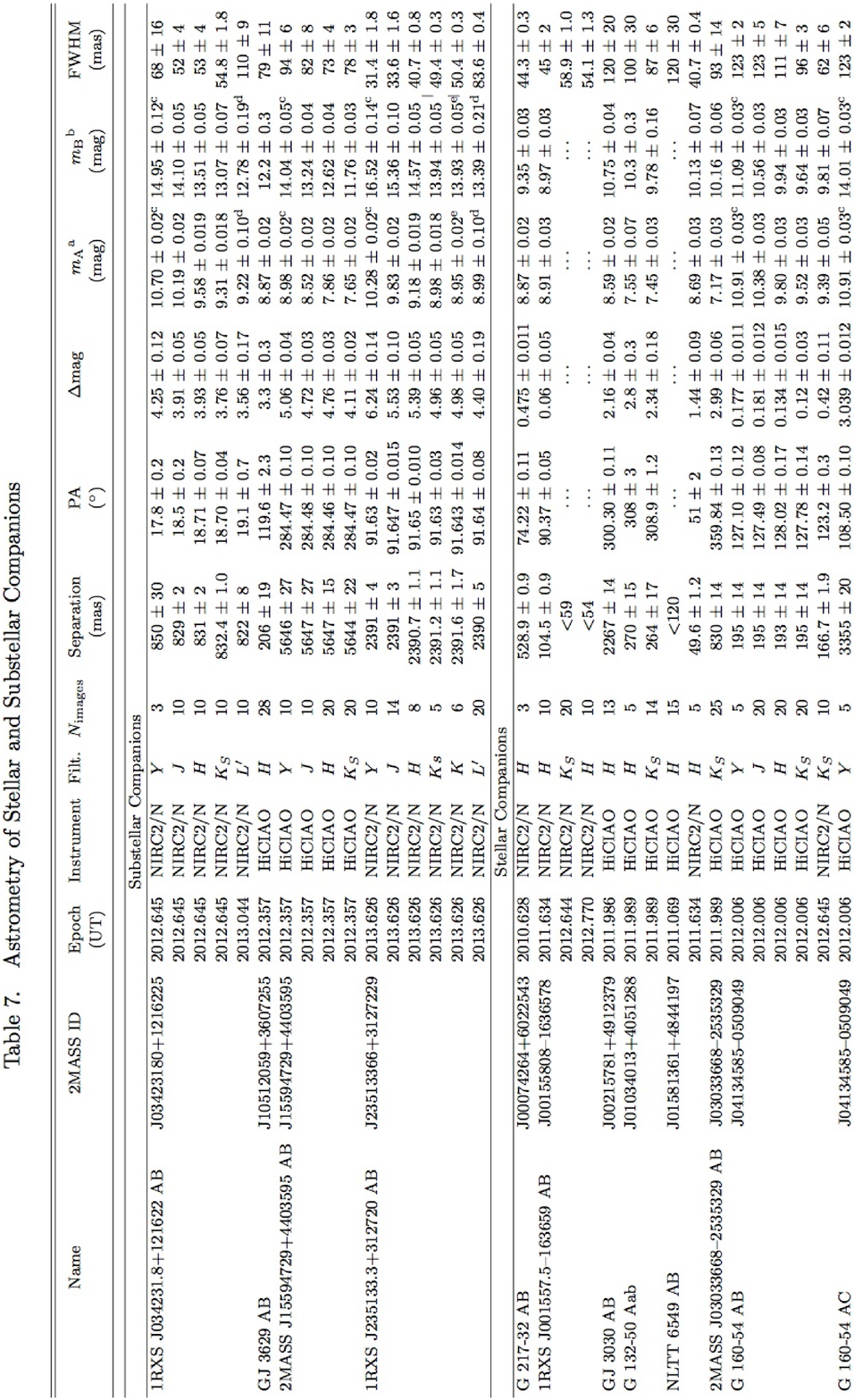}}
  \vskip 2in
  \caption{Table 7  \label{test}}
\end{figure}
\clearpage


\begin{figure}
  \vskip -.3in
  \hskip .3in
  \resizebox{6.3in}{!}{\includegraphics{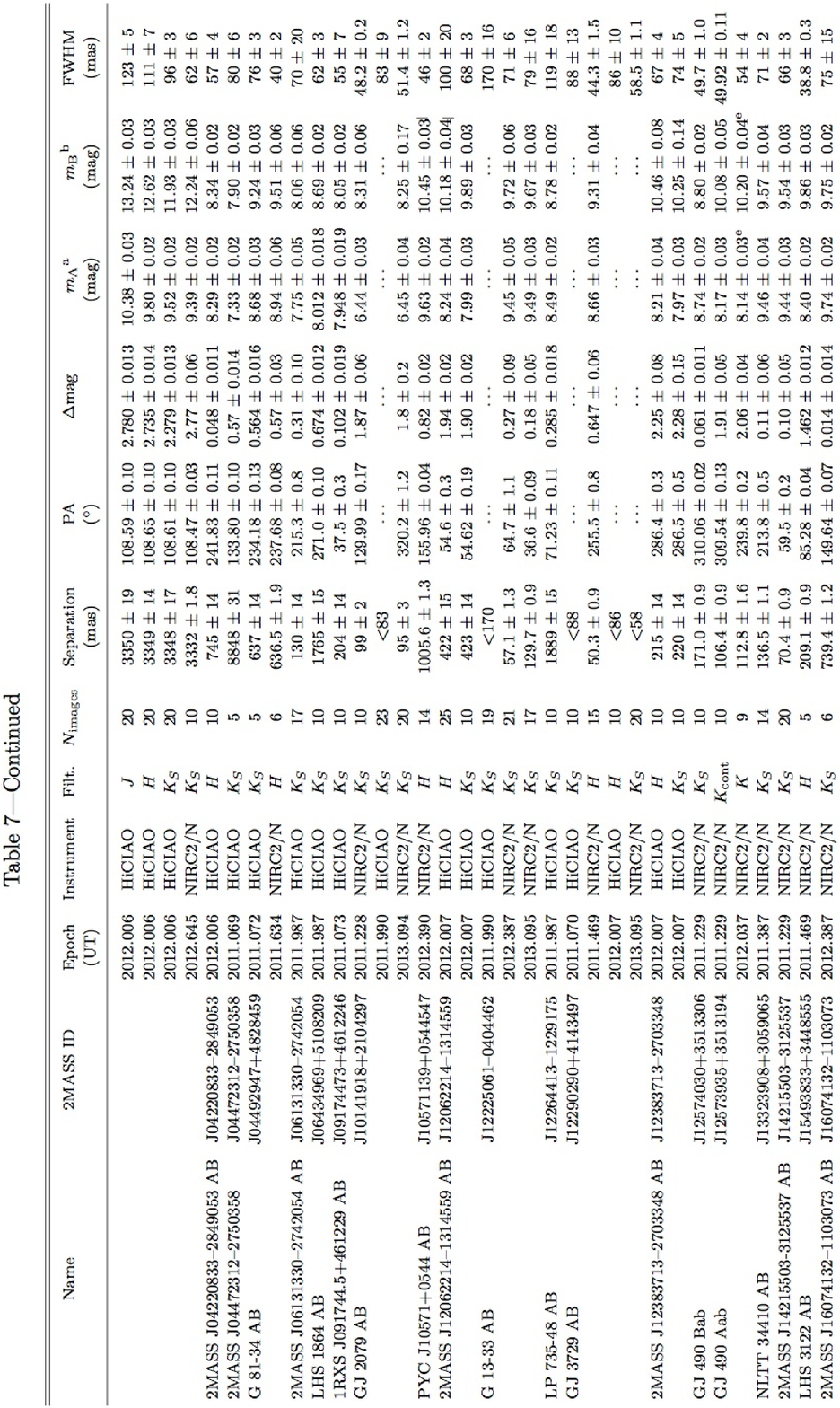}}
  \vskip 2in
  \caption{Table 7  \label{test}}
\end{figure}
\clearpage


\begin{figure}
  \vskip -.2in
  \hskip .3in
  \resizebox{5.in}{!}{\includegraphics{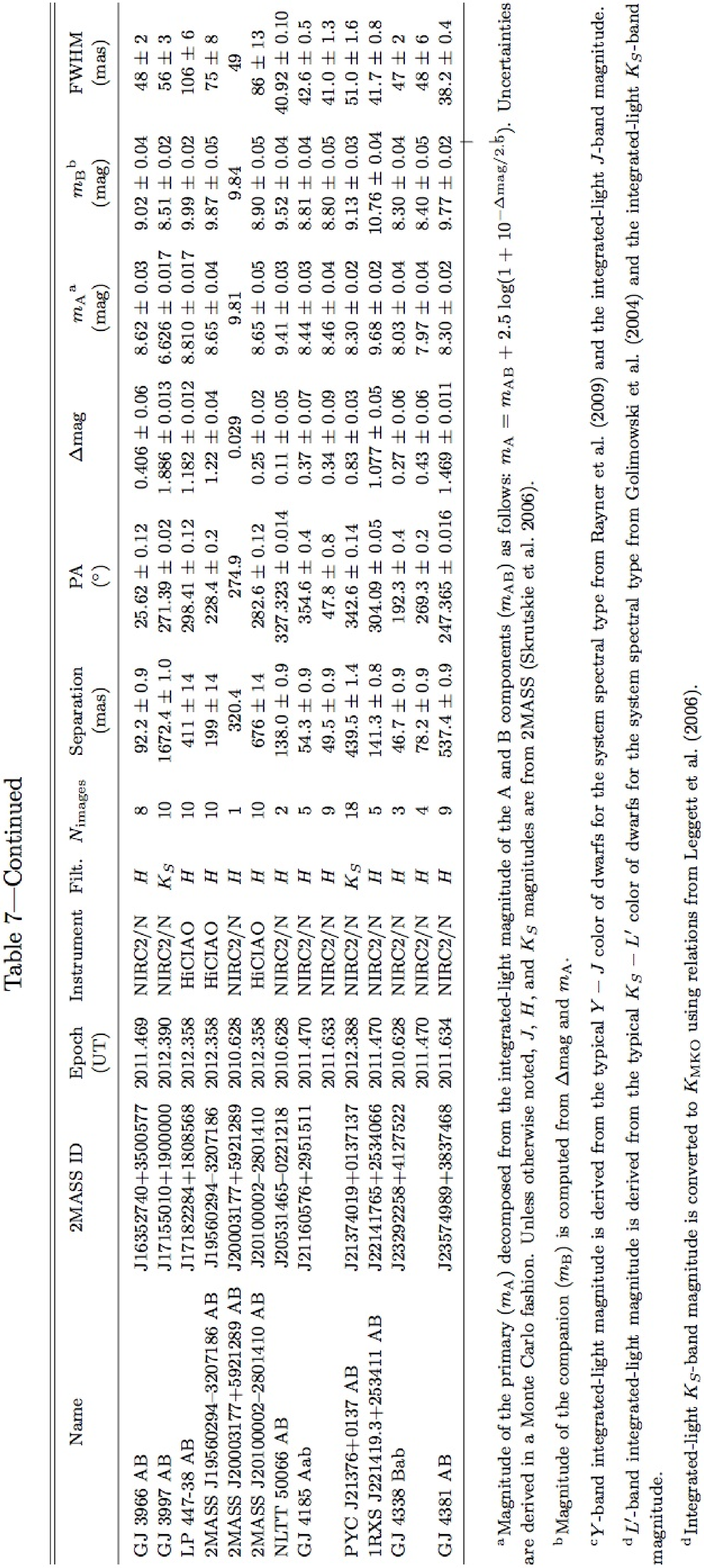}}
  \vskip 2in
  \caption{Table 7  \label{test}}
\end{figure}
\clearpage


\begin{figure}
  \vskip -.2in
  \hskip .3in
  \resizebox{6.3in}{!}{\includegraphics{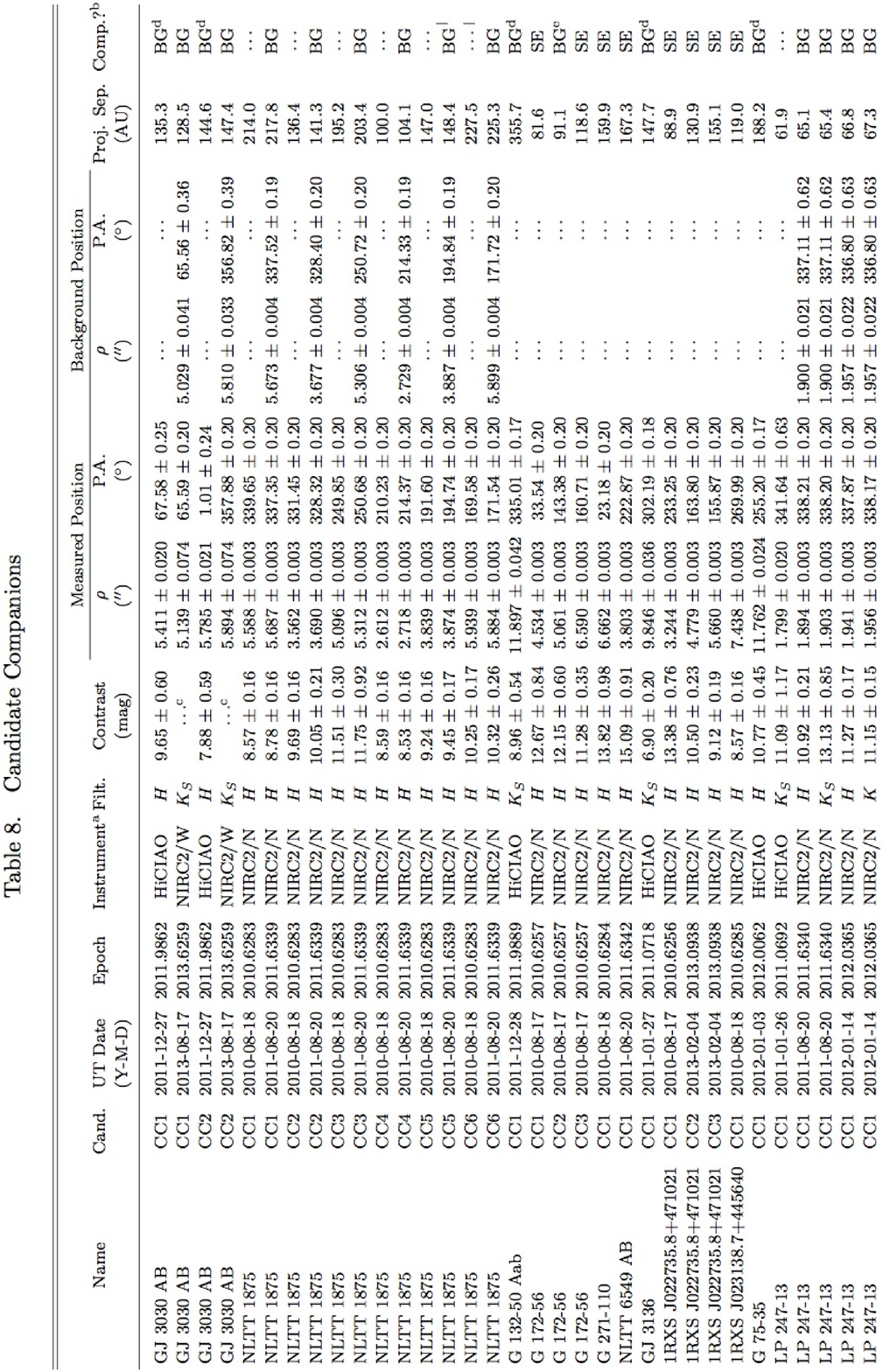}}
  \vskip 2in
  \caption{Table 8  \label{test}}
\end{figure}
\clearpage


\begin{figure}
  \vskip -.2in
  \hskip .3in
  \resizebox{6.3in}{!}{\includegraphics{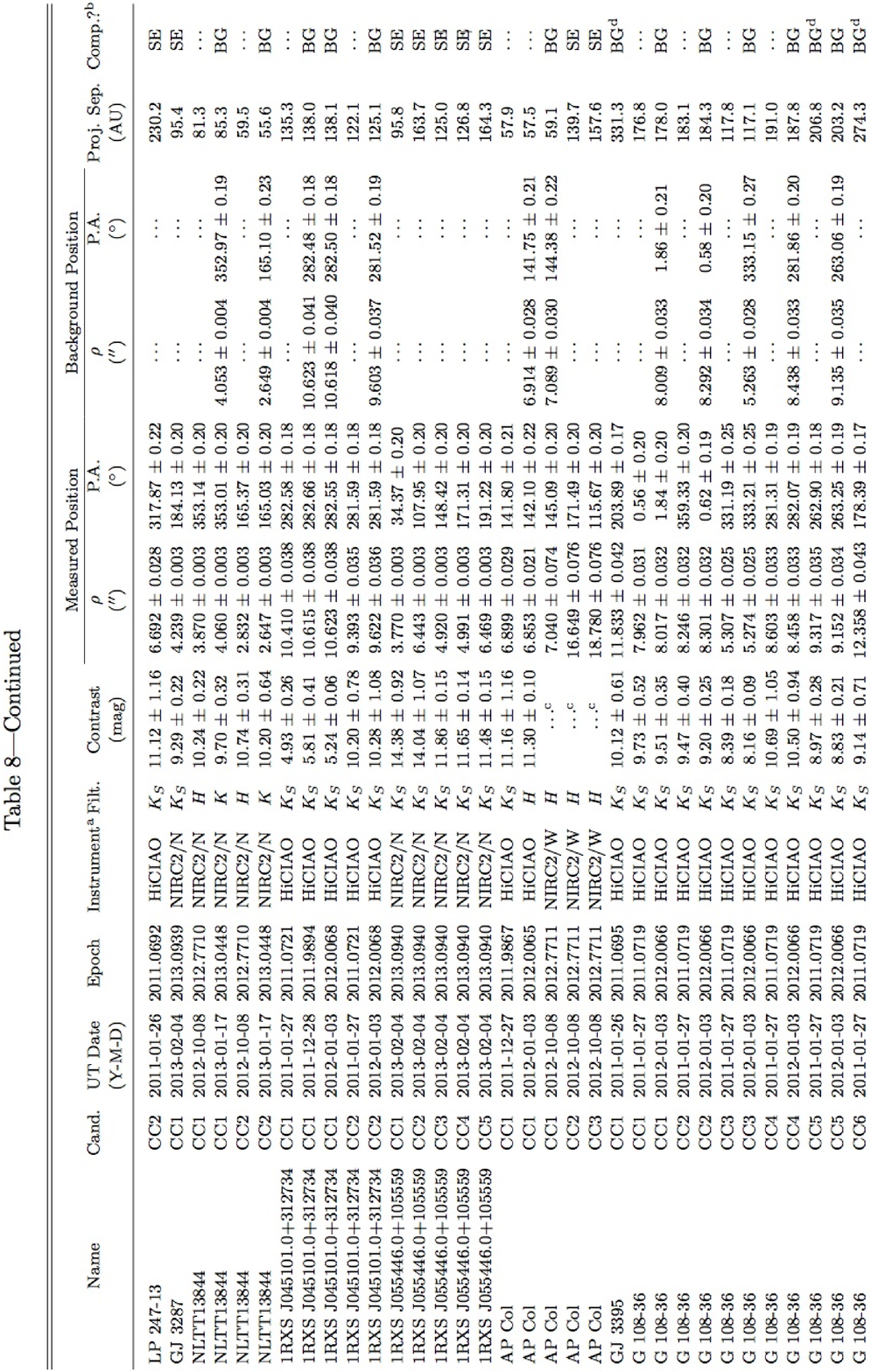}}
  \vskip 2in
  \caption{Table 8  \label{test}}
\end{figure}
\clearpage


\begin{figure}
  \vskip -.2in
  \hskip .3in
  \resizebox{6.3in}{!}{\includegraphics{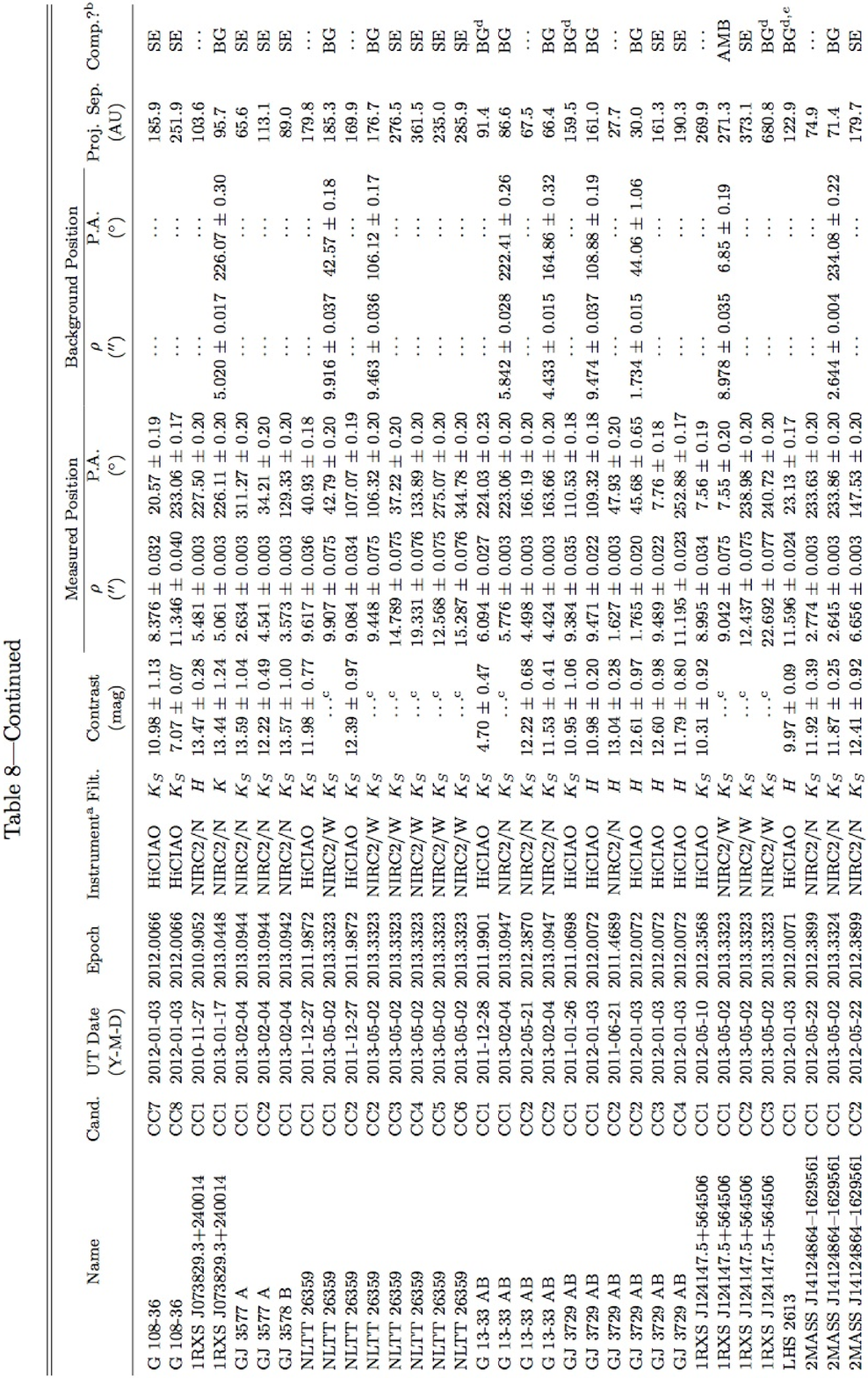}}
  \vskip 2in
  \caption{Table 8  \label{test}}
\end{figure}
\clearpage


\begin{figure}
  \vskip -.2in
  \hskip .3in
  \resizebox{6.3in}{!}{\includegraphics{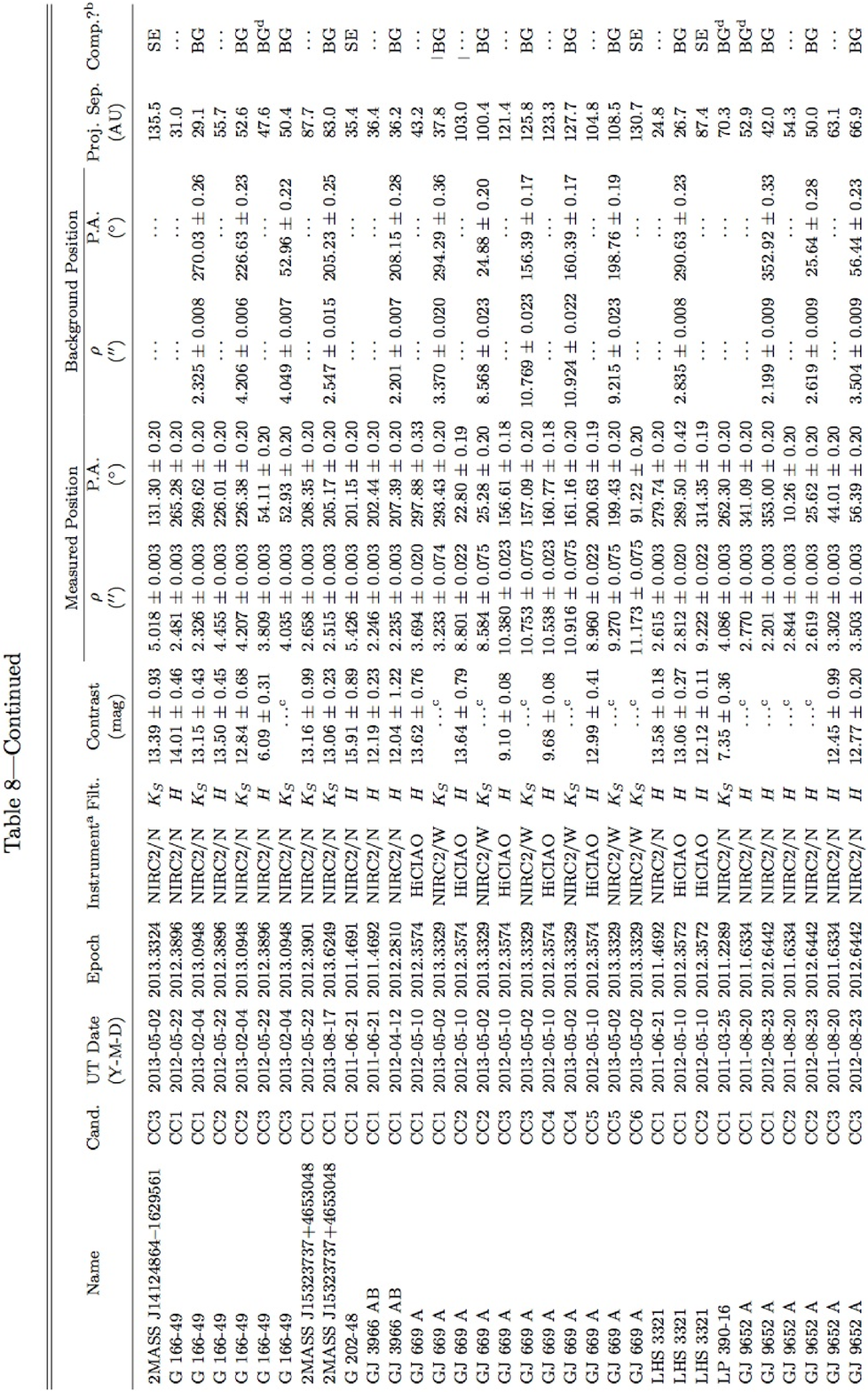}}
  \vskip 2in
  \caption{Table 8  \label{test}}
\end{figure}
\clearpage


\begin{figure}
  \vskip -.2in
  \hskip .3in
  \resizebox{6.7in}{!}{\includegraphics{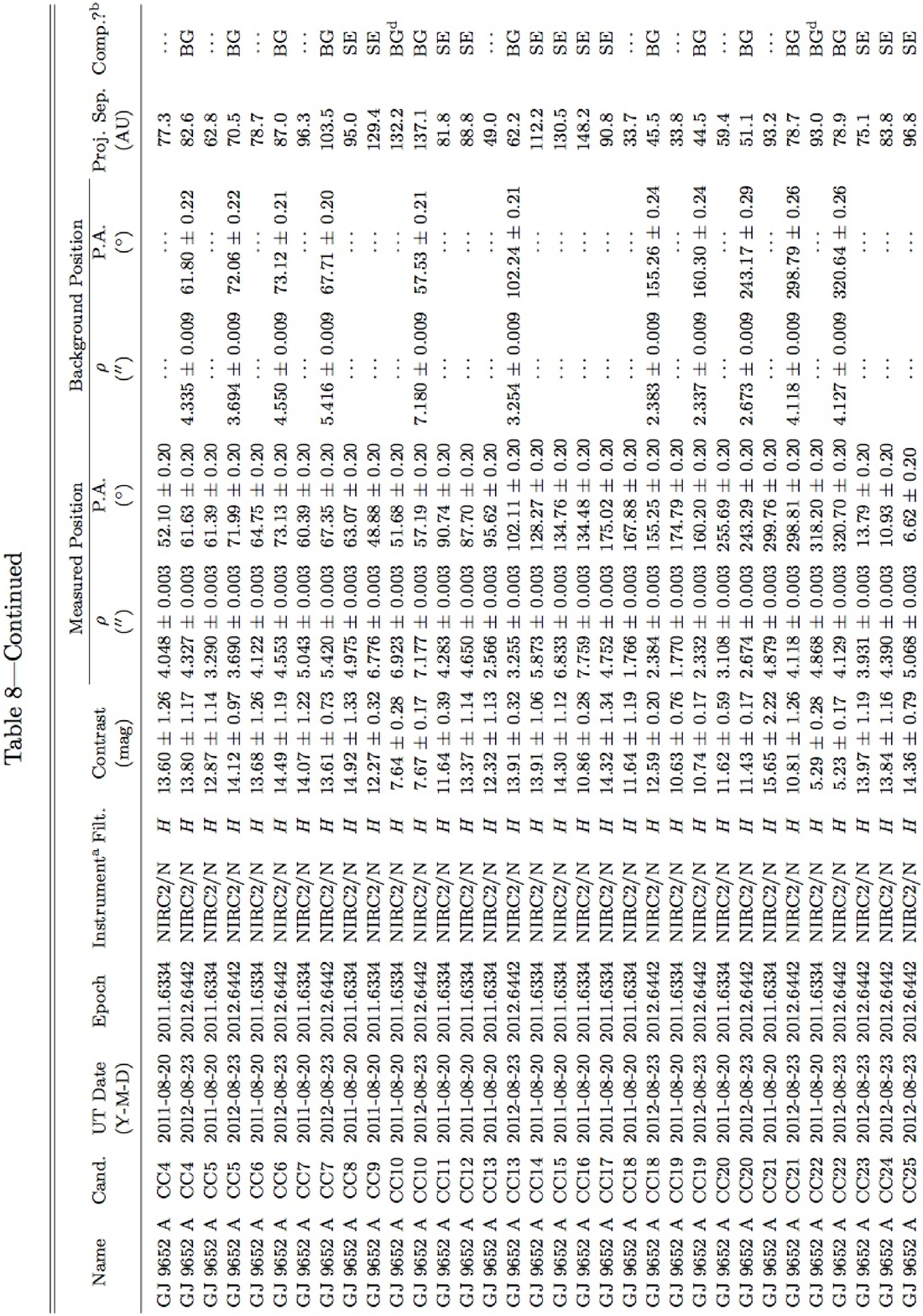}}
  \vskip 2in
  \caption{Table 8  \label{test}}
\end{figure}
\clearpage


\begin{figure}
  \vskip -.2in
  \hskip .3in
  \resizebox{6.3in}{!}{\includegraphics{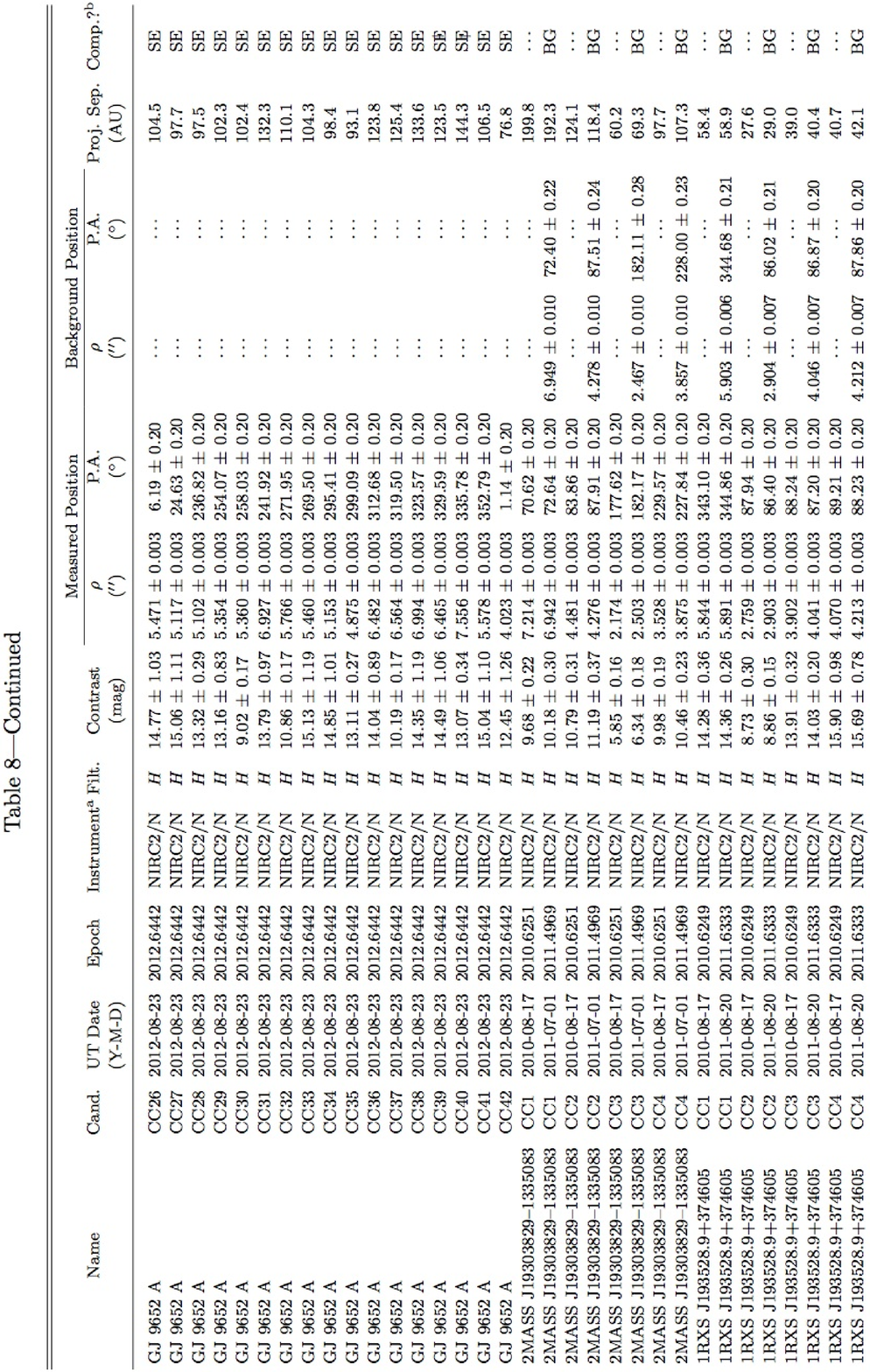}}
  \vskip 2in
  \caption{Table 8  \label{test}}
\end{figure}
\clearpage


\begin{figure}
  \vskip -.2in
  \hskip .3in
  \resizebox{6.3in}{!}{\includegraphics{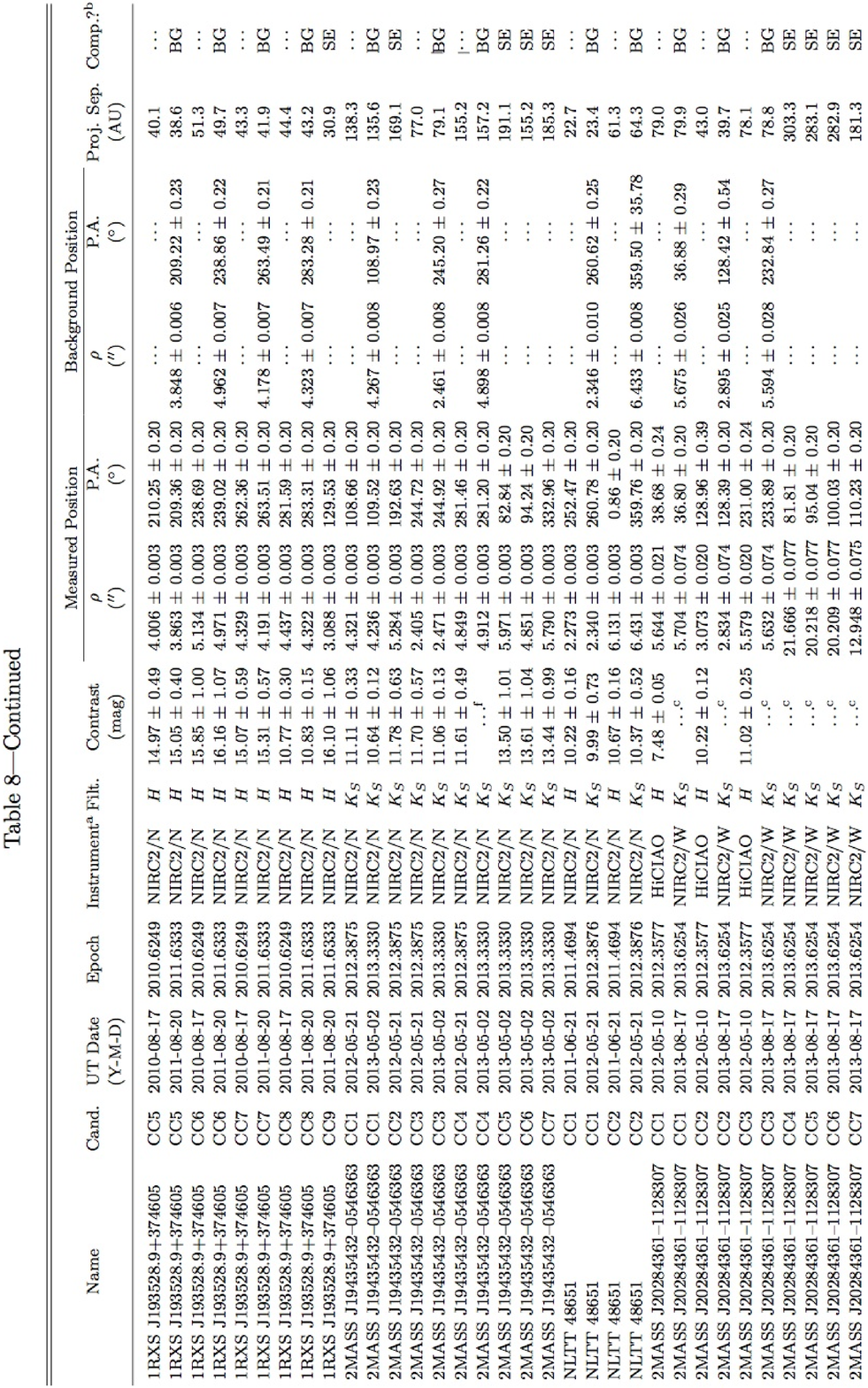}}
  \vskip 2in
  \caption{Table 8  \label{test}}
\end{figure}
\clearpage


\begin{figure}
  \vskip -.2in
  \hskip .3in
  \resizebox{6.5in}{!}{\includegraphics{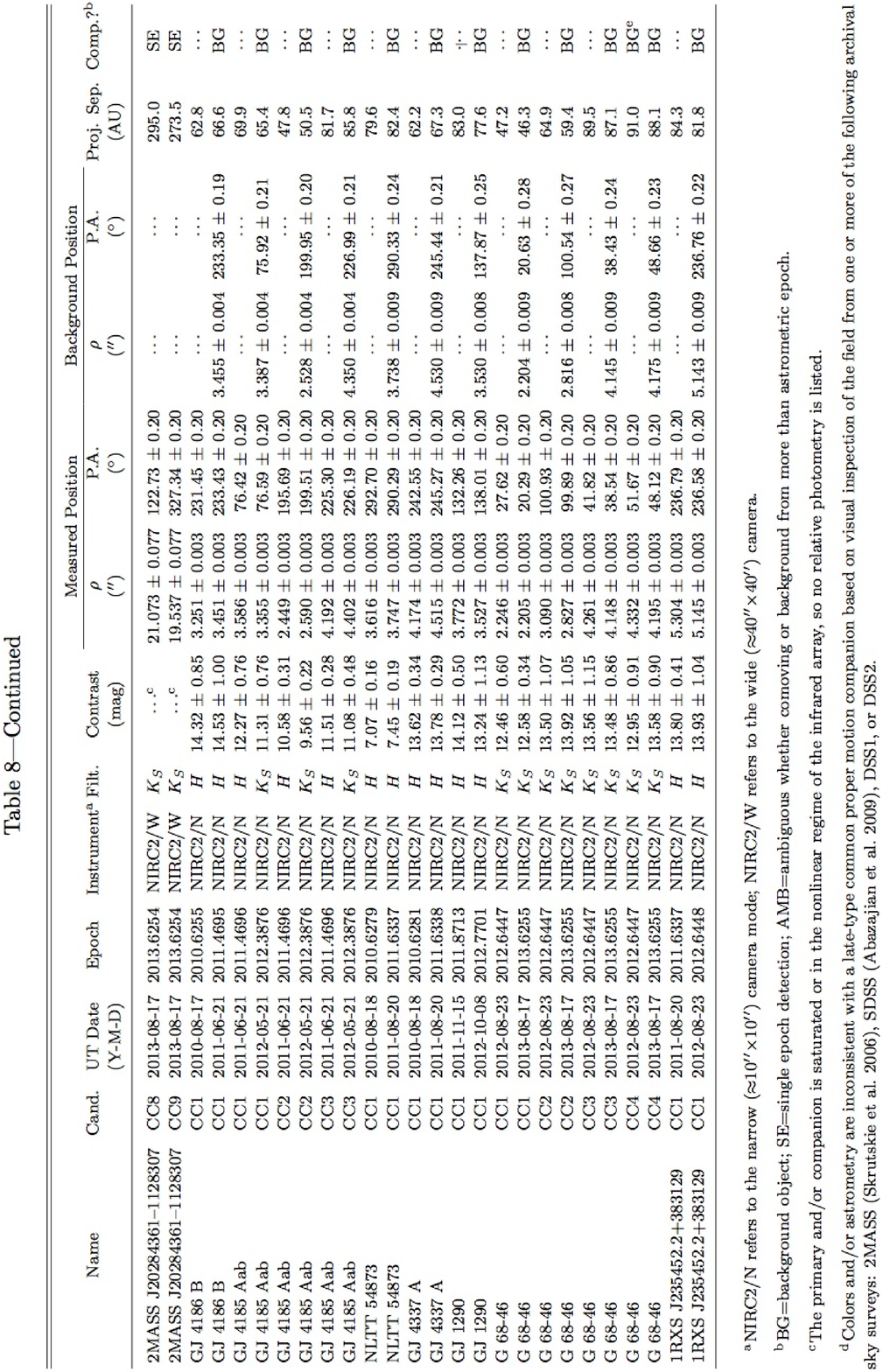}}
  \vskip 2in
  \caption{Table 8  \label{test}}
\end{figure}
\clearpage


\begin{figure}
  \vskip .2in
  \hskip 1.2in
  \resizebox{5.in}{!}{\includegraphics{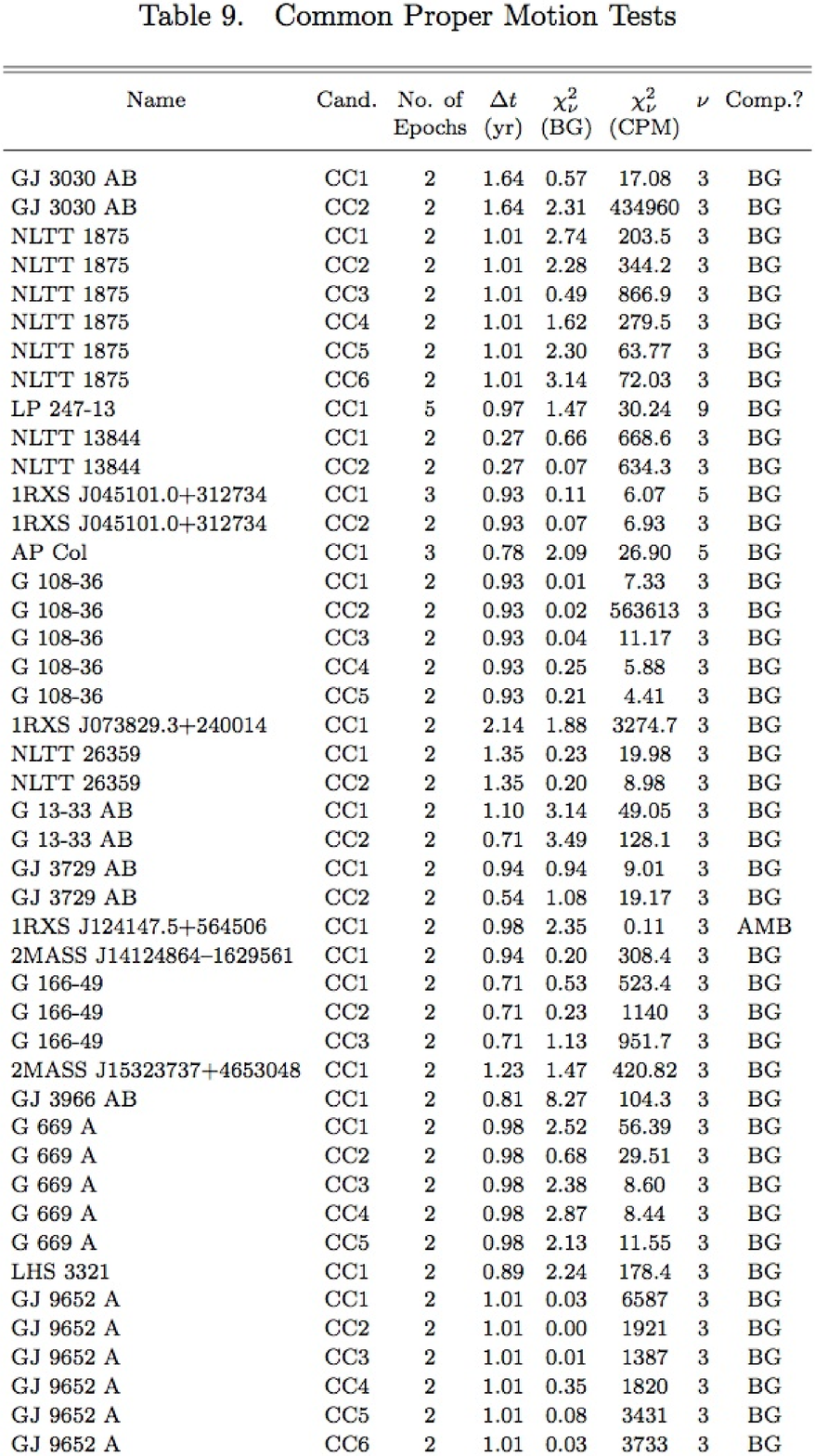}}
  \vskip 2in
  \caption{Table 9  \label{test}}
\end{figure}
\clearpage


\begin{figure}
  \vskip .2in
  \hskip 1.2in
  \resizebox{5.in}{!}{\includegraphics{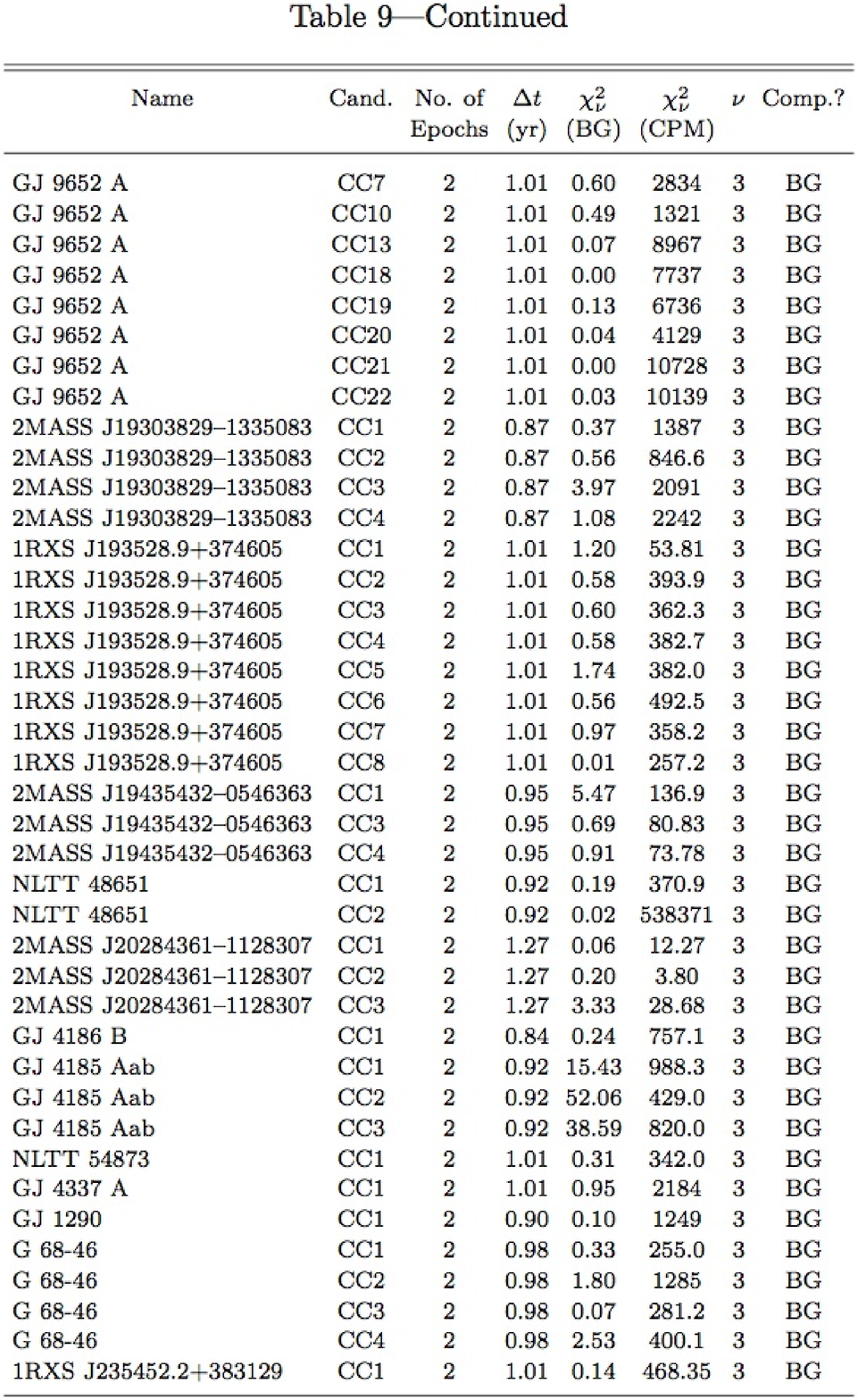}}
  \vskip 2in
  \caption{Table 9  \label{test}}
\end{figure}
\clearpage


\begin{figure}
  \vskip .2in
  \hskip .3in
  \resizebox{6.5in}{!}{\includegraphics{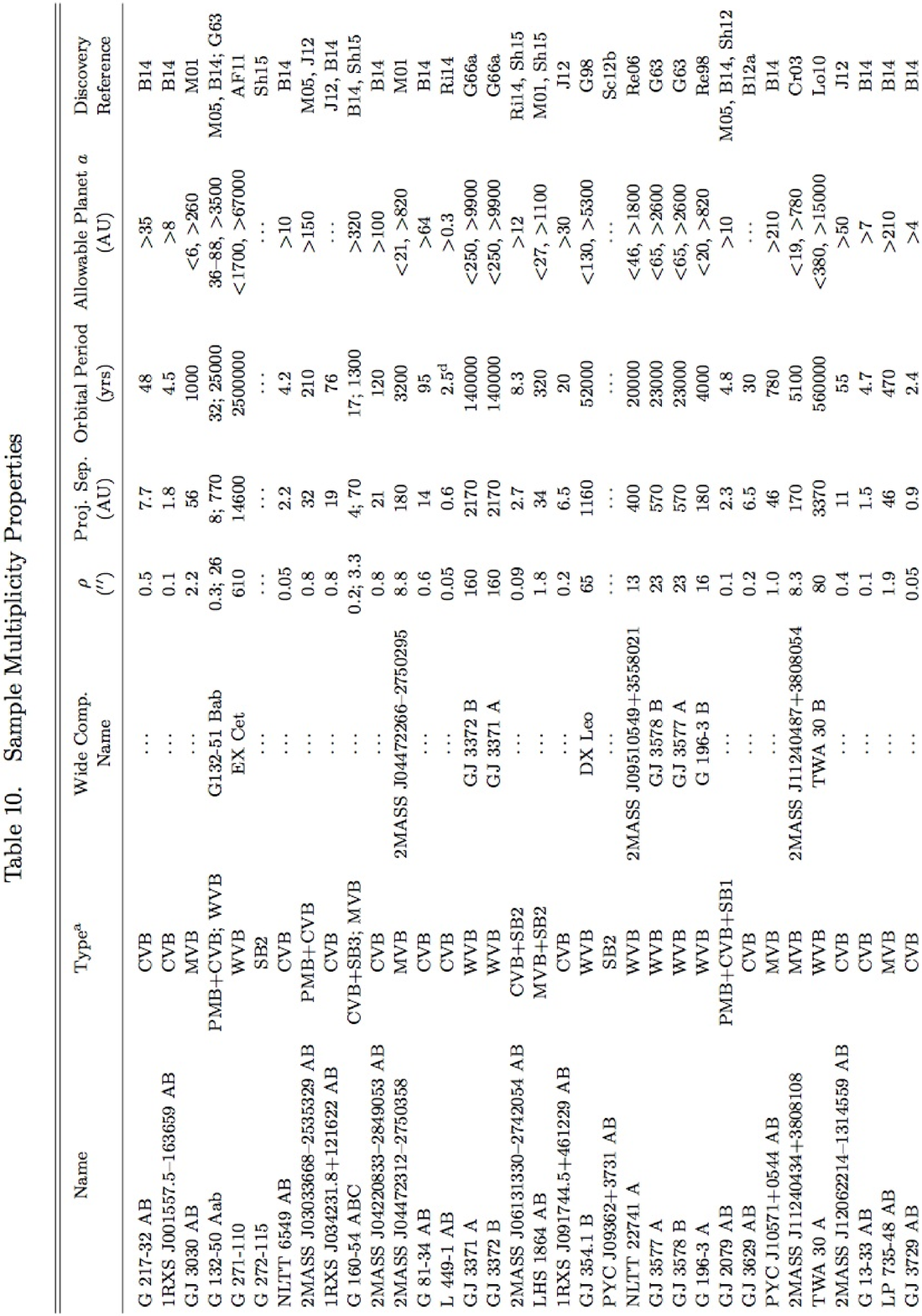}}
  \vskip 2in
  \caption{Table 10  \label{test}}
\end{figure}
\clearpage


\begin{figure}
  \vskip .2in
  \hskip .3in
  \resizebox{6.5in}{!}{\includegraphics{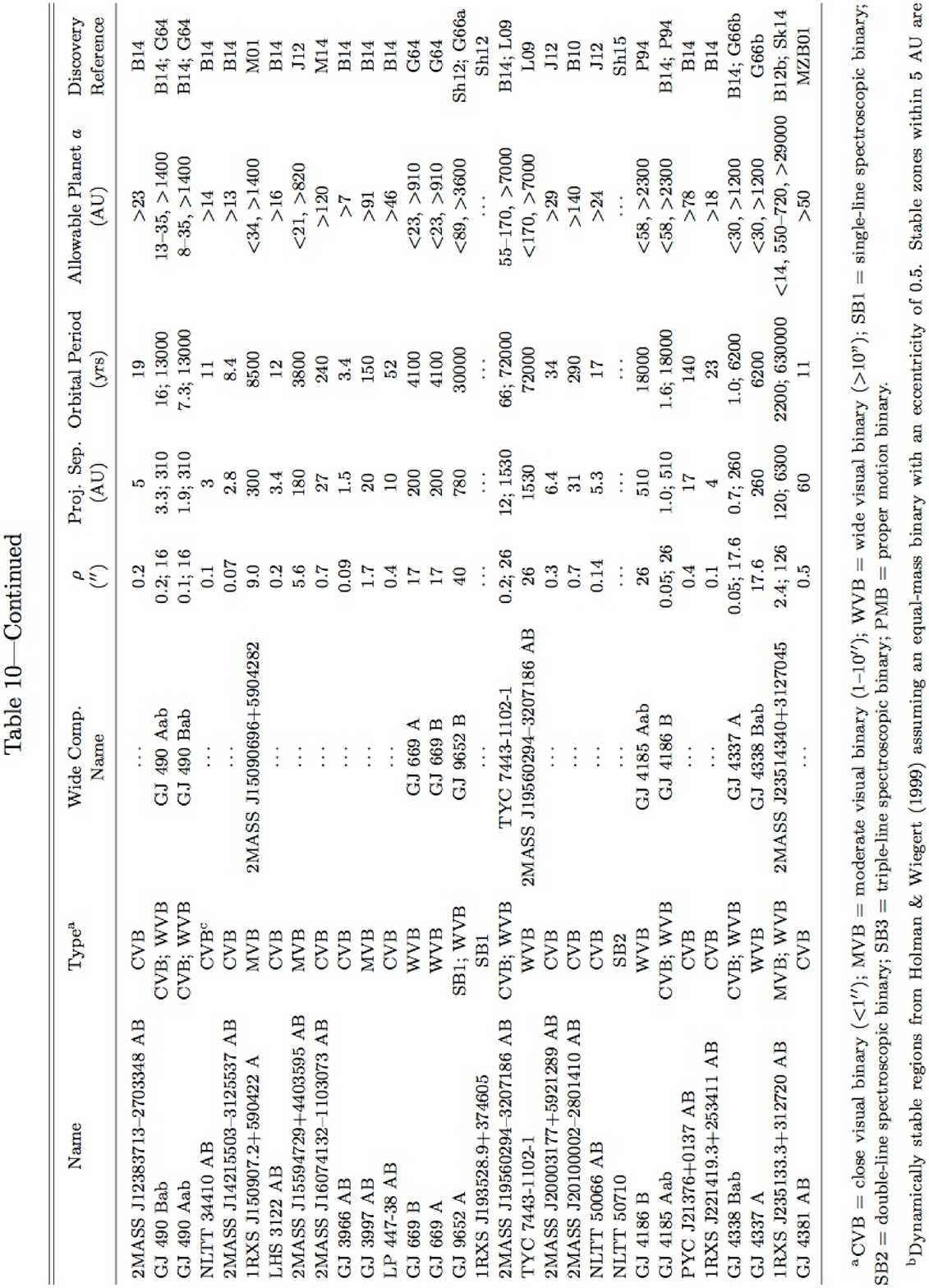}}
  \vskip 2in
  \caption{Table 10  \label{test}}
\end{figure}
\clearpage


\begin{figure}
  \vskip .2in
  \hskip .3in
  \resizebox{5.5in}{!}{\includegraphics{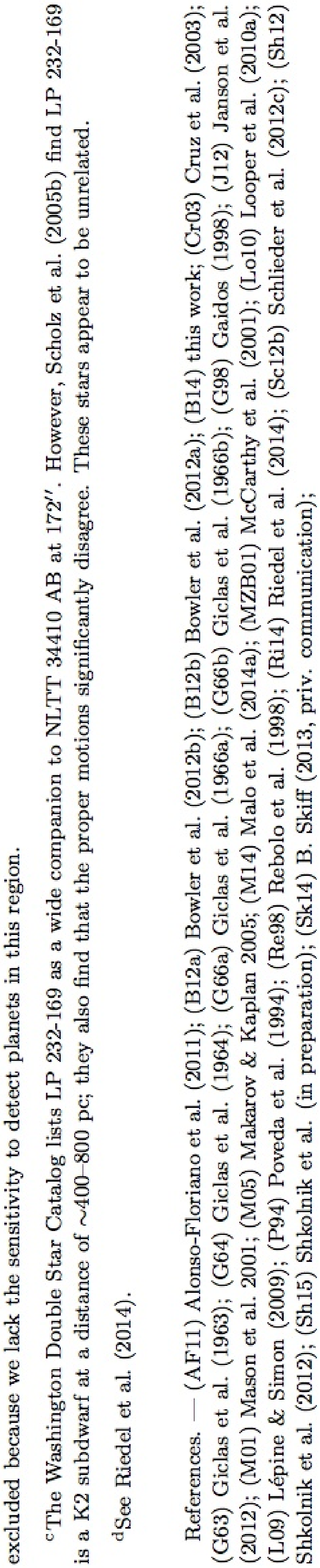}}
  \vskip 2in
  \caption{Table 10  \label{test}}
\end{figure}
\clearpage


\begin{figure}
  \vskip .3in
  \hskip .3in
  \resizebox{7in}{!}{\includegraphics{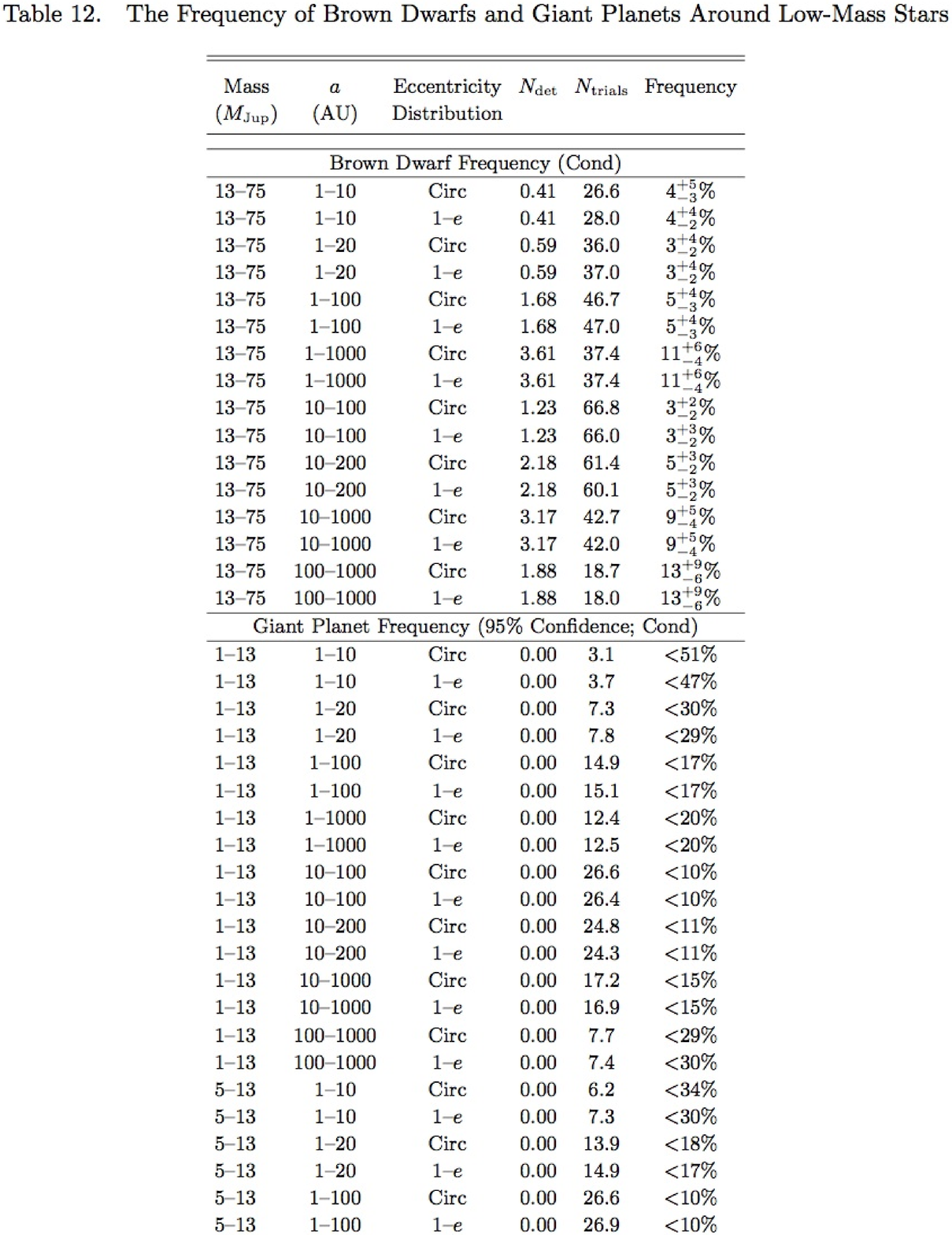}}
  \vskip 2in
  \caption{Table 12  \label{test}}
\end{figure}
\clearpage


\begin{figure}
  \vskip .3in
  \hskip 1.4in
  \resizebox{4.5in}{!}{\includegraphics{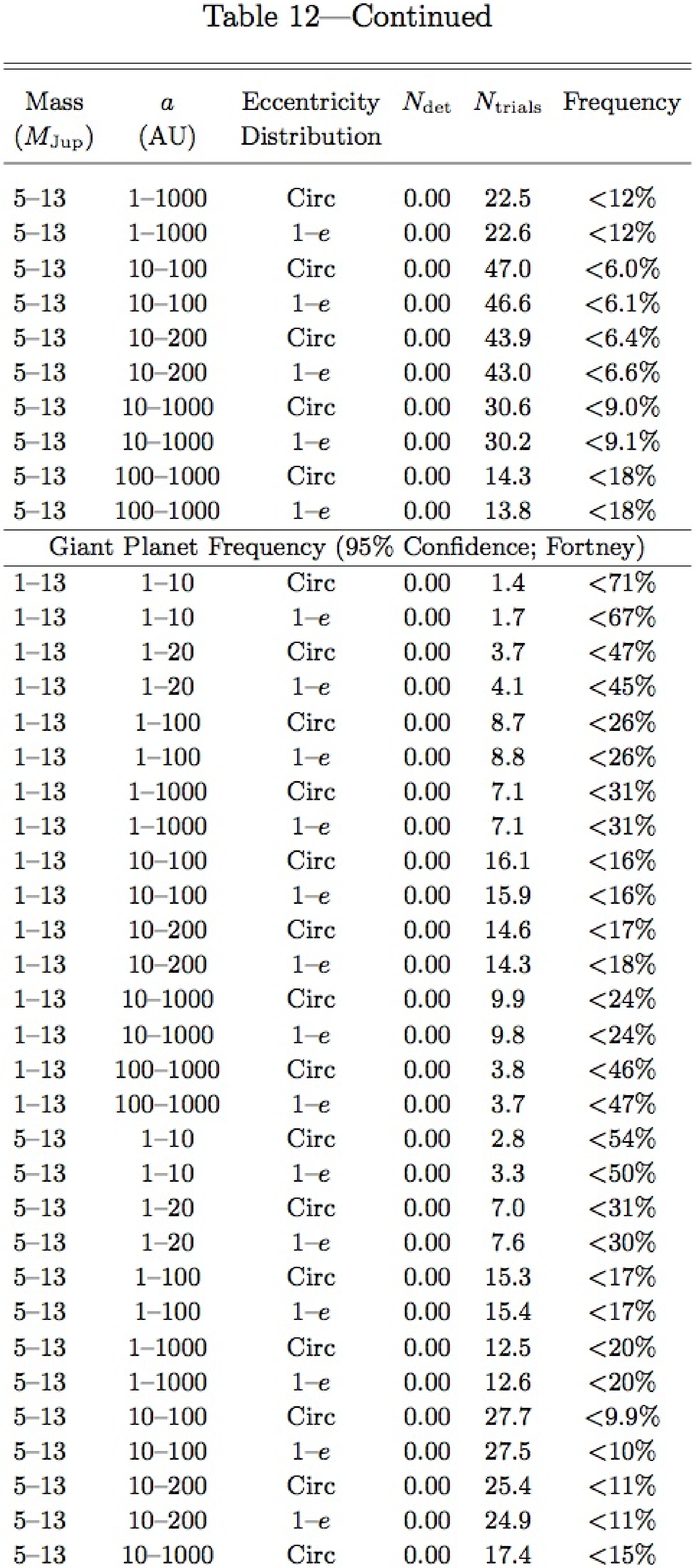}}
  \vskip 2in
  \caption{Table 12  \label{test}}
\end{figure}
\clearpage


\begin{figure}
  \vskip .3in
  \hskip 1.5in
  \resizebox{4.2in}{!}{\includegraphics{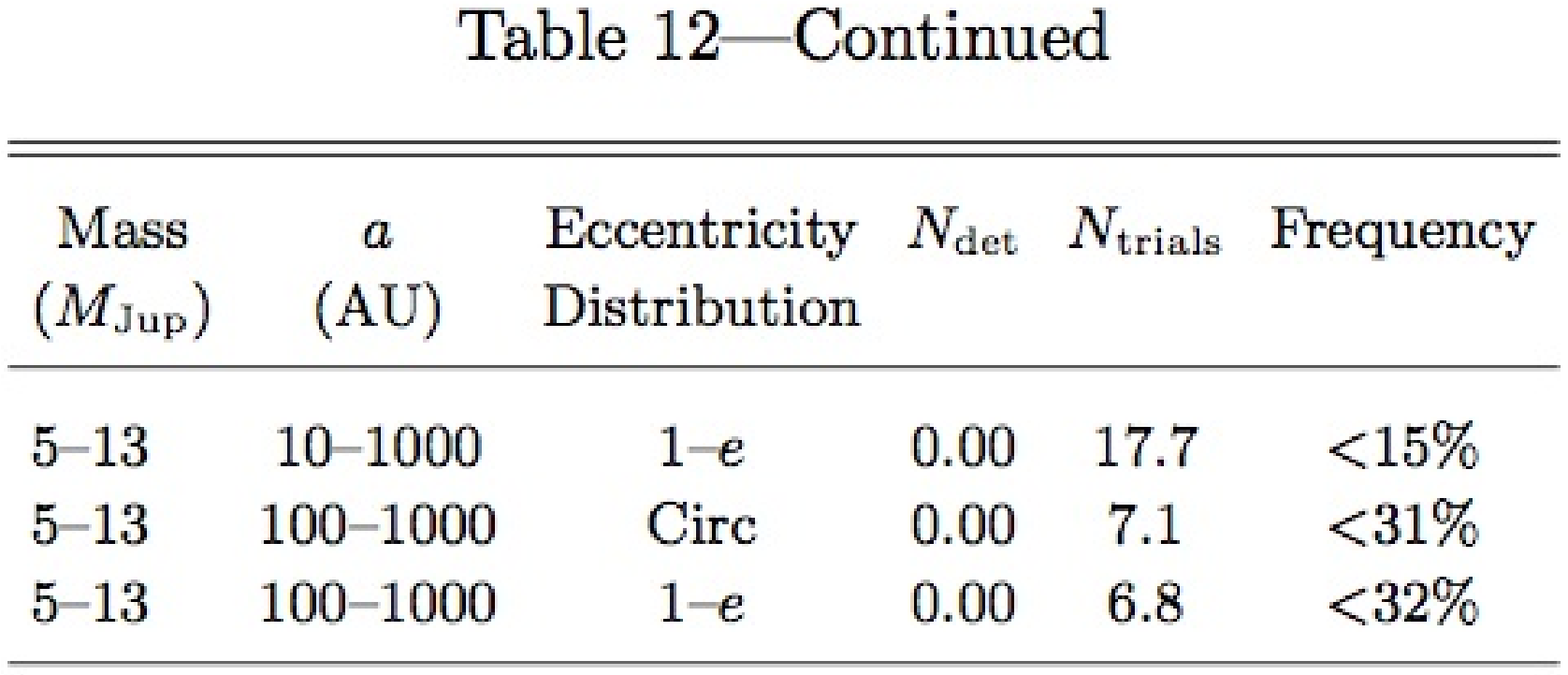}}
  \vskip 2in
  \caption{Table 12  \label{test}}
\end{figure}
\clearpage

\end{document}